\documentclass[10pt,a4paper,twocolumn,english,prb,aps,showpacs,floatfix,groupedaddress,superscriptaddress]{revtex4}

\usepackage{graphicx} 
\usepackage{epsfig} 
\usepackage[english]{babel}
\usepackage{amsmath} 
\usepackage{amssymb} 
\usepackage{amsfonts}
\usepackage{longtable} 
\usepackage{dcolumn} % Align table columns on decimal point
\usepackage{bm}
\usepackage{bbm}

\usepackage{subfigure}
\usepackage{tabularx}
\usepackage{color}

\DeclareMathOperator{\hc}{h.c.}  

\newcommand{\tr}[1]{{#1}}  

\begin{document}

\title{Self-consistent spin-wave theory for a frustrated Heisenberg model 
with biquadratic exchange in the columnar phase and its application to iron pnictides}

\author{Daniel Stanek} 
\email{stanek@fkt.physik.tu-dortmund.de}
\affiliation{Lehrstuhl f\"{u}r Theoretische Physik I, Technische Universit\"at
  Dortmund, Otto-Hahn Stra\ss{}e 4, 44221 Dortmund, Germany} 
\author{Oleg P. Sushkov}
\affiliation{School of Physics, University of New South Wales, Kensington
  2052, Sydney NSW, Australia} 
\author{G\"otz S. Uhrig}
\email{goetz.uhrig@tu-dortmund.de}
\affiliation{Lehrstuhl f\"{u}r Theoretische
  Physik I, Technische Universit\"at Dortmund, Otto-Hahn Stra\ss{}e 4, 44221
  Dortmund, Germany}

\date{\rm\today}

\begin{abstract}
  Recent neutron scattering studies revealed the three dimensional character
  of the magnetism in the iron pnictides and a strong anisotropy between the
  exchange perpendicular and parallel to the spin stripes. We extend 
  studies of the $J_1$-$J_2$-$J_c$  Heisenberg model with $S=1$ using
  self-consistent spin-wave theory. A discussion of two scenarios for the
  instability of the columnar phase is provided. The
  relevance of a biquadratic exchange term between in-plane nearest neighbors
  is discussed. We introduce mean-field decouplings for
  biquadratic terms using the Dyson-Maleev and the Schwinger boson
  representation. Remarkably their respective mean-field theories
  	 do not lead to the same results, even at zero temperature.
  	They are gauged in the N\'eel phase in comparison to
  exact diagonalization and series expansion. The $J_1$-$J_2$-$J_c$ model is analyzed
  under the influence of the biquadratic exchange $J_{\text{bq}}$ and a detailed description
  of the staggered magnetization and of the magnetic excitations is
  given. The biquadratic exchange increases the renormalization of the
  in-plane exchange constants which enhances the anisotropy between
  the exchange parallel and perpendicular to the spin stripes. Applying the
  model to iron pnictides, it is possible to
  reproduce the spin-wave dispersion for CaFe$_2$As$_2$ in the direction
  perpendicular to the spin stripes and perpendicular to the planes.
  Discrepancies remain in the direction parallel to the spin stripes
  which can be resolved by passing from $S=1$ to $S=2$. In
  addition, results for the dynamical structure factor within
  the self-consistent spin-wave theory are provided.
\end{abstract}


\maketitle

\section{Introduction}

\subsection{General Context}

\begin{figure}[htb]
  \centering 
  \includegraphics[width=.7\columnwidth]{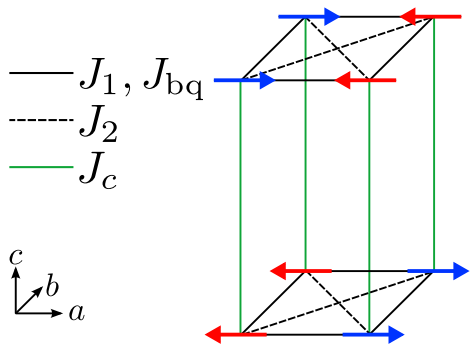}
  \caption{(Color online) Three dimensional collinear phase with antiferromagnetic
    interlayer exchange $J_c$ and biquadratic exchange $  J_{\text{bq}}$ between in-plane
    nearest neighbor sites.}
    \label{fig:model_col3D}
\end{figure}

Frustrated quantum antiferromagnetism 
has been a very active field of research over
the last 15 years which influences many related fields as well 
\cite{lacro11}. The frustration enhances the importance of
quantum effects because classical order is suppressed. Hence
new and unexpected phases may occur and govern the physics at low energies.
Among the models studied intensely is the $J_1$-$J_2$ model
on a square lattice
\cite{xu90,irkhi92,singh99b,kotov99b,sushk01,singh03,bisho08b,uhrig09a,majum10}
and its generalization to three dimensions
by stacking planes \cite{schma06,yao10,majum11a,majum11b,holt11}. 
Theoretically, two key issues are (i)
for which parameters classically ordered phases occur and (ii) whether
there exists a quantum disordered phase between the classically
ordered phases. The two long-range ordering patterns are
either the alternating order with staggered, N\'eel type sublattice 
magnetization or a columnar antiferromagnetic ordering where
the adjacent spins in one spatial direction (direction $a$ in one plane
of Fig.\ \ref{fig:model_col3D}) are aligned antiparallel
while they are aligned parallel in the other spatial
direction (direction $b$ in Fig.\ \ref{fig:model_col3D}).

The $J_1$-$J_2$ Heisenberg model
\begin{align}
  H&=J_1\sum\limits_{\langle i,j\rangle}\mathbf{S}_i\cdot\mathbf{S}_j+J_2\sum\limits_{\langle\langle i,j\rangle\rangle}\mathbf{S}_i\cdot\mathbf{S}_j,\label{eq:heisenberg_nnn}
\end{align} 
for $S=1/2$ and its ground states are of broad interest in
solid state physics. The ground state of the simple Heisenberg 
model with $J_2=0$ on the square lattice is the N\'eel order 
with staggered magnetization reduced by quantum fluctuations \cite{betts77,manou91,auerb94}. 
For $J_2\neq 0$, the ground state depends on the ratio $J_2/J_1$
of the couplings. On increasing $J_2/J_1$ a value is hit where the N\'eel 
phase becomes unstable towards a quantum disordered state. The
intermediate phase is stable in the range of $0.4\lesssim J_2/J_1\lesssim 0.6$
and is dominated by short-range singlet (dimer) formation
\cite{singh99b,sushk01}.  For $J_2/J_1>0.6$, the spins
arrange in a columnar pattern.  In the classical limit $S\rightarrow\infty$,
the transition between the N\'eel and columnar order occurs at $J_2/J_1=0.5$,
see Ref.\ \onlinecite{chand88}.

Singh \textit{et al}. \cite{singh03} studied the excitation
spectra of the columnar phase with series expansion and \tr{self-consistent} spin-wave
theory.  They calculated the spin-wave velocities which depend strongly on the
coupling ratio $J_2/J_1$. Gapless excitations are only found at $k=(0,0)$ and
$k=(1,0)$, while the modes at $k=(0,1)$ and $k=(1,1)$ are gapped because of
the order by disorder effect. \tr{We stress that the columnar phase
is very well described by self-consistent spin-wave theory  \cite{singh03,uhrig09a}
even in two dimensions and for $S=1/2$.
The stability of the N\'eel phase, however, is overestimated
by self-consistent spin-wave theory \cite{xu90,irkhi92} so that the intermediate
disordered phase is missed. This intermediate phase is seen by a direct
second order perturbative approach \cite{majum10} in $1/S$ but only for spatially 
anisotropic models with $J_{1a}\neq J_{1b}$ where $J_{1x}$ is the
nearest-neighbor exchange in $x$-direction with $x\in\{a,b\}$.
The intermediate phase is not the issue of the present paper and
we point out that it is to be expected that it is hardly relevant
in three dimensions and for $S\ge 1$ \cite{schma06,bisho08b,majum11a}.}

Note that we here and henceforth give all wave vectors in units
of $\pi/a$ where $a$ is the corresponding lattice constant.
This model was applied to the magnetic excitations
of undoped iron pnictides 
\cite{yao08,uhrig09a,yao10,apple10,schmi10,smera10,majum11a,majum11b}.

It was advocated by Chandra \textit{et al}. \cite{chand90}
that stripe order can occur at finite temperatures in isotropic frustrated 
Heisenberg models because the stripe order breaks a discrete symmetry,
namely rotation of the lattice by 90$^\circ$, which is not protected
by the Mermin-Wagner theorem \cite{mermi66}. Indeed the transition is
of Ising type. This result was corroborated
by classical \cite{weber03} and semi-classical \cite{capri04} numerics.
Its significance for the structural and the magnetic phase transitions in
the pnictides was noted in Refs.\ \onlinecite{xu08} and \onlinecite{uhrig09a}.

For completeness, we also mention the additional instability for ferromagnetic
couplings at $J_1\approx -2J_2$, where the system undergoes a 
phase transition from the collinear ordered state to a ferromagnetic state
\cite{shann06}.

Since the discovery of superconductivity upon doping of the iron-based
compound LaOFeAs \cite{kamih08}, the $J_1$-$J_2$ Heisenberg model
has been used to study the magnetism in the parent compounds of the iron
pnictides \cite{yao08,uhrig09a,yao10,apple10,holt11} \tr{although it does not
take the remaining itineracy of the charges into account}. The magnetic long-range order
is indeed a columnar phase where the spins at the iron sites show antiferromagnetic
order in $a$-direction and a ferromagnetic order in $b$-direction (see
Fig. \ref{fig:model_col3D}). Between the layers ($c$-direction), the spins are
also aligned antiparallel \cite{cruz08,diall09,zhao09}. The columnar
order of the spins is also supported by the results of band structure
calculations \cite{ren08,yin08}.

A related question still under debate is the reduced size of the local
magnetic moment. Neutron scattering studies find a reduced staggered
magnetic moment of (0.31-0.41)$\mu_\text{B}$ for LaOFeAs \cite{cruz08},
(0.8-0.9)$\mu_\text{B}$ for the 122 pnictide BaFe$_2$As$_2$
\cite{ewing08}, (0.90-0.98)$\mu_\text{B}$ for SrFe$_2$As$_2$
\cite{zhao08c} and 0.8$\mu_\text{B}$ for CaFe$_2$As$_2$
\cite{mcque08,zhao09}. In contrast, theoretical band
structure calculations determined much higher values, e.g., 
up to $2.3\mu_\text{B}$ for LaOFeAs \cite{yin08}.

One conceivable explanation is that the reduction of the local moment 
is due to a strong frustration within a localized model
\cite{si08,yao08,uhrig09a,ong09,yao10,holt11}.
This requires the system to be in the immediate
vicinity of a quantum phase transition. But it 
must be kept in mind that the local magnetic moment depends 
on matrix elements which are well known only in the limit of
localized electrons. Thus, it is possible that the significant reduction of
the staggered magnetization is due to electronic effects such as hybridization,
spin-orbit coupling  \tr{and the itineracy of the charges}
\cite{yildr08,wu08a,lv10,pulik10}. The bottomline
of this argument is that the value of the staggered
magnetization is not a stringent 
constraint for the applicable model for iron pnictides.

\subsection{Spatial Anisotropy of Exchange Couplings}

\label{sect:biquadratic}

Recently, it was shown that a frustrated Heisenberg model in the three
dimensional columnar phase \cite{holt11} can explain the spin-wave
dispersion of CaFe$_2$As$_2$ \cite{mcque08,diall09,zhao09} 
in the direction perpendicular to the spin stripes and between the layers. 
However, discrepancies at high energies are present in the direction 
parallel to the stripes. These discrepancies can be fitted by
a Heisenberg model which assumes that the nearest neighbor (NN) coupling $J_1$
depends on the spatial direction of the two coupled spins, i.e.,
one introduces $J_{1a} \gg J_{1b}$ \cite{zhao08a,diall09,zhao09,han09}.
This is very remarkable since the difference of the 
orthorhombic distortion between the 
lattice constants $a$ and $b$ of the columnarly ordered layers is less than 1\%
\cite{ewing08,diall09}, which by far does not give reasons for
the large spatial anisotropy of the magnetic exchange. This argument
is further supported by the observation that density functional calculations
achieve a good description of the pnictides if magnetic columnar order is
accounted for. But the consideration of the orthorhombic distortions
is only a minor point \cite{han09}.

Another possible explanation of the spatial anisotropy is the
possibility of orbital ordering \cite{singh09a}. But orbital ordering
is usually related to higher energies and should lead to clear experimental
signatures or theoretical signatures in density-functional calculations
which are so far not found.

So it appears that the magnetism itself generates the spatial anisotropy
although the original magnetic model is not anisotropic. Indeed, the
anisotropic order generates some anisotropic velocities in self-consistent
spin-wave theory \cite{uhrig09a,holt11}. But this effect is not sufficient
\cite{holt11} to account for $J_{1a}\approx 40$ meV and $J_{1b}\approx 0$ meV 
\cite{zhao09,han09}.

In order to identify a magnetic process which is able to generate the
observed spatial anisotropy one has to go beyond bilinear exchange.
This is possible in view of the larger spin value $S\ge 1$. 
Indeed, significant higher spin exchange processes 
are inevitable \cite{girar77a,mila00}.  To be specific, we will consider 
the biquadratic exchange 
\begin{align}
  H_\text{bq}&=-  J_{\text{bq}}\sum\limits_{\langle i,j\rangle}\left(\mathbf{S}_i\cdot\mathbf{S}_j\right)^2 
  \label{eq:Hbiquadratic}
\end{align}
with $  J_{\text{bq}}>0$. How does a term such as $H_\text{bq}$ influence
the magnetic excitations?
To obtain a rule of thumb we adopt a simple view 
and approximate
\begin{align}
\nonumber
  -  J_{\text{bq}}\sum\limits_{\langle i,j\rangle}\left(\mathbf{S}_i\cdot\mathbf{S}_j\right)^2 \approx -2  J_{\text{bq}}\sum\limits_{\langle i,j\rangle} \mathbf{S}_i\cdot\mathbf{S}_j \langle \mathbf{S}_i\cdot\mathbf{S}_j\rangle\\
  	+    J_{\text{bq}}\sum\limits_{\langle i,j\rangle} \langle \mathbf{S}_i\cdot\mathbf{S}_j\rangle^2.
  \label{eq:bq_simple}
\end{align}
In the above formula we do not list terms of the types 
$S_i^{(\gamma)}S_i^{(\delta)} \langle S_j^{(\gamma)}S_j^{(\delta)} \rangle$
or $\langle S_i^{(\gamma)}S_i^{(\delta)}\rangle \langle S_j^{(\gamma)}S_j^{(\delta)} \rangle$
because both would only contribute for $\gamma=\delta\in\{x,y,z\}$ and then
their sum over all spin components would merely yield trivial constants.
In the direction of alternating spin orientation
one has $\langle \mathbf{S}_i\cdot\mathbf{S}_j\rangle < 0$ so that the biquadratic
term effectively \text{strengthens} the bilinear
antiferromagnetic exchange: $J_{1a}^\text{eff}> J_1$. 
In contrast, in the direction of parallel spin orientation
one has $\langle \mathbf{S}_i\cdot\mathbf{S}_j\rangle > 0$ so that the biquadratic
term effectively \text{weakens} the bilinear
antiferromagnetic exchange: $J_{1b}^\text{eff} < J_1$. 
Hence a NN biquadratic exchange
appears to generate a spatial anisotropy purely from
the magnetic order. One of the two main goals
of the present paper is to substantiate this
argument phenomenologically.

If derived from extended standard Hubbard models
the biquadratic terms are of higher order in the intersite hoppings
than the bilinear exchange coupling which implies that it is 
actually small relative to the bilinear exchange \cite{mila00}. 
The exception is a situation where the bilinear terms are subject 
to near cancellation of antiferromagnetic and ferromagnetic contributions. 

In the pnictides, even the undoped systems are conducting bad metals.
It is known that close to the transition from localized to conducting systems
higher order processes start playing a significant role. For
instance, the cyclic exchange in one-band Hubbard models
becomes as large as 20\% of the NN exchange, see for instance
Ref.\ \onlinecite{hamer10} and references therein.

\begin{figure}[htb]
  \centering
  \includegraphics[width=\columnwidth]{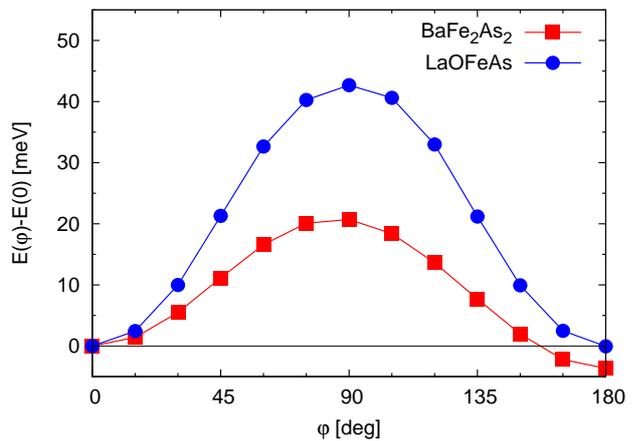}
  \caption{(Color online) Dependence of the ground state energy on the relative angle $\varphi$ 
  between 	the two sublattice magnetizations on the two
  intercalated N\'eel lattices, after Ref.\ \onlinecite{yares09}.}
  \label{fig:intro_angle_dep}
\end{figure}

In addition, evidence for biquadratic exchange has already appeared in
calculations based on the self-consistent local spin-density approximation 
(LSDA) by Yaresko \textit{et al.} \cite{yares09}. Results
for the dependence of the ground state energy of two intercalated
N\'eel ordered sublattices on the angle $\varphi$ between their
sublattice magnetizations are displayed in Fig.\ \ref{fig:intro_angle_dep}. 
With bilinear exchanges no dependence
is expected at all. The results for BaFe$_2$As$_2$ and LaOFeAs are are reproduced in
Fig. \ref{fig:intro_angle_dep}. The dependence of
$E(\varphi)=A\cdot\sin^2\varphi$ on $\varphi$ is compelling evidence for a
biquadratic NN exchange because a dependence $\propto\sin^2\varphi$ does not occur in the frustrated $J_1$-$J_2$~Heisenberg model. Appropriate values for the biquadratic exchange are determined from the maximum $E(\varphi=90^\circ)$ in
Fig. \ref{fig:intro_angle_dep} for $S=1$. The obtained values are
\begin{subequations}
\begin{align}
    J_{\text{bq}}&=21.5\ \text{meV} \ \text{for LaOFeAs}, \\
    J_{\text{bq}}&=10.1\ \text{meV} \ \text{for BaFe$_2$As$_2$},
\end{align}
\end{subequations}
which are indeed sizeable in view of $J_1$ of the order $40$ meV.

In general, the electronic situation in the iron pnictides is very complex 
\cite{wu08a}; up to five bands are important \cite{yildr08,yin08}. 
The appropriate electronic description
has not yet been identified \cite{grase10,wu08a,schic11} and thus it is 
presently not possible to discuss the relative size of a biquadratic
exchange reliable. Hence we adopt here the phenomenological
approach to take the microscopic arguments as evidence
for the existence of such a biquadratic exchange. Next, the aim
will be to derive the size of $  J_{\text{bq}}$ from fits to experiment.

\subsection{Methods}

Based on our previous work \cite{uhrig09a,holt11}, we study the
$J_1$-$J_2$-$J_c$ model in the three-dimensional phase
with columnar spin order. First, we
discuss the ``critical'' value of $x=J_1/J_2$ where the columnar phase ceases
to exist and the corresponding two scenarios for this instability. 
Furthermore, we extend the
$J_1$-$J_2$-$J_c$ model by the biquadratic exchange discussed above. 
To this end, an appropriate mean-field decoupling has to be
introduced which is able to tackle biquadratic exchange as well. 

To establish a reliable mean-field approach we employ the Dyson-Maleev as well as the Schwinger bosons representation. To gauge the resulting decoupling schemes
both are applied to the two dimensional N\'eel phase of
a NN bilinear Heisenberg model plus biquadratic
exchange for $S=1$. Then, the successful decoupling is  applied to the
three dimensional columnar phase with biquadratic exchange for $1\le S\le 2$. 
General results are discussed before we  apply the model to CaFe$_2$As$_2$.

\section{Scenarios for the $J_1$-$J_2$-$J_c$ model}

\label{sect:scenarios}

Here we discuss two fundamental scenarios for the instability
of spatially anisotropic phases of magnetic long-range order.
We do not consider first-order transitions to other phases;
\tr{in particular we do not aim to discuss the existence of possible
intermediate disordered phases between N\'eel and columnar phase.}
Instead we focus on how the columnar phase can become unstable
towards fluctuations. \tr{We stress that the self-consistent spin-wave
theory is expected to yield reliable results for $S\ge 1>1/2$ and dimension
$d=3>2$ since its results are already very good \cite{singh03,uhrig09a} 
for $S=1/2$ and $d=2$.}

It is generally agreed that the staggered magnetization in the columnar
phase of two-dimensional $J_1$-$J_2$ model can take any possible value 
provided an adequate fine-tuning of the coupling ratio $x=J_1/J_2$ is performed
\cite{uhrig09a,smera10}.  Here we point out that this is no longer
true in the presence of an interlayer coupling $J_c$.
The quantitative aspects, though not the qualitative ones,
depend strongly on the spin $S$.

Passing from two to three dimensions the additional coupling between the columnarly
ordered planes cuts off the logarithmic divergence of the Goldstone modes \cite{smera10}. Thus, in three dimensions
the staggered magnetization may no longer adopt any arbitrary value $\ge 0$
even if the ratio $x=J_1/J_2$ is increased towards 2. Instead,
there can be  a finite minimum value of the staggered magnetization 
whose value depends on the relative interlayer coupling $\mu=J_c/J_1$.

\begin{figure}[ht]
  \centering
  \includegraphics[width=.99\columnwidth]{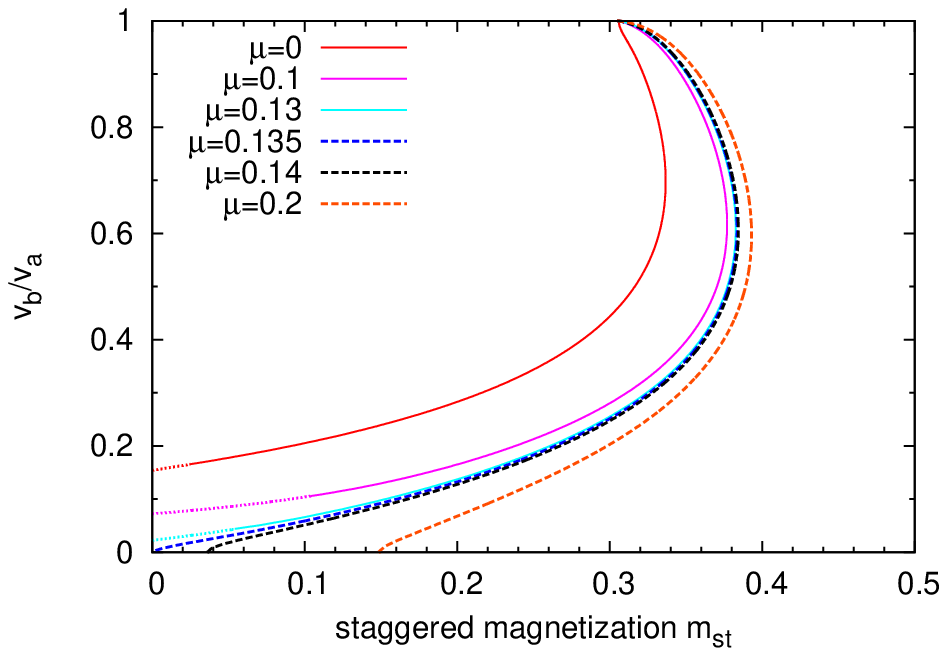}
  \includegraphics[width=.99\columnwidth]{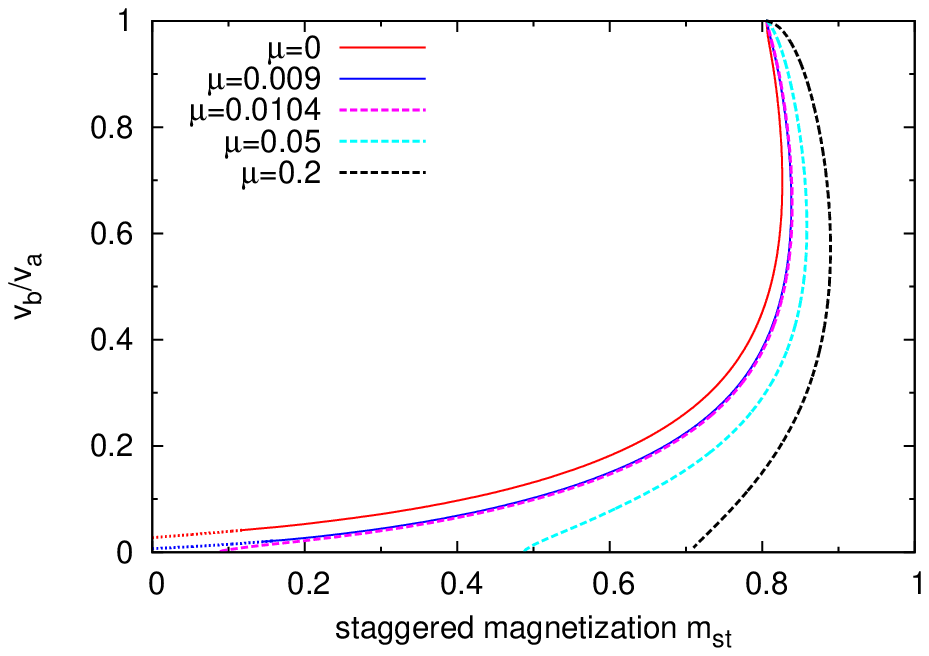}
   \caption{(Color online) Ratio $v_b/v_a$ as a function of $m_\text{st}$ for $S=1/2$ (upper panel)
   	and  $S=1$ (lower panel). The curves either end on the ordinate, i.e,
   	for zero staggered magnetization, or on the abscissa, i.e., for zero
   	spin-wave velocity $v_b < v_c < v_a$. Thus, two qualitatively
   	different instabilities emerge naturally. The solid lines
    correspond to the magnetization-driven scenario (vanishing magnetization)
    while the dashed lines correspond to the velocity-driven scenario 
    	(vanishing spin-wave velocity). The dotted lines are extrapolations in the
    magnetization-driven scenario.} 
    \label{fig:scenarios}
\end{figure}

There are two possible outcomes upon $x\to 2$.
If the value of $\mu$ is not too large, it is still possible to drive the
magnetization to zero while all the three different spin-wave
velocities remain finite. But for larger values of $\mu$ the magnetization
remains finite while the smallest of the spin-wave velocities vanishes 
\cite{holt11}. Thus, we face two qualitatively different scenarios. 
They can easily be distinguished in plotting the ratio $v_b/v_a$ 
versus the staggered magnetization $m_\text{st}$. 
The results for $S=1/2$ and $S=1$ are shown in Fig.\ \ref{fig:scenarios}.

In the first scenario (solid lines), the ratio $v_b/v_a$ always stays positive and the instability of the phase is marked by the staggered magnetization $m_\text{st}$ going to zero. Thus, we call this scenario the magnetization-driven scenario because the breakdown
of the columnar phase is driven by the vanishing magnetic long-range
order. The numerical evaluation of the three dimensional integrals
becomes very difficult close to $m_\text{st}=0$ so that the solid
curves in Fig.\ \ref{fig:scenarios} can only be computed down to
some small finite value of the staggered magnetization. Beyond
this point they are extrapolated as depicted by the dotted curves.
In this way the phase diagrams of the $J_1$-$J_2$-$J_c$ Heisenberg model 
in Ref.\ \onlinecite{holt11} were determined.

The dashed lines in Fig.\ \ref{fig:scenarios} correspond to the second
scenario.  We call it the velocity-driven scenario because the vanishing of
the spin-wave velocity $v_b$ marks the instability of the columnar phase.
If $x$ were taken infinitesimally higher, $v_b^2$ would become negative
so that no physically meaningful mean-field model would be found.
The staggered magnetization $m_\text{st}$ stops at this point
at a finite positive value.  This finite positive value of the staggered
magnetization may appear surprising because it signifies that the
 magnetic long-range order still
persists even though the columnar phase ceases to exist
because its excitations become unstable.

In the case of $S=1/2$, the transition from the magnetization- to the
velocity-driven scenario occurs for $\mu_\text{mv} \approx 0.134$. 
Thus, the magnetization-driven scenario applies in the range $0\le\mu\lesssim 0.134$. The precise value of $\mu$ where the magnetization-driven scenario
switches over to the velocity-driven scenario corresponds to the
curve in the $v_b/v_a$-$m_\text{st}$ plane which ends at the origin.
If the interlayer coupling is increased beyond $\mu_\text{mv}$,
the velocity-driven scenario applies.

For $S=1$, the change from one scenario to the other
occurs roughly at $\mu_\text{mv}\approx 0.01$, see lower panel
of Fig.\ \ref{fig:scenarios}. Thus, the
magnetization-driven scenario for $S=1$ persists in the range of
$0\le\mu\lesssim 0.01$, which is a much smaller region compared to the case for
$S=1/2$. We stress, however, that there is no 
qualitative difference between $S=1/2$ and $S=1$.
For even larger spin values the magnetization-driven scenario
will only apply for a quickly decreasing range of interplane couplings
$\mu=J_c/J_1$. This is due to the fact that the 
instability of the columnar phase shifts for increasing spin
very quickly to where it occurs classically \cite{chand88}, i.e., for $x=2$.
Since larger and larger spins correspond to a more and more classical
behavior it is not astounding that one has to draw nearer and 
nearer to the logarithmic divergence of the quantum
fluctuations in order for them to become important.
Hence $S\to\infty$ implies an exponential convergence 
$x\to 2$ and $\mu_\text{mv}\to 0$.

We emphasize that the full discussion of the instability of the columnar
phase requires the consideration of first order transitions as well.
This is beyond the scope of the present paper because we do not know
the phase into which the columnar phase becomes unstable. 
\tr{It may be a disordered phase, but in view of the larger spin $S\ge 1$ and
the dimension $d=3$ we expect that an intermediate phase
exists only for a very small parameter region.} 

From the knowledge for $S=1/2$ and $S=1$ in two and three dimensions 
\cite{singh99b,kotov99b,sushk01,singh03,schma06,bisho08b,majum10,majum11a}
we \tr{presume} that the magnetization-driven scenarios eventually becomes a
weak first order transition \tr{to an intermediate phase existing
only within a small parameter region}. 
Further, we \tr{expect} that the velocity-driven
scenario becomes a strong first order transition \tr{to the N\'eel phase}.

\section{Biquadratic Exchange in the 2D N\'eel Phase}

In this section, we study the effects of a biquadratic exchange \eqref{eq:Hbiquadratic}
on the simple NN Heisenberg model ($J_2=0$)
with $S=1$ on the square lattice. The aim is to establish
a reliable mean-field description.

We apply the Schwinger bosons as well as the Dyson-Maleev
representation to the model and introduce the corresponding 
mean-field approximation.  
Our aim is to study the influence of the biquadratic exchange on the
dispersion of the spin-waves. The findings are checked
against results obtained by exact diagonalization and series expansion.  

The Hamiltonian under study reads
\begin{align}
  H=&J\sum\limits_{\langle i,j\rangle} \mathbf{S}_i\cdot\mathbf{S}_j-  J_{\text{bq}}\sum\limits_{\langle i,j\rangle}
  \left(\mathbf{S}_i\cdot\mathbf{S}_j\right)^2, \label{eq:H_Neel}
\end{align}
where $J,  J_{\text{bq}}>0$. The brackets $\left< i,j\right>$ indicate the summation over
NN sites.

\subsection{Dyson-Maleev representation}

First, we apply the Dyson-Maleev transformation
\cite{dyson56a,dyson56b,malee58} to the Hamiltonian \eqref{eq:H_Neel}. On
sublattice $A$, the expression of the spin operators in terms of bosonic
operators reads
\begin{subequations} \label{eq:DM_subA}
  \begin{align}
    S^+_i&=b^\dagger_i\left(2S-\hat{n}^{\phantom{\dagger}}_i\right) \\
    S^-_i&=b_i^{\phantom{\dagger}} \\
    S^z_i&=-S+\hat{n}_i^{\phantom{\dagger}} .
  \end{align}
\end{subequations}
After a $\pi$-rotation of the spins around the $x$-axis, the transformation on
sublattice $B$ is given by
\begin{subequations} \label{eq:DM_subB}
  \begin{align}
    S^+_i&=b^\dagger_i  \\
    S^-_i&=\left(2S-\hat{n}^{\phantom{\dagger}}_i\right)b_i^{\phantom{\dagger}} \\
    S^z_i&=-S+\hat{n}_i^{\phantom{\dagger}} .
  \end{align}
\end{subequations}

The complete Hamiltonian in the Dyson-Maleev representation is given in the
Appendix. In preparation of the decoupling  we introduce the expectation values
\begin{subequations}
  \begin{align}
    n&:=\left\langle b^\dagger_ib_i^{\phantom{\dagger}}\right\rangle=\left\langle
      b^\dagger_jb^{\phantom{\dagger}}_j\right\rangle \\
    a&:=\left\langle b^\dagger_ib_j^\dagger\right\rangle=\left\langle
      b^{\phantom{\dagger}}_ib^{\phantom{\dagger}}_j\right\rangle ,
  \end{align}
\end{subequations}
where $i,j$ are NN sites with $i\in A$ and $j\in B$.  All other
expectation values vanish because of the conservation of the total
$S^z$ component $\sum_i S^z_i$. For simplicity, we assume
that the expectation values are real. The high-order terms are decoupled 
according to Wick's theorem \cite{fette71}.  Neglecting all constant 
terms the mean-field Hamiltonian is given by
\begin{align} 
H^\text{MF}&=\widetilde{J}_\text{DM}(\alpha)
\left(S-\alpha\right)\sum\limits_{\langle i,j\rangle}\left(\hat{n}_i+
\hat{n}_j+b^\dagger_ib^\dagger_j+b^{\phantom{\dagger}}_ib^{\phantom{\dagger}}_j\right),
  \label{eq:HMF_Neel_DM}
\end{align}
where the subscript DM stands for Dyson-Maleev and should
not be confused with Dzyaloshinskii-Moriya. We define
\begin{align}
  \begin{split}
    \widetilde{J}_\text{DM}(\alpha)&:=J+\frac{  J_{\text{bq}}}{S-\alpha}
    \left[2S^3-2S^2(1+5\alpha)\right. \\
    &\left.+S(18\alpha^2+8\alpha+1)-12\alpha^3-9\alpha^2-2\alpha\right] 
    %\label{eq:Jeff_afm}
      \label{eq:JDM_definition}
  \end{split}
\end{align}
where the new parameter 
\begin{align}
\alpha &:=n+a
\end{align}
 has been introduced for convenience. The
parameter $\alpha$ has to be determined self-consistently, see below.  The mean-field
Hamiltonian \eqref{eq:HMF_Neel_DM} can easily be transformed into momentum
space where it can be  diagonalized using a standard Bogoliubov transformation.  The
diagonalized Hamiltonian reads
\begin{align}
  H^\text{MF}&=\sum\limits_{\mathbf{k}\in\text{BZ}}\omega_\mathbf{k}^{\phantom{\dagger}}\beta^\dagger_\mathbf{k}
  \beta^{\phantom{\dagger}}_\mathbf{k}+E^\text{MF}.
\end{align}
We stress that the full Brillouin zone (BZ), i.e., 
$-\pi< k_\gamma\le \pi$ for each component $\gamma\in\{a,b\}$, is used here and hereafter.
This is done for simplicity because full translational invariance holds 
and there is only one mode per $k$ point;
but it does not have a definite value of its total $S_z$ component.
The dispersion is given by
\begin{subequations}
	\begin{align}
  	\omega_\mathbf{k}&=2\widetilde{J}_\text{DM}(\alpha)\left(S-\alpha\right) \widetilde\omega_\mathbf{k}
  	\\
    \widetilde\omega_\mathbf{k}^2 &= 4-\left(\cos k_a+\cos k_b\right)^2
  \end{align}
  \end{subequations}
and the ground state energy by 
\begin{subequations}
  \begin{align} E^\text{MF}&= 
  \widetilde{J}_\text{DM}(\alpha)\left(S-\alpha\right) \widetilde E^\text{MF}\\	
  \widetilde E^\text{MF} &=
  	\sum\limits_{\mathbf{k}\in\text{BZ}} \left(\widetilde\omega_\mathbf{k} -2\right).
  \end{align}
  \label{eq:MFenergy}
\end{subequations}

The gapless excitations at $k=(1,1)$ which is the magnetic ordering vector of the
N\'eel phase and at $k=(0,0)$ are the expected Goldstone modes, because the
ground state of the system has broken symmetry \cite{lange66}. 
The existence of these gapless
excitations is guaranteed by the systematic expansion in $1/S$.
We draw the reader's attention to the fact that this argument
ensures massless modes only for the systematic expansion, i.e.,
for a particular value of $\alpha$. But it turns out that
the vanishing of the energy of the Goldstone modes does not
depend on the precise numbers of the expectation values so that
it does not matter whether the expansion is performed systematically
or self-consistently.

The self-consistent equation for the  parameter $\alpha$ is
found by comparing \eqref{eq:HMF_Neel_DM} with the mean-field
ground state energy \eqref{eq:MFenergy} per site yielding
$\alpha = (1/2)\widetilde E^\text{MF}/N$ if $N$ is the number of sites.
Hence we have in the thermodynamic limit $N\to \infty$
\begin{subequations}
\begin{align}
    \alpha&=\frac{1}{4}\frac{1}{\left(2\pi\right)^2}\iint\limits_\text{BZ}\mathrm{d}^2k
    \; (\widetilde \omega_\mathbf{k} -2) \\
    &=-0.0789737105.
     \label{eq:Neel_alpha_DM}
\end{align}
\end{subequations}
One integration
in \eqref{eq:Neel_alpha_DM} can be done analytically, the second one with any
computer algebra system. Note that due to the simplicity of the 
system no real self-consistency needs to be determined; $\alpha$ can
be computed directly.
 
\subsection{Schwinger boson representation}

Here we apply the Schwinger boson representation \cite{auerb94} to the
Hamiltonian \eqref{eq:H_Neel}.  The spin operators are expressed as
\begin{subequations} 
\label{eq:schwinger_bosons}
  \begin{align}
    S^+_i&=a^\dagger_i b^{\phantom{\dagger}}_i \\
    S^-_i&=b^\dagger_i a^{\phantom{\dagger}}_i \\
    S^z_i&=\frac{1}{2}\left(a^\dagger_i a_i^{\phantom{\dagger}}+b^\dagger_i
      b_i^{\phantom{\dagger}}\right).
  \end{align}
\end{subequations}
The constraint
\begin{align}
  a^\dagger_i a^{\phantom{\dagger}}_i + b^\dagger_i
  b_i^{\phantom{\dagger}}&=2S \label{eq:sb_constraint}
\end{align}
restricts the infinite bosonic Hilbert space to the relevant physical subspace 
of the spin $S$. In this way, the Hamiltonian \eqref{eq:H_Neel} is rewritten as
\begin{align}
  \begin{split}
    H&=-\frac{J}{2}\sum\limits_{\langle i,j\rangle} \left(A^\dagger_{ij}A^{{\phantom{\dagger}}}_{ij}-2S^2\right)\\
    &\quad-\frac{  J_{\text{bq}}}{4}\sum\limits_{\langle i,j\rangle} \left(4S^4-4S^2A^\dagger_{ij}A^{{\phantom{\dagger}}}_{ij}+
    \left(A^\dagger_{ij}A^{{\phantom{\dagger}}}_{ij}\right)^2\right),
  \end{split}
  \label{eq:HamiltonSB}
\end{align}
where the bond operators
\begin{subequations}
  \begin{align}
    A^\dagger_{ij}&=a^\dagger_ia^\dagger_j+b^\dagger_ib^\dagger_j \\
    A^{{\phantom{\dagger}}}_{ij}&=a^{{\phantom{\dagger}}}_ia^{{\phantom{\dagger}}}_j+b^{{\phantom{\dagger}}}_ib^{{\phantom{\dagger}}}_j
  \end{align}
\end{subequations}
are used. They connect adjacent sites on the different sublattices
with $i\in A$ and $j\in B$.

The mean-field approximation in terms of the bond operators is based on an
expansion in the inverse number of boson flavors $1/\mathcal{N}$
\cite{auerb94,auerb88,arova88}. Thus, we intermediately extend to
$\mathcal{N}$ Schwinger boson flavors to justify the approximation.  For
$\mathcal{N}$ flavors, the bilinear term reads
\begin{align}
  \mathbf{S}_i\cdot\mathbf{S}_j&=-\frac{1}{\mathcal{N}}A^\dagger_{ij}A^{\phantom{\dagger}}_{ij}+S^2
\end{align}
with the generalized bond operators
\begin{subequations}
  \begin{align}\label{eq:sbN_bond_afm}
    A^\dagger_{ij}&=\sum\limits_{m=1}^\mathcal{N}a^\dagger_{i,m}a^\dagger_{j,m} \\
    A^{{\phantom{\dagger}}}_{ij}&=\sum\limits_{m=1}^\mathcal{N}a^{\phantom{\dagger}}_{i,m}a^{\phantom{\dagger}}_{j,m}
  \end{align}
\end{subequations}
and the constraint
\begin{align}
  \sum\limits^\mathcal{N}_{m=1}a^\dagger_{i,m}a^{{\phantom{\dagger}}}_{i,m}&=\mathcal{N}S.\label{eq:sbN_constraint}
\end{align}
We define the expectation value
\begin{align}
  A:=\langle A^{{\phantom{\dagger}}}_{ij}\rangle=\langle A^{\dagger}_{ij}\rangle\
  \ge 0,
\end{align}
which is proportional to $\mathcal{N}$ according to the generalized constraint
\eqref{eq:sbN_constraint}.  Hence, the bilinear term is decoupled according to
\begin{align}
\begin{split}
    \left.\mathbf{S}_i\cdot\mathbf{S}_j\right|_\text{MF}& \approx
    -\frac{1}{\mathcal{N}}\left(AA^\dagger_{ij}+\hc\right),
    \label{eq:neel_sb_bldec}
  \end{split}
\end{align}
where we only keep the leading order
$\mathcal{O}\left((1/\mathcal{N})^0\right)$ omitting constants.

In complete analogy, the quartic term of bond operators is decoupled in leading
order by
\begin{align}
  \frac{1}{\mathcal{N}^2}\left(A^\dagger_{ij}A^{{\phantom{\dagger}}}_{ij}\right)\cdot
  \left(A^\dagger_{ij}A^{{\phantom{\dagger}}}_{ij}\right)
  &\approx
  \frac{2A^2}{\mathcal{N}^2}\left(A^\dagger_{ij}A+A A^{{\phantom{\dagger}}}_{ij}\right).
\end{align}
The factor 2 is a consequence of
Wicks's theorem, as there are two possibilities to contract the operators
$A^\dagger_{ij}$ or $A^{{\phantom{\dagger}}}_{ij}$.  Thus, the mean-field
decoupling of the entire biquadratic term reads
\begin{align}
\left.\left(\mathbf{S}_i\cdot\mathbf{S}_j\right)^2\right|_\text{MF}
&\approx
-\frac{2S^2}{\mathcal{N}}A\left(1-\frac{A^2}{\mathcal{N}S^2}\right)
\left(A^\dagger_{ij}+A^{{\phantom{\dagger}}}_{ij}\right). 
\label{eq:neel_sb_bq_dec}
\end{align}
Since $A\propto\mathcal{N}S$,  $A^2>\mathcal{N} S^2$ holds so that
the biquadratic term yields a positive contribution in total.
Due to the prefactor $-  J_{\text{bq}}$, cf.\ Eq.\ \eqref{eq:HamiltonSB}, the
biquadratic term enhances the bilinear one \eqref{eq:neel_sb_bldec}.
 
Returning to $\mathcal{N}=2$, the mean-field Hamiltonian is finally given by
\begin{align}
H^\text{MF}&=- B \sum\limits_{\langle i,j\rangle}\left(A^\dagger_{ij}+\hc\right)+
\lambda\sum\limits_i\left(a^\dagger_ia_i^{{\phantom{\dagger}}}+b^\dagger_ib_i^{{\phantom{\dagger}}}\right),
\label{eq:HMF_Neel_SB}
\end{align}
where
\begin{align}
	B &:= {A\widetilde{J}_\text{SB}(A)}/{2}\\
  \widetilde{J}_\text{SB}(A)&:=J-2  J_{\text{bq}} S^2\left(1-{A^2}/{(2S^2)}\right).
  \label{eq:JSB_definition}
\end{align}
The Lagrange multiplier $\lambda$ is introduced to enforce the
constraint \eqref{eq:sb_constraint} on average. 
In the symmetry broken phase at zero temperature, 
$\lambda$ is fixed to the value
\begin{align}
  \lambda&=4B=2\widetilde{J}_\text{SB}\left(A\right)A
\end{align}
in order to retrieve massless Goldstone bosons.
In momentum space, the mean-field Hamiltonian \eqref{eq:HMF_Neel_SB} is
diagonalized by a standard Bogoliubov transformation leading to the
dispersion and the ground state energy
\begin{subequations}
  \begin{align}
    \omega_\mathbf{k}&=\sqrt{\lambda^2-\left(4B\left(\cos k_a+\cos
          k_b\right)\right)^2}
 \\
    E^\text{MF}&=\sum\limits_{\mathbf{k}\in\text{BZ}}
    		\left(\omega_\mathbf{k}-\lambda\right).
  \end{align}
\end{subequations}

The self-consistency equations for the parameters $A$ and 
$\lambda$ are given by
\begin{subequations}
  \begin{align}
    2S&=\frac{1}{N}\frac{\partial E^\text{MF}}{\partial\lambda} \\
    -4A &=\frac{1}{N}\frac{\partial
      E^\text{MF}}{\partial B}\, .
  \end{align}
\end{subequations}
These two equations can be combined to yield $A$ in the symmetry broken phase
\begin{align}
  A&=2S+1-\frac{1}{2}\iint\limits_\text{BZ} \frac{\mathrm d^2k}{\left(2\pi\right)^2}
  \sqrt{4-\left(\cos k_a+\cos k_b\right)^2}
  \end{align}
in the thermodynamic limit. 
Note that the massless Goldstone modes are again guaranteed
by the systematic expansion which is here done in the inverse
number of flavors $1/\mathcal{N}$. As for the Dyson-Maleev
mean-field approach the expansion can also be done self-consistently
without spoiling the Goldstone theorem because the precise
number of the expectation value $A$ does not matter for
this qualitative aspect. For $S=1$, the calculated value is
\begin{align}
  A&=2.1579474210
\end{align}
which corresponds to $2(1-\alpha)$ from the Dyson-Maleev calculation.
Thus the obtained dispersions are identical for zero biquadratic exchange
because then $\widetilde J_\text{SB}=\widetilde J_\text{DM}$ holds.

\subsection{Results}

\begin{figure}[ht]
  \centering
  \includegraphics[width=\columnwidth]{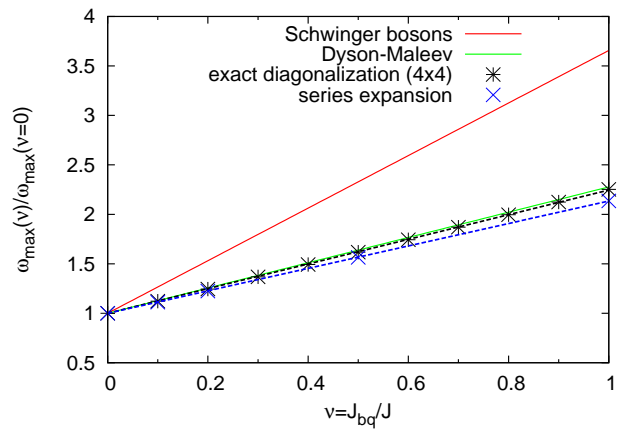}
  \caption{(Color online) Influence of the biquadratic exchange on the excitation energies in
    the 2D N\'eel phase ($S=1$). The dashed lines are fits to the
    corresponding data points. Note the almost perfect straight lines obtained 
    	in exact diagonalization and in series expansion.} 
    \label{fig:2DNeel_comparison}
\end{figure}

We are interested in the influence of the biquadratic exchange on the
excitation energies. Thus the ratios of the dispersions with and without
biquadratic exchange is an appropriate quantity. Since the 
shape of the dispersions is the same for the Dyson-Maleev and the 
Schwinger boson representation, no more quantities need to be compared.
This ratio depends linearly on the coupling ratio $\nu:=  J_{\text{bq}}/J$,
see Eqs.\ \eqref{eq:JDM_definition} and   \eqref{eq:JSB_definition}.
This behavior is depicted in Fig.\ \ref{fig:2DNeel_comparison} and the 
corresponding slopes are given in Tab.\ \ref{tab:2DNeel_comparison}.

 \begin{table}[ht]
  \centering
  \begin{tabular}{c||c}
    \textbf{method} & \textbf{slopes} \\ \hline\hline
    Schwinger bosons & 2.65674 \\
    Dyson-Maleev & 1.27708 \\
    exact diagonalization (4x4) & 1.24505 \\
    series expansion & 1.13423    
  \end{tabular}
  \caption{(Color online) Slopes for the influence of the biquadratic exchange on the
    maximum excitation energy relative to the one without biquadratic exchange,
    	see Fig.\ \ref{fig:2DNeel_comparison} for the data from which the slopes 
    		are fitted for exact diagonalization and series expansion.} 
    \label{tab:2DNeel_comparison}
\end{table}

The mean-field results for the 2D N\'eel phase with
biquadratic exchange are checked  to the results obtained by other methods.
In detail, the excitation energies can be calculated by an exact
diagonalization of the Hamiltonian \eqref{eq:H_Neel} on a finite 4x4 lattice
\cite{Raas2009} and by series expansion \cite{oitma06,Oitmaa2009}.  For
both methods, the slope is determined by a linear fit, see 
Fig.\ \ref{tab:2DNeel_comparison}. All results are
compared in Tab.\ \ref{tab:2DNeel_comparison} and displayed in Fig.
\ref{fig:2DNeel_comparison}.  The data points are the maximum excitation energies
normalized to the maximum without biquadratic exchange.

The Schwinger bosons result immediately catches one's eye because the slope
is far too large by a factor greater than two. In contrast,
the Dyson-Maleev result matches very well with the exact diagonalization.
Of course, the exact diagonalization data is hampered by finite size
effects even though they are only moderate because we
consider the ratio of the maximum of the dispersion.
The slope of the series expansion, which is essentially exact, is slightly
smaller. The deviation of
the Dyson-Maleev slope from the series expansion is in the range of
10-15\%, which is well acceptable for a mean-field approach
applied to complicated terms made from eight bosonic operators. 
Extending the model to three dimensions, we expect the result to improve
because mean-field approximations generally  become
the more accurate  the higher the number of dimensions. Therefore, we choose
 to employ to the Dyson-Maleev representation and the corresponding mean-field 
 approximation in the study of the columnar phase.

Discrepancies of the Schwinger boson mean-field theory to other treatments
appeared already without biquadratic exchange when the theory was introduced 
by Auerbach and Arovas \cite{auerb88,arova88}. In their calculations, the  
mean-field free energy exceeded previous results by a factor of 2 and the 
sum rule of the dynamic structure factor exceeded its exact value
 by a factor of $3/2$.  
 
 In the mean-field
approximation for  Schwinger bosons, we treat both Schwinger bosons on
a site as independent. Thus, the constraint
\eqref{eq:sb_constraint} is violated because a change in the
occupation of the $a$ boson should always be connected to a change of the occupation
of the $b$ boson on the same lattice site. Hence, we
suggest that the additional factors appear because of the missed suppression
of fluctuations of the boson number due to the constraint.
Since the biquadratic exchange can be roughly 
viewed as a bilinear exchange multiplied by a NN expectation
value, see Eq.\ \eqref{eq:bq_simple}, it is comprehensible that the 
slope of the Schwinger boson mean-field result is too large
by about a factor of 2 because the NN expectation value
is overestimated by this factor \cite{auerb88}.

\section{3D columnar phase with biquadratic exchange} \label{sec:col3D_bq}

As explained in the Introduction we advocate that a localized
spin model for the pnictides has to comprise a biquadratic
term in order to account for the spatial anisotropy of 
the spin-wave velocities measured by inelastic
neutron scattering (INS) and computed by density functional theory.
Thus, we set out to investigate the Hamiltonian
\begin{align}
  \begin{split}
    H&=J_1\sum\limits_{\langle i,j\rangle}\mathbf{S}_i\cdot\mathbf{S}_j+J_2\sum\limits_{\langle\langle i,j\rangle\rangle}\mathbf{S}_i\cdot\mathbf{S}_j \\
    &\quad\quad +J_c\sum\limits_{[i,j]}\mathbf{S}_i\cdot\mathbf{S}_j -  J_{\text{bq}}\sum\limits_{\langle i,j\rangle}\left(\mathbf{S}_i\cdot\mathbf{S}_j\right)^2, 
    \label{eq:H_col3D_bq}
  \end{split}
\end{align}
where a NN in-plane biquadratic exchange ($  J_{\text{bq}}>0$) is  introduced. The
brackets $\left<i,j\right>$ and $\left<\left<i,j\right>\right>$ correspond to
in-plane nearest and next-nearest neighbor sites, while the bracket $[i,j]$
indicates the exchange between interplane nearest neighbor sites. The very 
small orthorhombic distortion in the columnarly ordered layers is neglected.  
Justified by the checks in the previous Section, we employ the 
Dyson-Maleev representation and the ensuing mean-field approximation. 
In order to establish our notation we provide a brief
derivation of the spin-wave dispersion and the self-consistency equations.
For further details the reader is referred to Ref.\ \onlinecite{stane10}.

For the mean-field decoupling, the following parameters are needed:
\begin{subequations}
	\label{eq:col3d_ev}
  \begin{itemize}
  \item average occupation number per lattice site
    \begin{align} n&:=\left\langle
        b^\dagger_ib_i^{\phantom{\dagger}}\right\rangle=\left\langle
        b^\dagger_jb^{\phantom{\dagger}}_j\right\rangle,
    \end{align}
  \item in-plane NN antiparallel spin orientation perpendicular to
    the spin stripes ($a$-direction)
    \begin{align} a_1&:=\left\langle
        b^\dagger_ib_j^\dagger\right\rangle=\left\langle
        b^{\phantom{\dagger}}_ib^{\phantom{\dagger}}_j\right\rangle
    \end{align}
  \item interlayer NN  antiparallel spin orientation ($c$-direction)
    \begin{align} a_c&:=\left\langle
        b^\dagger_ib_j^\dagger\right\rangle=\left\langle
        b^{\phantom{\dagger}}_ib^{\phantom{\dagger}}_j\right\rangle
    \end{align}
  \item in-plane  next-nearest neighbor (NNN) antiparallel spin orientation
    \begin{align} a_2&:=\left\langle
        b^\dagger_ib_j^\dagger\right\rangle=\left\langle
        b^{\phantom{\dagger}}_ib^{\phantom{\dagger}}_j\right\rangle
    \end{align}
  \item in-plane NN  parallel spin orientation parallel to the spin
    stripes ($b$-direction)
    \begin{align} f&:=\left\langle
        b^\dagger_ib_j^{\phantom{\dagger}}\right\rangle=\left\langle
        b^\dagger_jb^{\phantom{\dagger}}_i\right\rangle
    \end{align}
  \end{itemize}
\end{subequations}

It turns out to be advantageous to introduce the combined
 parameters $\alpha_1$, $\alpha_2$, $\alpha_c$
and $\beta$ according to
\begin{subequations}
\label{eq:alpha_def}
  \begin{align} \alpha_1&:=n+a_1 \\ \alpha_c&:=n+a_c \\ \alpha_2&:=n+a_2 \\
    \beta&:=n-f.
  \end{align}
\end{subequations}
The values of these parameters are determined by the self-consistency
equations given below.

Thereby, the mean-field decoupling of the Hamiltonian \eqref{eq:H_col3D_bq}
reads
\begin{subequations}
\label{eq:H_col3D_MF}
\begin{align}
  H&=H_\perp+H_\parallel+H_c+H_{\text{NNN}} 
\\ 
  H_\perp &=J_2x_{1a}\left(S-\alpha_1\right)\sum\limits_{\langle i,j\rangle}\left(\hat{n}^{\phantom{\dagger}}_i+
  \hat{n}^{\phantom{\dagger}}_{j}+b^\dagger_ib^\dagger_{j}+b^{\phantom{\dagger}}_ib^{\phantom{\dagger}}_{j}\right)
  \label{eq:hamilton_perp}
  \\ 
  H_\parallel&=-J_2x_{1b}\left(S-\beta\right)\sum\limits_{\langle i,j\rangle}\left(\hat{n}_i^{\phantom{\dagger}}+
  \hat{n}^{\phantom{\dagger}}_{j}-b^\dagger_ib^{\phantom{\dagger}}_{j}-b^\dagger_{j}b^{\phantom{\dagger}}_i\right)
   \label{eq:hamilton_para}
  \\
  H_c&=J_2x\mu\left(S-\alpha_c\right)\sum\limits_{[i,j]}\left(\hat{n}^{\phantom{\dagger}}_i+
  \hat{n}^{\phantom{\dagger}}_{j}+b^\dagger_ib^\dagger_{j}+b^{\phantom{\dagger}}_ib^{\phantom{\dagger}}_{j}\right)
  \label{eq:hamilton_c}	
  \\
  H_\text{NNN}&=J_2\left(S-\alpha_2\right)\sum\limits_{\langle\langle i,j\rangle\rangle}\left(\hat{n}^{\phantom{\dagger}}_i+
  \hat{n}^{\phantom{\dagger}}_{j}+b^\dagger_ib^\dagger_{j}+b^{\phantom{\dagger}}_ib^{\phantom{\dagger}}_{j}\right)
  	  \label{eq:hamilton_NNN}	
\end{align}
\end{subequations}
with $x:=J_1/J_2$, $\mu:=J_c/J_1$, $\nu:=  J_{\text{bq}}/J_1$ and
%\begin{widetext}
  \begin{subequations}
  \label{eq:x1ab}
    \begin{align}
      x_{1a} &:=x+\frac{x\nu}{S-\alpha_1}
      \big[2S^3-2S^2(1+5\alpha_1)+
     \nonumber\\
      &\qquad S(18\alpha_1^2+8\alpha_1+1)-12\alpha_1^3-9\alpha_1^2-2\alpha_1\big] 
      \label{eq:BQ_x1a}
      \\
      x_{1b} &:=x-\frac{x\nu}{S-\beta}\big[2S^3-2S^2\left(1+5\beta\right)+
      \nonumber\\
      	&\qquad S(18\beta^2+8\beta)-12\beta^3-9\beta^2-\beta\big].
      \label{eq:BQ_x1b}
    \end{align}
  \end{subequations}
%\end{widetext}
The mean-field Hamiltonian
\eqref{eq:H_col3D_MF} is easily transformed into momentum space and
diagonalized by a standard Bogoliubov transformation. Thereby, the spin-wave
dispersion 
\begin{align}
  \omega_\mathbf{k}&:=2\sqrt{A_\mathbf{k}^2-B_\mathbf{k}^2} \label{eq:3Dbq_dispersion}
\end{align}
is obtained with
%\begin{widetext}
  \begin{subequations} 
  \label{eq:BQ_ABbar}
    \begin{align}
      A_\mathbf{k}&:=J_2[2(S-\alpha_2)+x_{1a}(S-\alpha_1)+ 
      \nonumber\\
			& \qquad x\mu(S-\alpha_c)+x_{1b}(S-\beta)(\cos k_b-1)] 
          \\
      B_\mathbf{k}&:=J_2[x_{1a}(S-\alpha_1)\cos k_a+ x\mu(S-\alpha_c)\cos k_c+
      \nonumber\\
        	&\qquad  2(S-\alpha_2)\cos k_a\cos k_b].
    \end{align}
  \end{subequations}
%\end{widetext}
The components $k_a$, $k_b$ and $k_c$ of the momentum vector $\mathbf{k}$ are
directed along the crystal axes as shown in Fig.\ \ref{fig:model_col3D}. 
The dispersion is gapless at $k=(0,0,0)$ and
$k=(1,0,1)$ corresponding to the required Goldstone modes \cite{uhrig09a,holt11}. 
Note that this feature is again guaranteed
by the systematic expansion in the inverse spin $1/S$. As for the N\'eel phase
the expansion can also be done self-consistently
without spoiling the Goldstone theorem because the precise
numbers of the expectation values do not matter for
this qualitative aspect. But we consider this a highly non-trivial aspect
in view of the four different quantum corrections occuring here.

For small momenta, the dispersion
\eqref{eq:3Dbq_dispersion} can be expanded
\begin{align}
  \omega_\mathbf{k}&\approx\sqrt{v_a^2k_a^2+v_b^2k_b^2+v_c^2k_c^2}
\end{align}
and one obtains the spin-wave velocities
%\begin{widetext}
  \begin{subequations}
    \begin{align}
      v_a^2 &=(2J_2)^2 [2(S-\alpha_2) + x_{1a}(S-\alpha_1)+
      x\mu(S-\alpha_c)]\times
      \nonumber\\
      &	[2(S-\alpha_2)+ x_{1a}(S-\alpha_1)]
      \\
      v_b^2 &=(2J_2)^2[2(S-\alpha_2)+ x_{1a}(S-\alpha_1)+ 
        x\mu (S-\alpha_c)]\times
      \nonumber\\
      &[2(S-\alpha_2) -x_{1b}(S-\beta)]
      \\
      v_c^2 &=(2J_2)^2 [2(S-\alpha_2)+ x_{1a}(S-\alpha_1)+
      x\mu(S-\alpha_c)]\times
      \nonumber\\
      &x\mu(S-\alpha_c).
    \end{align}
  \end{subequations}
%\end{widetext}

The parameters of the quantum corrections are determined from the self-consistency
equations in the thermodynamic limit
\begin{subequations} \label{eq:BQ_sce}
  \begin{align}
    \alpha_1&=\frac{1}{2}\frac{1}{\left(2\pi\right)^3}\iiint\limits_{\text{BZ}}\mathrm d^3k\ \frac{A_\mathbf{k}-B_\mathbf{k}\cos k_a}{\sqrt{A_\mathbf{k}^2-B_\mathbf{k}^2}}-\frac{1}{2} \\
    \alpha_c&=\frac{1}{2}\frac{1}{\left(2\pi\right)^3}\iiint\limits_{\text{BZ}}\mathrm d^3k\ \frac{A_\mathbf{k}-B_\mathbf{k}\cos k_c}{\sqrt{A_\mathbf{k}^2-B_\mathbf{k}^2}}-\frac{1}{2} \\
    \alpha_2&=\frac{1}{2}\frac{1}{\left(2\pi\right)^3}\iiint\limits_{\text{BZ}}\mathrm d^3k\ \frac{A_\mathbf{k}-B_\mathbf{k}\cos k_a\cos
      k_b}{\sqrt{A_\mathbf{k}^2-B_\mathbf{k}^2}}-\frac{1}{2} \\
    \beta&=-\frac{1}{2}\frac{1}{\left(2\pi\right)^3}\iiint\limits_{\text{BZ}}\mathrm d^3k\ \frac{A_\mathbf{k}\left(\cos
        k_b-1\right)}{\sqrt{A_\mathbf{k}^2-B_\mathbf{k}^2}}-\frac{1}{2},
  \end{align}
\end{subequations}
where the integrations run over the full Brillouin zone (BZ)
as before. 
The above set of equations is solved by numerical integration and 
by four-dimensional root finding.

The staggered magnetization $m_\text{st}$ is defined as
$m_{\text{st}}:=\left<S^z_i\right>(-1)^\sigma$ where $\sigma=0$ for $i\in A$
and $\sigma=1$ for $i\in B$. In the thermodynamic limit, it reads
\begin{align}
  m_\text{st}&=S+\frac{1}{2}-\frac{1}{2}\frac{1}{\left(2\pi\right)^3}\iiint\limits_{\text{BZ}}\mathrm d^3k\
  \frac{A_\mathbf{k}}{\sqrt{A_\mathbf{k}^2-B_\mathbf{k}^2}}. \label{eq:BQ_mst}
\end{align}

\subsection{General Results for $S=1$}

We discuss the qualitative aspects for $S=1$ and $\mu=0.25$, 
which is a generic value for the relative interlayer coupling. 
The values of the relative biquadratic exchange $\nu$ 
are chosen in the range of $0.1$ to $0.7$ as motivated in Sect.\ 
\ref{sect:biquadratic}. For
comparison, we also show the results without biquadratic exchange.

\begin{figure}[ht]
  \centering
  \includegraphics[width=\columnwidth]{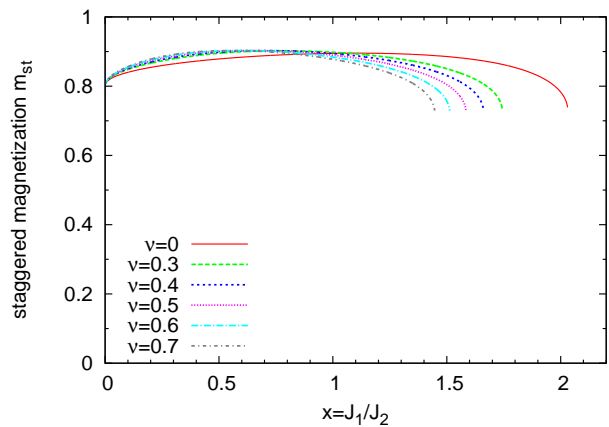}
  \caption{(Color online) Staggered magnetization as a function of $x$ for $S=1$ and
    $\mu=0.25$. All curves are in the velocity-driven scenario 
    \tr{which we presume to signal a strong first order transition to the N\'eel phase.
    The magnetization remains finite at the endpoint where the columnar
    phase ceases to exist, for details see Sect.\ \ref{sect:scenarios}.}}
  \label{fig:BQ_mst}
\end{figure}

From the staggered magnetization in Fig.\ \ref{fig:BQ_mst}, we conclude
that the biquadratic exchange destabilizes the columnar phase and
drives the critical point towards lower values of $x=J_1/J_2$. There is only a
negligible influence on the maximum value of $m_\text{st}$, which is also
shifted further left.

\begin{figure*}[ht]
  \centering {\includegraphics[width=\columnwidth]{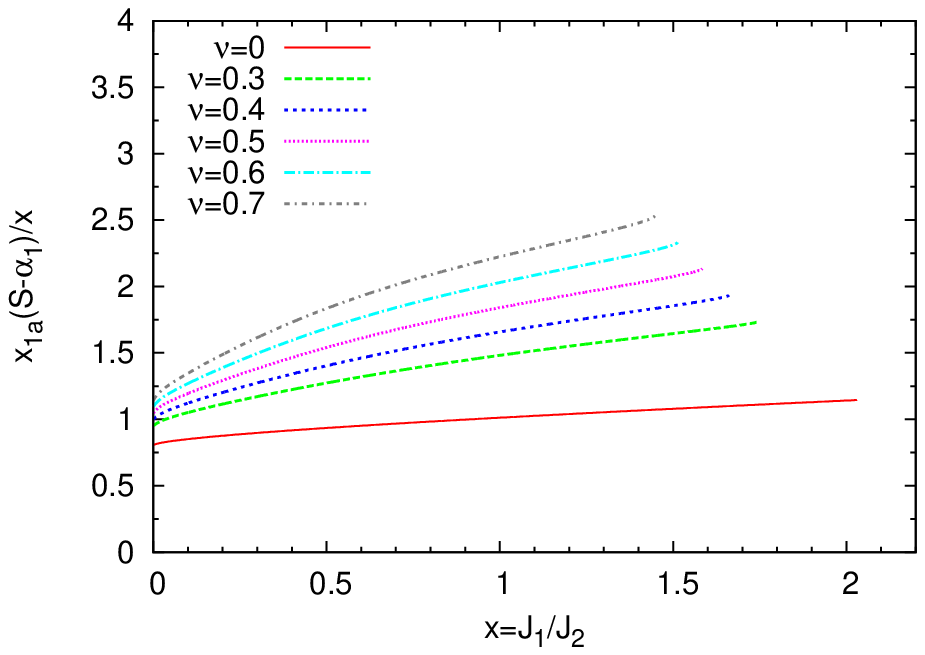}}
  {\includegraphics[width=\columnwidth]{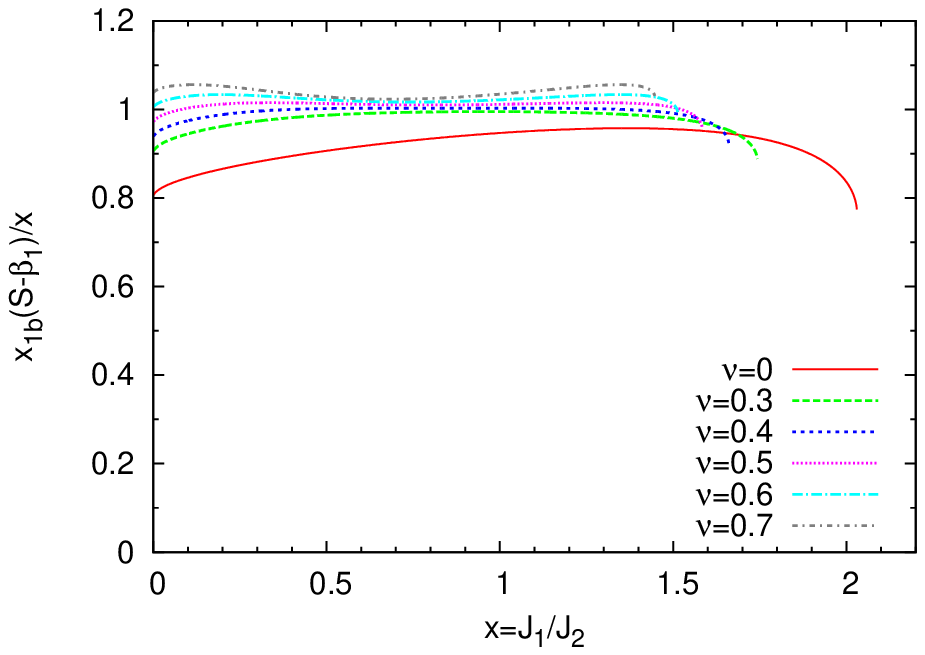}} \\
  {\includegraphics[width=\columnwidth]{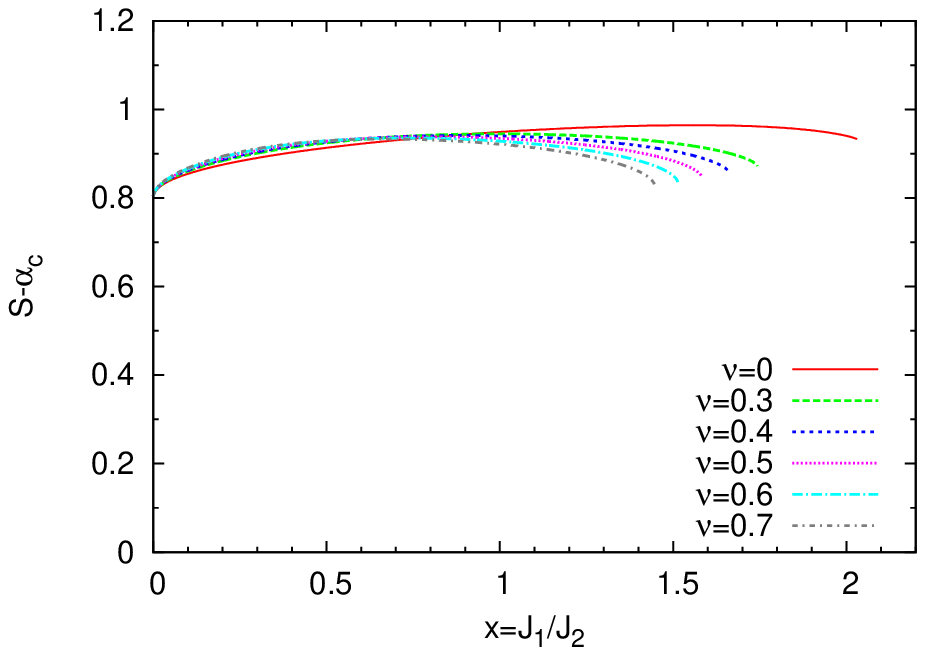}}
  {\includegraphics[width=\columnwidth]{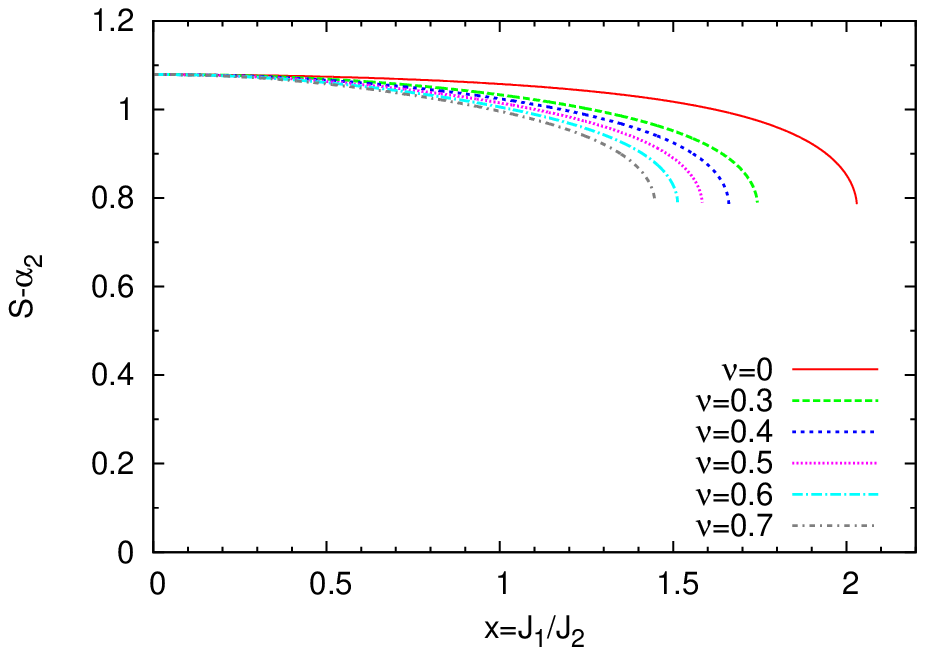}}
  \caption{(Color online) Renormalization of the effective exchange parameters
  by quantum fluctuations as function of $x$  for $S=1$ and $\mu=0.25$. 
  	All curves are in the velocity-driven scenario.}
  \label{fig:BQ_renorm}
\end{figure*}

Important is the effect of the biquadratic exchange on the
renormalization of the quantum correction parameters (for definitions see
Eqs.\ (\ref{eq:alpha_def}, \ref{eq:H_col3D_MF}, \ref{eq:x1ab})) shown in
Fig.\ \ref{fig:BQ_renorm}. The effective exchange perpendicular to the spin stripes
is significantly strengthened by factors up to 2, see upper left
panel of Fig.\ \ref{fig:BQ_renorm}. In contrast, the 
effective  exchange parallel to the spin stripes stays almost constant except for small
frustrations and in vicinity of the end point, see upper right
panel of Fig.\ \ref{fig:BQ_renorm}. Because the biquadratic
exchange in the Hamiltonian \eqref{eq:H_col3D_bq} is restricted to in-plane
NN sites, the only effect on the renormalization of the
interlayer and NNN exchange is the shift of the critical
point to smaller values of $x$, see lower panels
of Fig.\ \ref{fig:BQ_renorm}.  Hence, we indeed find the expected effect
that an NN biquadratic exchange enhances the spatial anisotropy.
This validates our qualitative considerations in Sect.\ \ref{sect:biquadratic}.

\begin{figure}[ht]
  \centering {\includegraphics[width=\columnwidth]{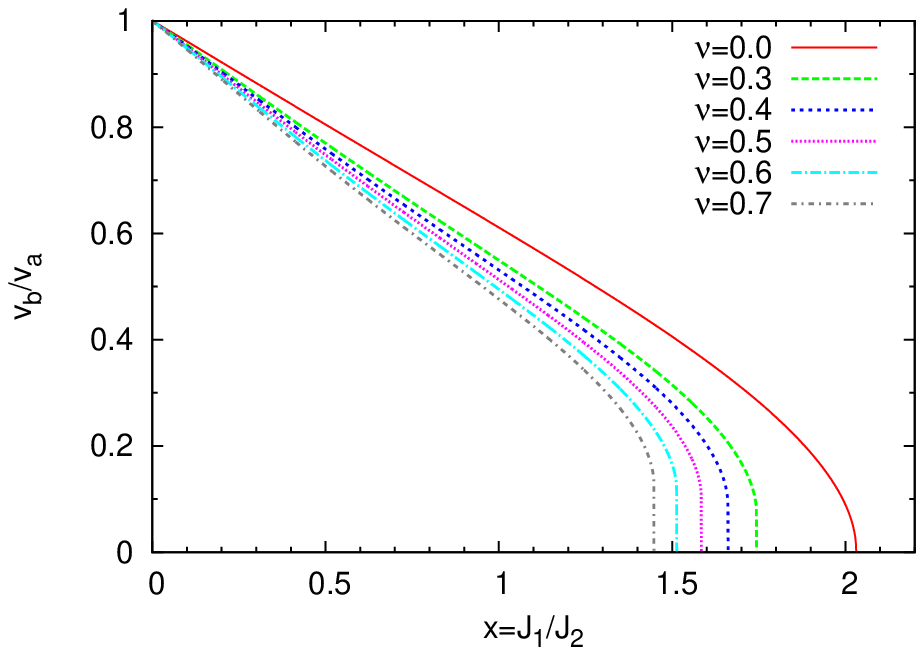}}
  {\includegraphics[width=\columnwidth]{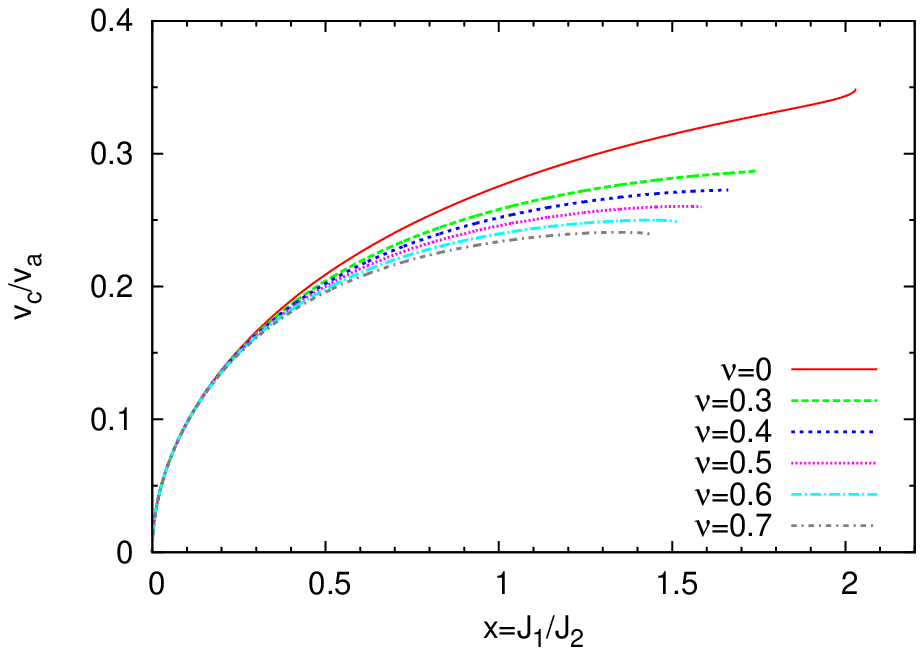}}
  \caption{(Color online) Ratios of the spin-wave velocities as a function of $x$ for $S=1$
    and $\mu=0.25$. All curves are in the velocity-driven scenario, i.e.,  the ratio
    $v_b/v_a$ vanishes at the end point. This is at the limit of the resolution of 
    	the numerical evaluation of the three-dimensional integrals.}
  \label{fig:BQ_swv}
\end{figure}

The ratios of the spin-wave velocities depicted in Fig.\ \ref{fig:BQ_swv} are
weakened by increasing biquadratic exchange because the strengthening of the effective
exchange perpendicular to the stripes leads to
a greater spin-wave velocity $v_a$ and thus to a smaller ratio $v_b/v_a$. 
In general, the qualitative behavior of
the ratios is similar to the one observed without biquadratic exchange.

All in all, a biquadratic exchange acting on in-plane NN sites
increases the anisotropy of the exchange parallel and perpendicular to the
spin stripes. The growing anisotropy with increasing biquadratic exchange
$\nu=  J_{\text{bq}}/J_1$ is caused by the strong renormalization of the exchange
perpendicular to the spin stripes. The exchange parallel to the spin stripes
experiences a marginal strengthening and is weakened only in proximity to the
critical point. Hence, even a large biquadratic exchange does not induce an effective
ferromagnetic exchange $J_{1b}^\text{eff}\lessapprox 0$ in $b$-direction.

\begin{figure}[ht]
  \centering {\includegraphics[width=\columnwidth]{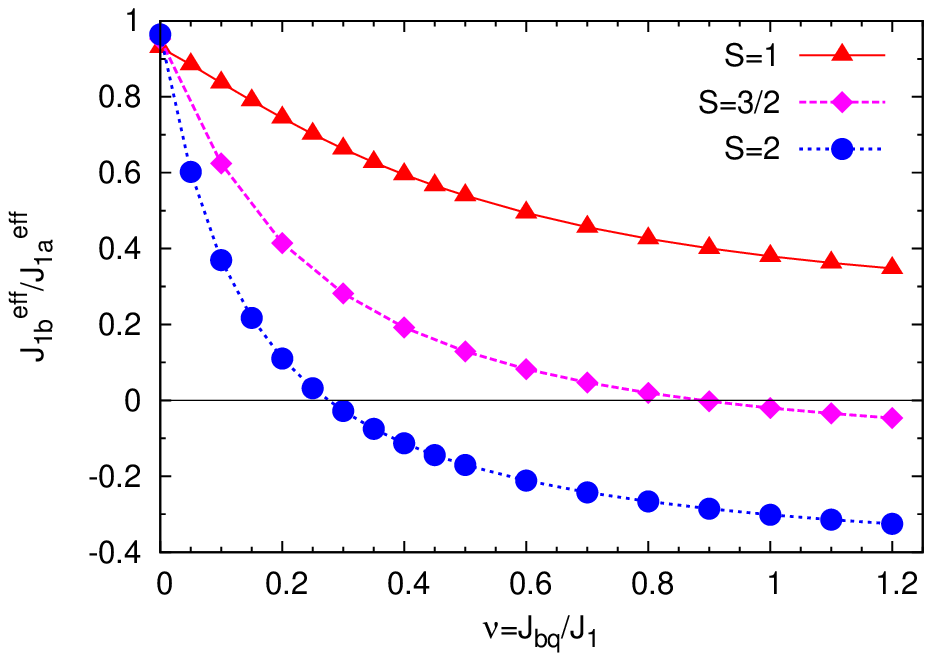}}
  \caption{(Color online) Relative anisotropy, cf.\ definitions \eqref{eq:eff_coupling_defs}, as function
  of the relative biquadratic exchange $\nu$ in the two-dimensional model for
  $S=1$, $3/2$ and $S=2$.}
  \label{fig:relative_anisotropy}
\end{figure}

To render this point quantitatively we define the effective exchange couplings
\begin{subequations}
\label{eq:eff_coupling_defs}
\begin{align}
J_{1a}^\text{eff} &:= J_2 x_{1a}(1-\alpha_1/S)\\
 J_{1b}^\text{eff} &:= J_2 x_{1b}(1-\beta/S)\\
 J_{c}^\text{eff} &:= J_c (1-\alpha_c/S)\\
  J_2^\text{eff} &:= J_2(1-\alpha_2/S)
 \end{align}
 \end{subequations}
according to the mean-field Hamiltonians 
(\ref{eq:hamilton_perp}, \ref{eq:hamilton_para}, \ref{eq:hamilton_c}, \ref{eq:hamilton_NNN}).
These definitions enable a direct comparison of the effective spatial anisotropy to
the ones determined in fits based on linear spin-wave theory as they are used
in experiment \cite{diall09,zhao09}, for the analysis of density-functional theory \cite{han09}
and in other theoretical analyses \cite{schmi10,yao10}
The relative anisotropy is shown in Fig.\ \ref{fig:relative_anisotropy} for
the two-dimensional model, i.e., for $J_c=0$. The results for finite three-dimensional
coupling are qualitatively the same. 

Strikingly, the size of the spin really matters
due to the different relative importance of quantum fluctuations. 
For $S=1$ the self-consistency procedure prevents the occurrence of negative
effective couplings in $b$ direction -- even for very large biquadratic exchange.
This can be understood by inspecting Eq.\ \eqref{eq:BQ_x1b}. For large
$\nu$ the expectation value $\beta$ goes to zero. Hence the influence
of the square bracket multiplying $\nu$ decreases more and more 
because the terms independent of $\beta$, i.e., $2S^3-2S^2$, cancel for $S=1$.
Hence no zero or negative effective coupling along the spin stripes occurs.
Only for $S>1$ a zero or even negative effective coupling along the direction
of parallel spins is possible. We will come back to this point when comparing with
experimental data.

\section{Application to the Iron Pnictides}

Here we discuss the applicability of our model to iron pnictides.  We consider
$S=1$, $S=3/2$ and $S=2$. The former value is suggested by the relatively low
values for the staggered magnetizations measured
\cite{cruz08,ewing08,zhao08c,mcque08,zhao09}, see also list in Ref.\ \onlinecite{schmi10}. 
Further support for this value
comes from the successful studies of two-band models
\cite{raghu08,lee10a,lee10b}. Another interesting support is provided by
advanced Gutzwiller calculations of the distribution of local charges
in five-band models which are consistent with $S=1$ \cite{schic11}.
On the other hand, simple chemical electron count implies 
that the iron is doubly positive charged so that the d-shell
contains four holes and Hund's rule implies
$S=2$. Indeed, recent findings for closely related iron compounds
showed that static local moments up to $2.2 - 3.3\mu_\text{B}$ can occur
\cite{bao11a,li11a,pomja11a} or even larger ($4.7\mu_\text{B}$) \cite{ryan11a}.
Furthermore, we stress that the value of the staggered static magnetization
per site represents only a lower bound for the local moment.
Moreover, the observable static moment also depends on electronic
matrix elements \cite{uhrig09a} influenced by the complicated electronic situation.
In view of these ambiguities, it is certainly worthwhile to consider
also $S=3/2$ and $S=2$.

Experimentally, the magnetic dispersion is best known
for the 122 compound CaFe$_2$As$_2$ \cite{mcque08,diall09,zhao09}.
Hence we fit the dispersion obtained by self-consistent mean-field theory of the
three-dimensional model \eqref{eq:H_col3D_bq} with biquadratic
exchange to the measured spin-wave dispersion. Then we are able
to draw conclusions regarding the quality of the agreement and the
plausibility of the exchange values obtained.

\begin{table}
  \centering
  \begin{tabular}{c||c|c|c|c|c}
    $\nu$ & 0.3 & 0.4 & 0.5 & 0.6 & 3.0\\ \hline\hline
    $x$ & 0.645 & 0.616 & 0.589 & 0.565 & 0.284 \\ 
    $\mu$ & 0.297 & 0.314 & 0.332 & 0.349 & 0.793 \\ \hline
    $J_1$ & 18.9 & 17.9 & 16.9 & 16.0 & 7.1 \\ 
    $J_c$ & 5.6 & 5.6 & 5.6 & 5.6 & 5.6 \\
    $J_2$ & 29.4 & 29.0 & 28.7 & 28.4 & 24.9\\
    $  J_{\text{bq}}$ & 5.7 & 7.1 & 8.5 & 9.6 & 21.2\\ \hline
    $J^\text{eff}_{1a}$ & 25.4 & 26.3 & 27.1 & 27.9 & 36.2 \\
    $J^\text{eff}_{1b}$ & 18.8 & 17.9 & 17.0 & 16.3 & 8.0\\
    $J^\text{eff}_{c}$ & 5.3 & 5.3 & 5.3 & 5.3 & 5.3\\
    $J^\text{eff}_2$ & 31.1 & 30.7 & 30.3 & 29.9 & 25.8
  \end{tabular}
  \caption{Fit parameters of the model \eqref{eq:H_col3D_bq} in the three dimensional columnar
  	 phase for given values of the relative biquadratic exchange $\nu=  J_{\text{bq}}/J_1$. 
  	 	The parameters are determined by fits to the experimental spin-wave velocities in
    CaFe$_2$As$_2$ \cite{zhao09}. The exchange constants $J^\text{eff}_i$ are the exchange 
    constants which would provide the same results for a model without biquadratic exchange 
    in linear spin-wave theory, 
    see the definitions \eqref{eq:eff_coupling_defs}.}
    \label{tab:BQ_results}
\end{table}

\begin{figure*}
  \centering {\includegraphics[width=\columnwidth]{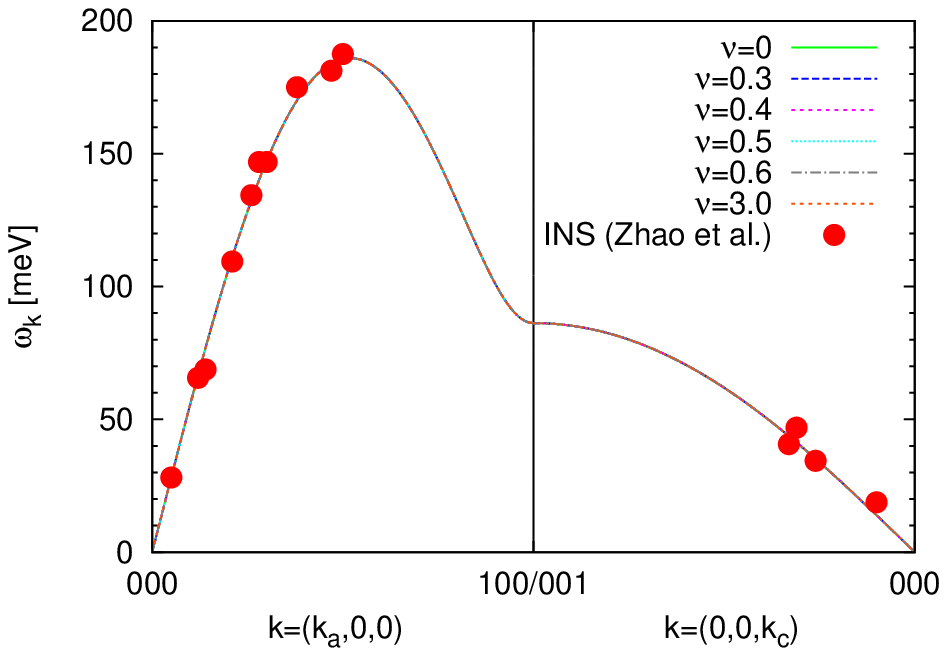}}
  {\includegraphics[width=\columnwidth]{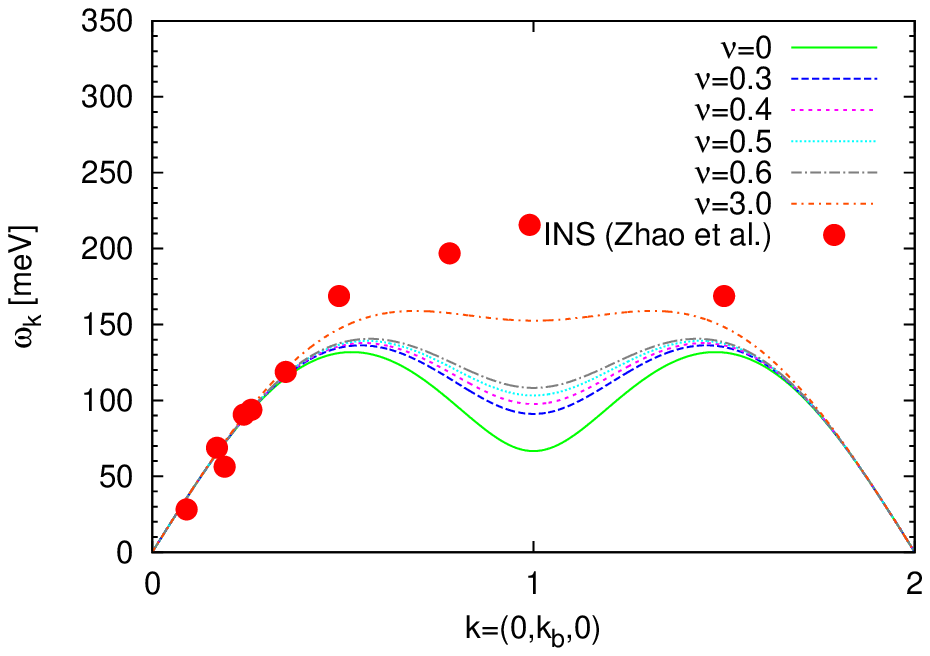}}
  \caption{(Color online) Spin-wave dispersion \eqref{eq:3Dbq_dispersion} 
  	resulting from the self-consistent mean-field theory of the model 
   \eqref{eq:H_col3D_bq} for $S=1$ in the three dimensional columnar
  	 phase with biquadratic exchange for
    various values of $\nu=  J_{\text{bq}}/J_1$. The dispersions (lines) are plotted for
    the parameters given in Tab.\ \ref{tab:BQ_results}.
     The red dots are experimental data extracted from
    INS for CaFe$_2$As$_2$ from Ref.\ \onlinecite{zhao09}. Wave
    vectors are given in units of $\pi$ over lattice constants assuming
    	an orthorhombic crystal. Note that along the path in the BZ shown
    	in the left panel all fits collapse in one curve. Significant 
    		differences occur only along the path shown in the right panel.}
  \label{fig:BQ_dispersions}
\end{figure*}

The results of fits to Zhao's data are given in Tab.\  \ref{tab:BQ_results}
for plausible values of $  J_{\text{bq}}$. In order to fix $  J_{\text{bq}}$ independently an 
additional fourth piece of information is required from experiment, 
for instance the energy at $k=(0,1,0)$. This energy was only determined by
Zhao \textit{et al.} \cite{zhao09}. The curves shown in  Fig.\
\ref{fig:BQ_dispersions} illustrate that perfect agreement
is possible for small energies and perpendicular to the 
spin stripes, i.e., the direction of parallel spins. 
 We refrain from showing the results for
Diallo's experimental data. The results are very similar, despite the smaller
interlayer coupling derived from Diallo's INS data. The interested
reader is referred to Ref.\ \onlinecite{stane10}.

\begin{figure*}
  \centering {\includegraphics[width=\columnwidth]{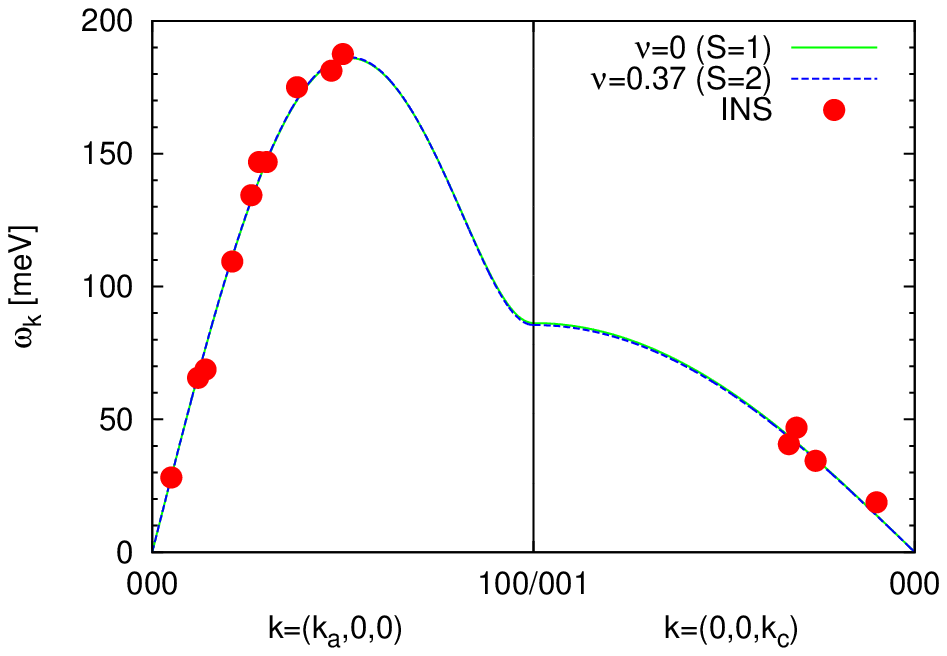}}
  {\includegraphics[width=\columnwidth]{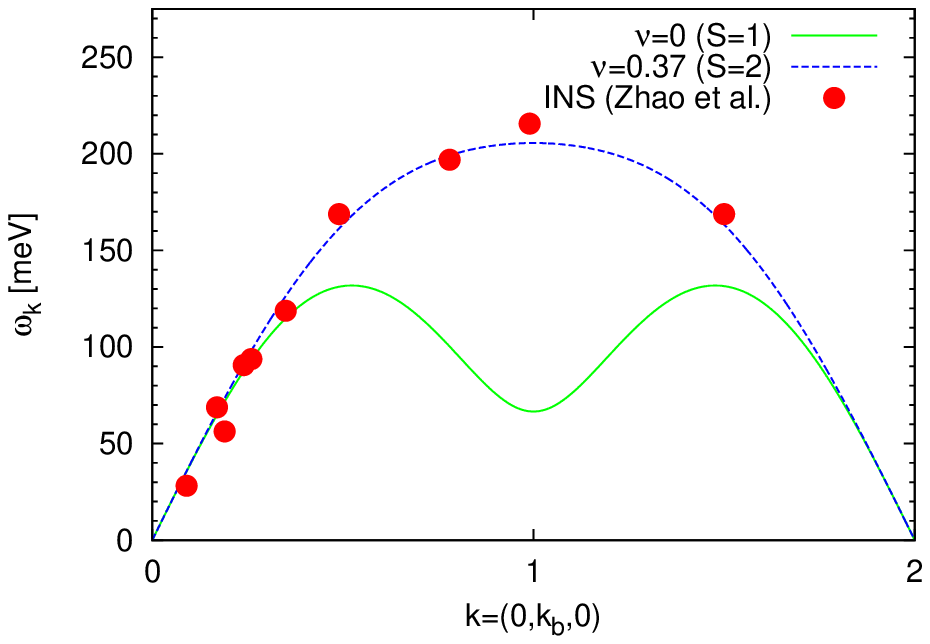}}
  \caption{(Color online) Same as in Fig.\ \ref{fig:BQ_dispersions}
  	but for $S=3/2$ and $S=2$, the parameters are given in Tab.\
  		\ref{tab:S_higher}.}
  \label{fig:BQ_dispersions_higher_S}
\end{figure*}

But even for an artificially large biquadratic exchange $\nu=3$ 
the qualitative behavior of the dispersion along the spin stripes
is not described satisfactorily, see right panel of Fig.\ \ref{fig:BQ_dispersions}.
At first glance, this comes as a surprise since naively a sufficiently
large biquadratic exchange should induce an arbitrary spatial anisotropy.
But the self-consistency prevents this, see Fig.\ \ref{fig:relative_anisotropy}
and Eq.\ \eqref{eq:BQ_x1b} and the discussion of them.

\begin{table}
  \centering
  \begin{tabular}{c||c|c|c|c||c|c|c|c}
 $S$ & $x$ & $\mu$ & $\nu$ & $J_2$ & $J_{1a}^{\text{eff}}$ &
 $J_{1b}^{\text{eff}}$ & $J_{c}^{\text{eff}}$ & $J_{2}^{\text{eff}}$ \\ \hline\hline
$3/2$ & 0.52 & 0.56 & 1.20 & 12.4 & 50.4 & -5.7 & 5.2 & 18.8 \\ \hline
2 & 0.78 & 0.37 & 0.37 & 9.3 & 50.4 & -5.7 & 5.2 & 18.8  \\ \hline \hline
Zhao \textit{et al.} & - & - & - & - & 49.9 & -5.7 & 5.3 & 18.9
  \end{tabular}
  \caption{Fit parameters of the model \eqref{eq:H_col3D_bq} in the three dimensional columnar
  	 phase for $S=3/2$ and $S=2$. The parameters are determined by fits to the experimental 
  	 dispersions in CaFe$_2$As$_2$ \cite{zhao09}. Exchange couplings in meV. For
    comparison, the exchange constants $J_i^\text{eff}$ obtained in linear spin
    wave theory by Zhao \textit{et al.} \cite{zhao09} are given.}
    	\label{tab:S_higher}
\end{table}

The understanding of the quantitative effect of the biquadratic term
tells us that only a larger spin value may lead to the experimentally
observed anisotropy. Indeed, for $S=2$ and for $S=3/2$ we achieve a perfect
fit, see Tab.\ \ref{tab:S_higher}. But for $S=3/2$ the required value of $  J_{\text{bq}}$ appears unreasonably large relative to $J_1$.
The resulting dispersion is shown in Fig.\ \ref{fig:BQ_dispersions_higher_S} for $S=2$. 
The dispersion obtained for $S=3/2$ (not shown)
look essentially the same. Note the excellent agreement obtained without
assuming any spatial anisotropy in the model itself. It is
the magnetic long-range directional Ising-type order which induces the strong
spatial anisotropy. We judge the fit parameters necessary
for $S=2$ to be perfectly reasonable, in particular the moderate value
of the biquadratic exchange of 37\%.

The above finding provides an interesting piece of information
for the description of the magnetic excitations in the pnictides 
by a model of localized spins. But it leaves open the issue
why the observed staggered moments are much lower than $4\mu_\text{B}$
which is the value one would expect for $S=2$. Here we can only
speculate that the \tr{complicated local electronic levels and the remaining 
itinerant character of the charges} are the physical reasons for this.
Also, the issue of line broadening due to Landau damping is
not included in the present model \cite{diall09,knoll10b,goswa11}.

All in all, the additional biquadratic exchange influences the
dispersion parallel to the spin stripes and strengthens the anisotropy of
the effective in-plane NN exchange constants. However, for $S=1$ it is not
possible to reproduce the whole experimental dispersion measured by Zhao
\textit{et al}.. This is possible for $S=3/2$ for parameters
which do not appear to be realistic. For $S=2$, the dispersion
can be described very well for realistic parameters. 
The question why the observed magnetic moments are
much smaller than one would expect for $S=2$ remains 
unresolved at present. \tr{The local electronic situation and the
residual itineracy are likely candidates to reduce the magnetic moment.}

\section{Dynamic Structure Factor}

Besides the dispersion it is the dynamic structure factor $S(\mathbf{k},\omega)$
which matters for the understanding of experimental observations. In addition
to the information about the energies of the collective excitations, 
$S(\mathbf{k},\omega)$ contains information about the relevant matrix elements.
In the Dyson-Maleev representation, the inelastic part of the dynamical
structure factor for $N$ spins at $T=0$ reads
\begin{align}
  S_0^{xx}\left(\mathbf{k},\omega\right)&=
  N\pi\left(S-n\right)\frac{A_\mathbf{k}-B_\mathbf{k}}{2\sqrt{A_\mathbf{k}^2-B_\mathbf{k}^2}}
  \delta\left(\omega-\omega_\mathbf{k}\right), 
  \label{eq:BQ_Sxx}
\end{align}
where $n$, $A_\mathbf{k}$ and $B_\mathbf{k}$ are defined in
Sect.\ \ref{sec:col3D_bq}.  Because of
the rotational symmetry about $S^z$ , $S^{yy}_T(\mathbf{k},\omega)$ is identical to
$S^{xx}_T(k,\omega)$. In the limit of $\mathbf{k}\rightarrow (0,0,0)$,
the dynamical structure factor \eqref{eq:BQ_Sxx} vanishes because
$A_\mathbf{k}=B_\mathbf{k}$. In the limit of $\mathbf{k}\rightarrow (1,0,1)$,
the dynamical structure factor \eqref{eq:BQ_Sxx} diverges because
$A_\mathbf{k}=-B_\mathbf{k}$. Note that in both cases we have
$\omega_\mathbf{k}\rightarrow 0$.  Similar results have been
derived within linear spin-wave theory in Ref.\ \onlinecite{yao10}.

\begin{figure*}
  \centering \subfigure[ $E=48\pm6$ meV,
  $k_c=3$]{\includegraphics[width=.5\columnwidth]{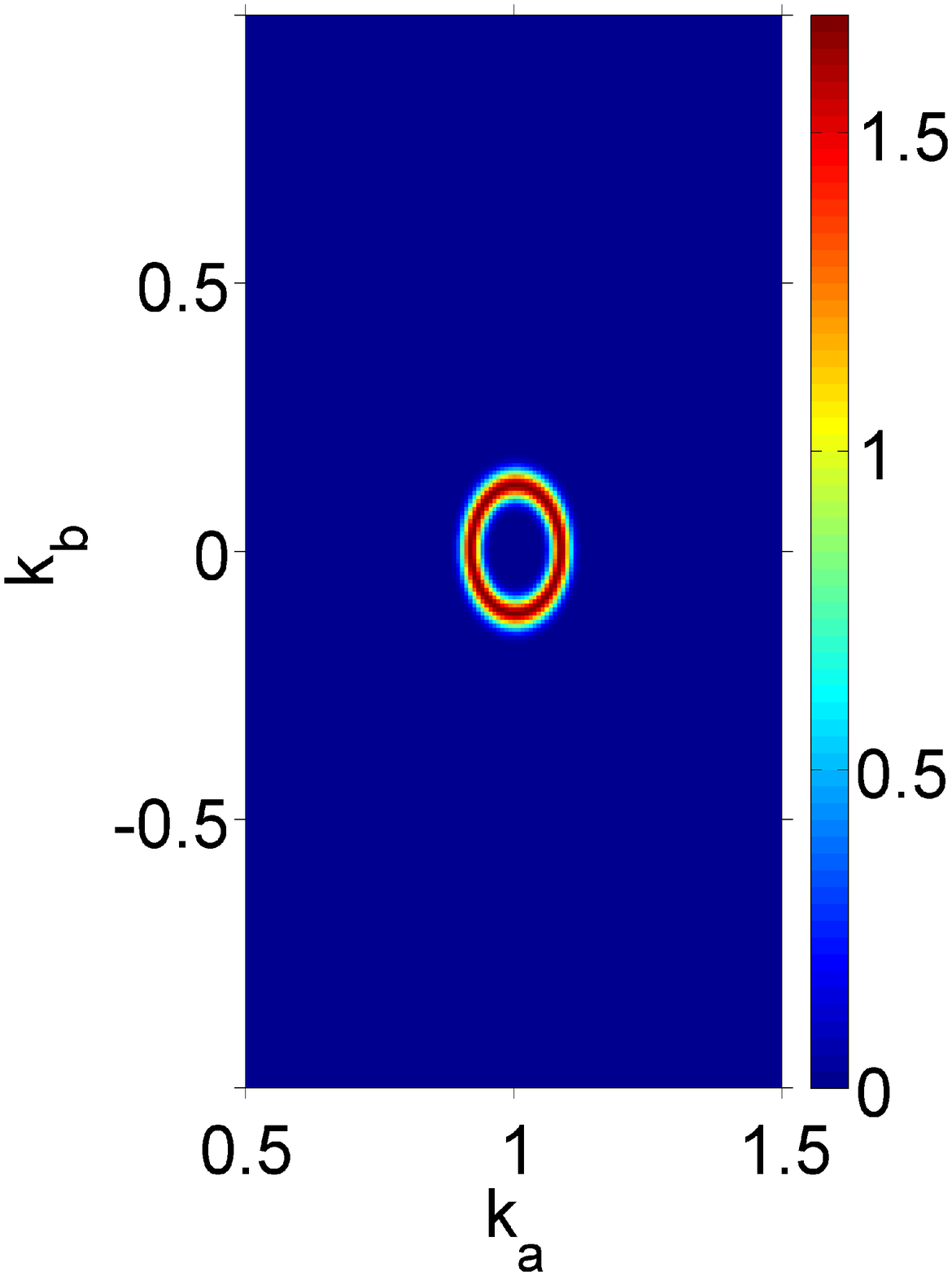}} \subfigure[
  $E=65\pm4$ meV,
  $k_c=3$]{\includegraphics[width=.5\columnwidth]{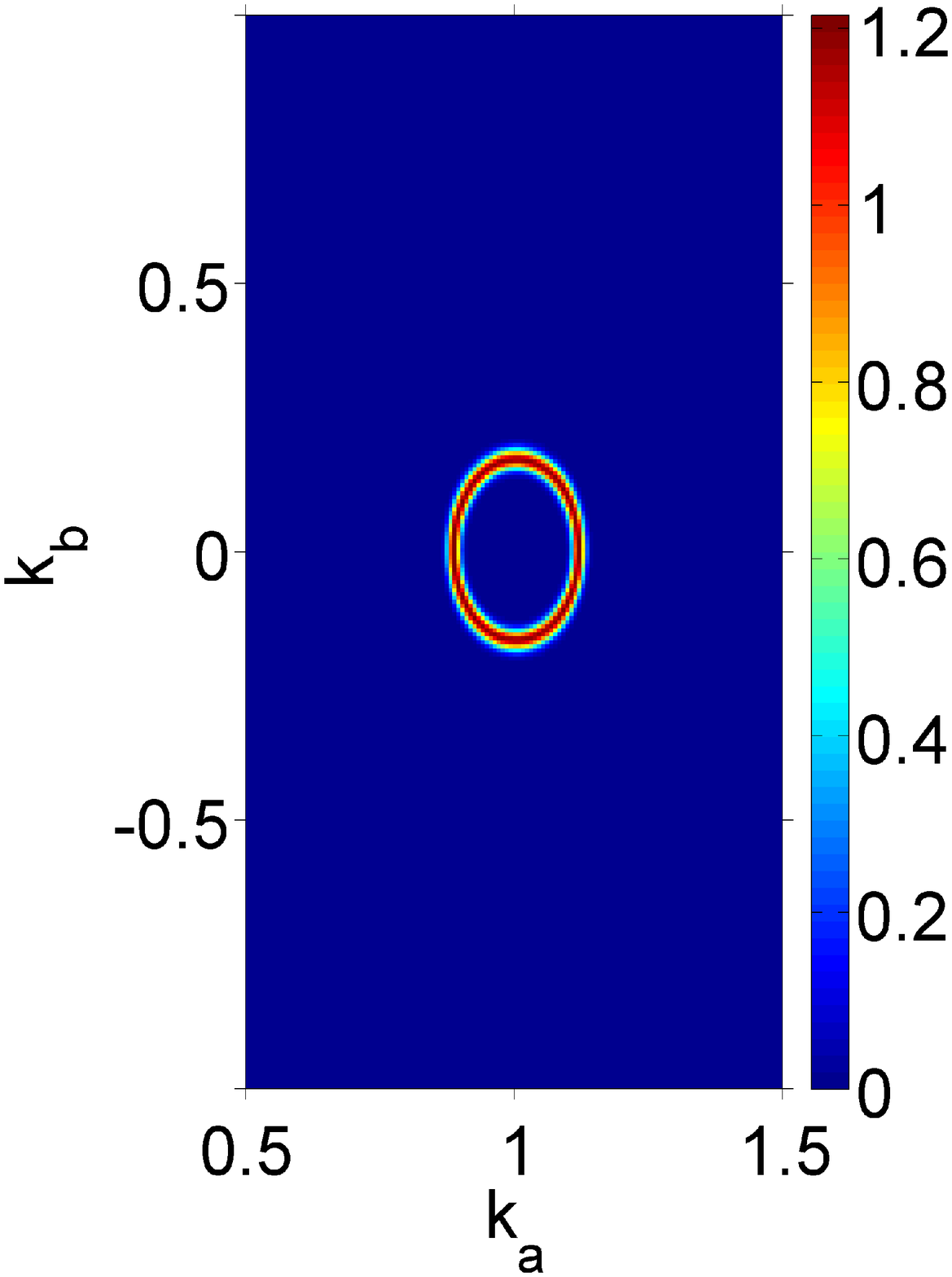}} \subfigure[
  $E=100\pm10$ meV,
  $k_c=3.5$]{\includegraphics[width=.5\columnwidth]{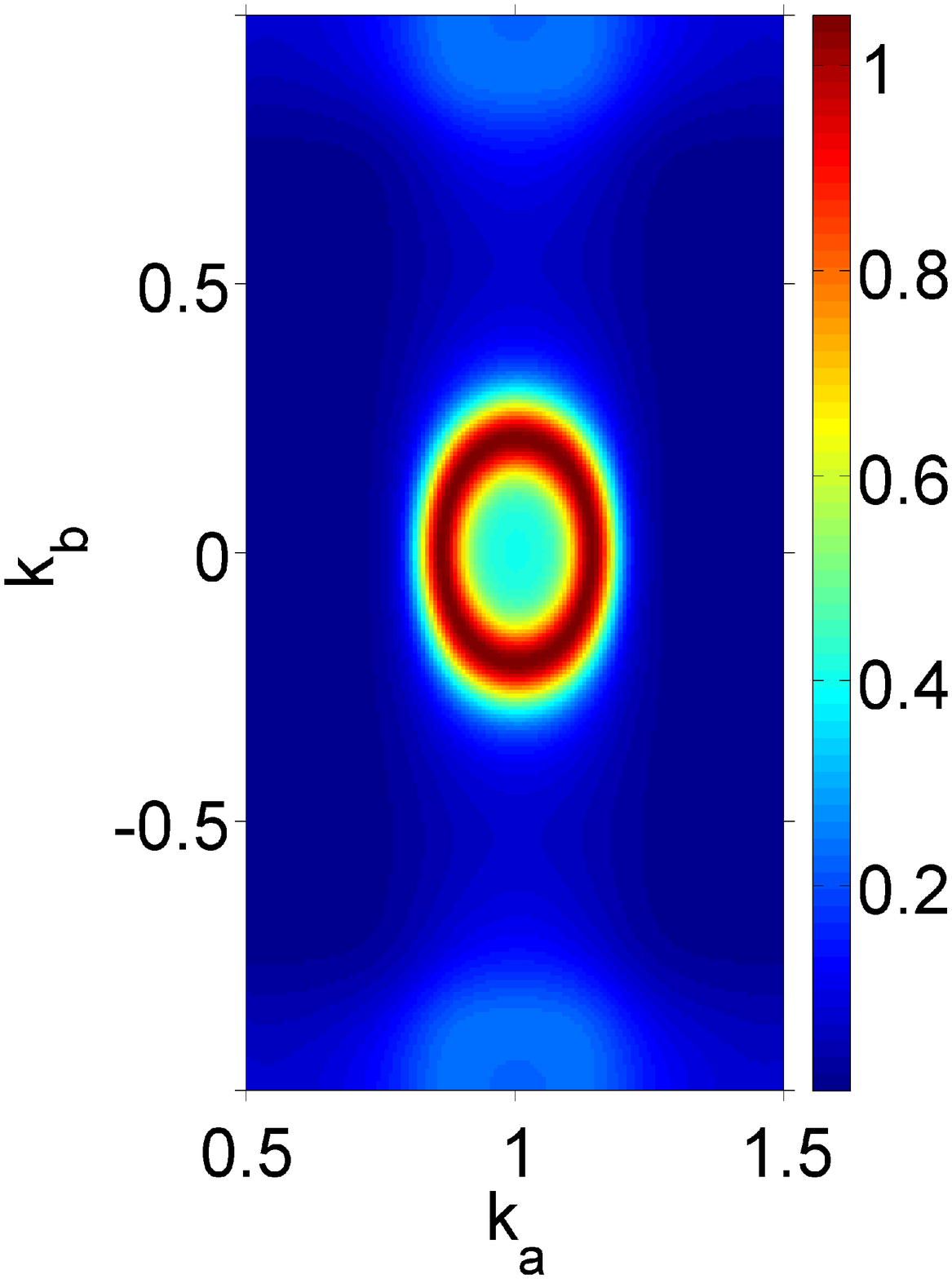}}
  \subfigure[ $E=115\pm10$ meV, $k_c=4$]{\includegraphics[width=.5\columnwidth]{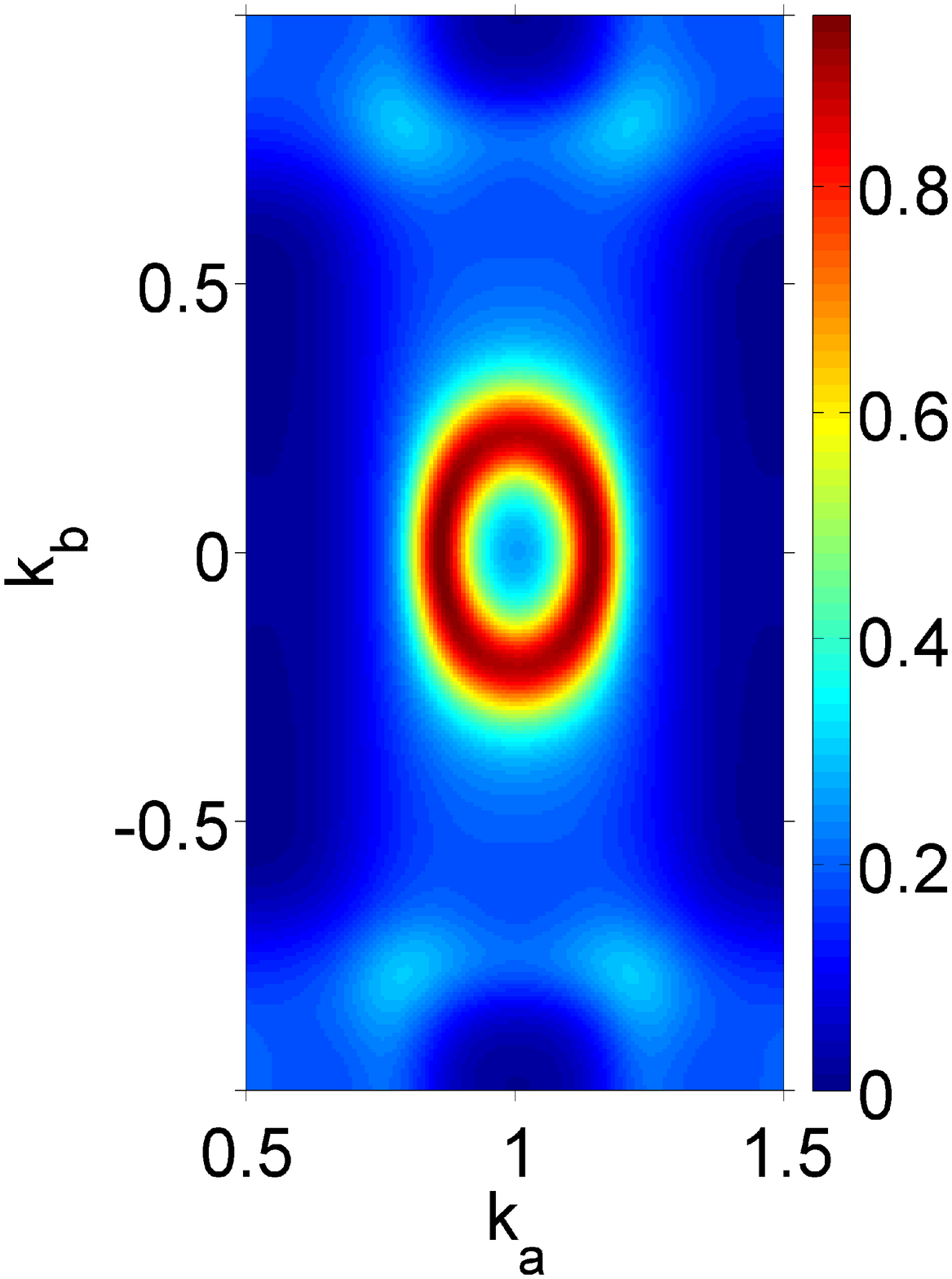}} \\
  \subfigure[ $E=137\pm15$ meV,
  $k_c=4$]{\includegraphics[width=.5\columnwidth]{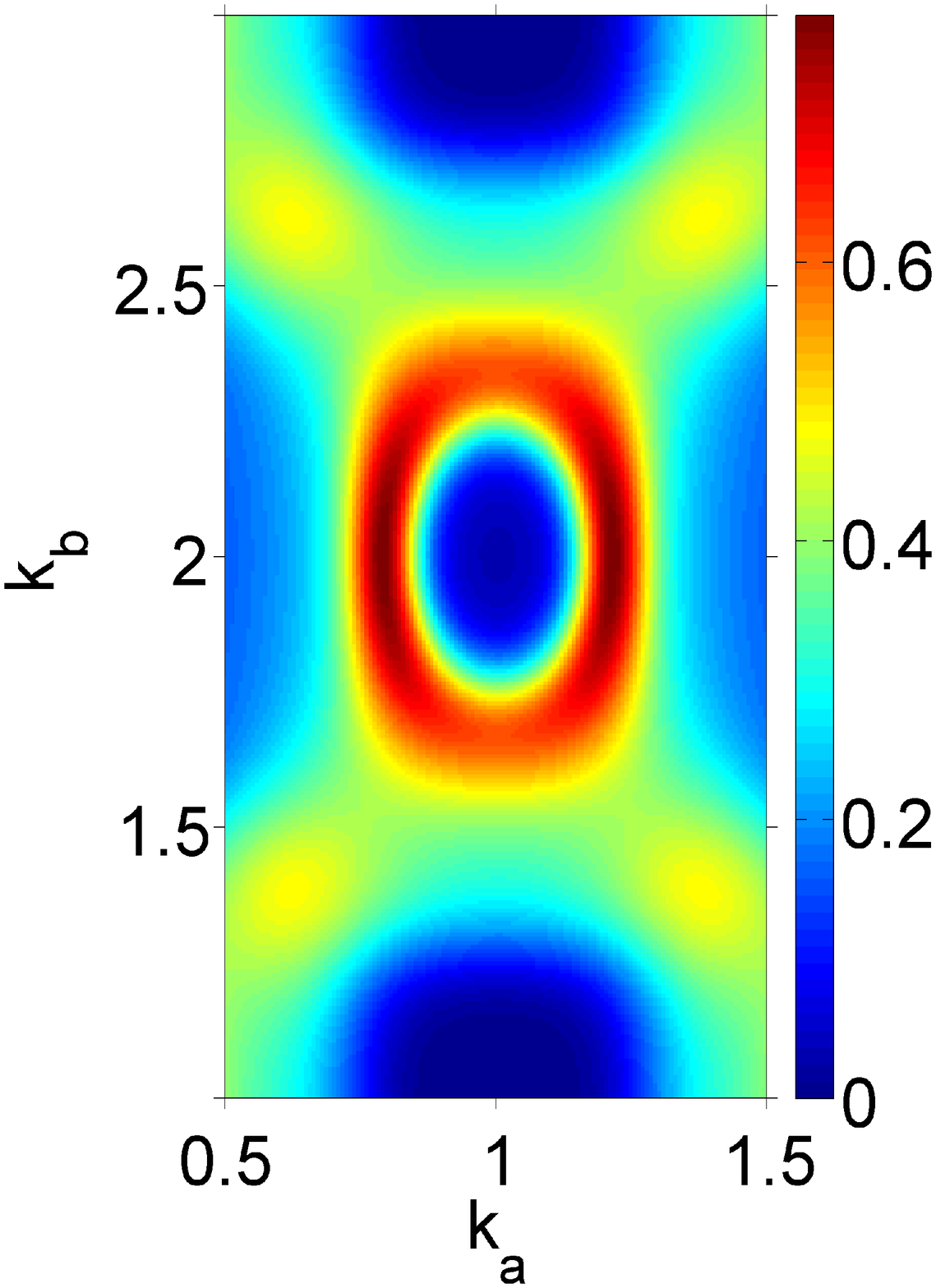}} \subfigure[
  $E=135\pm10$ meV,
  $k_c=4.5$]{\includegraphics[width=.5\columnwidth]{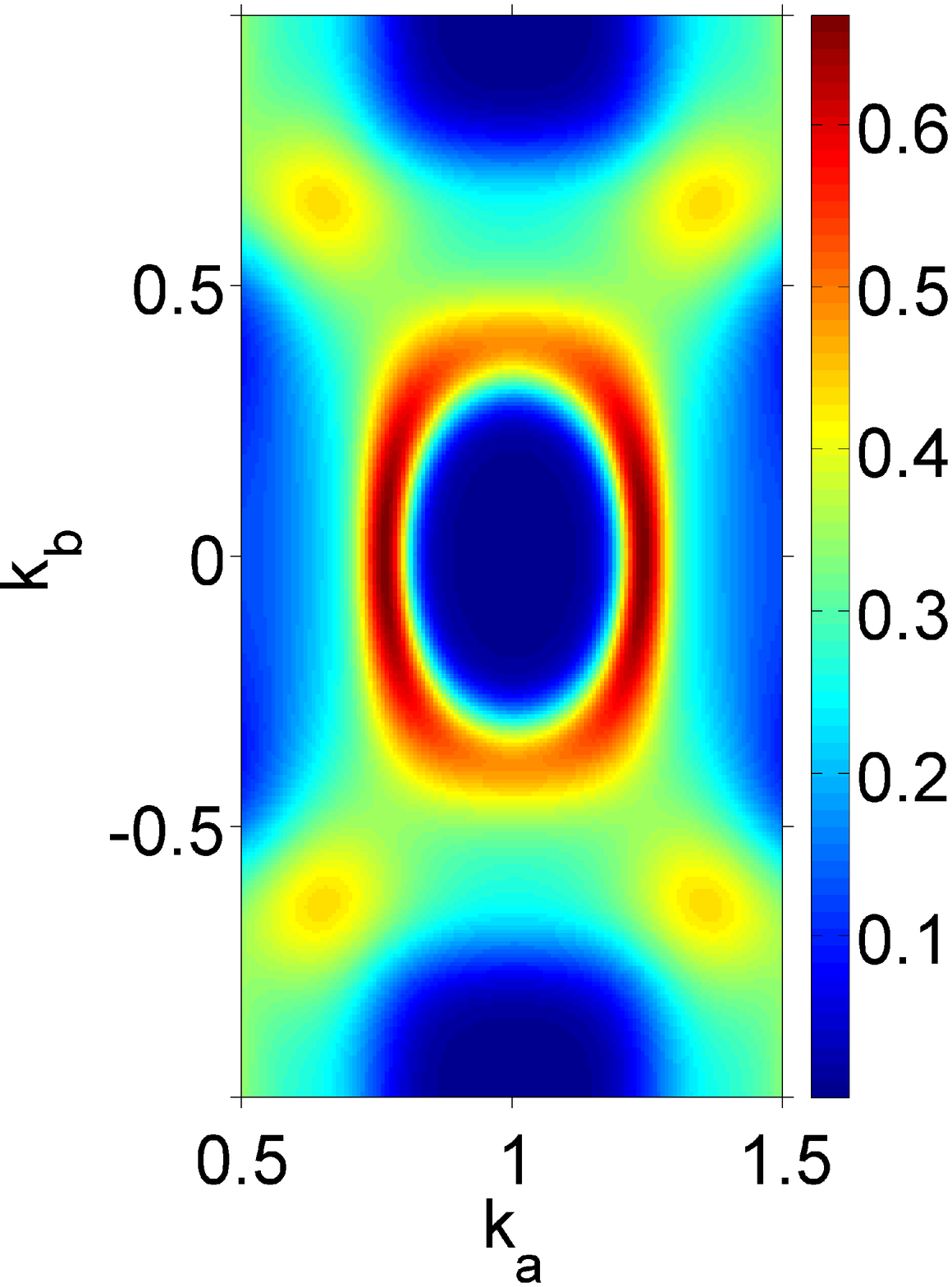}}
  \subfigure[ $E=144\pm15$ meV,
  $k_c=5$]{\includegraphics[width=.5\columnwidth]{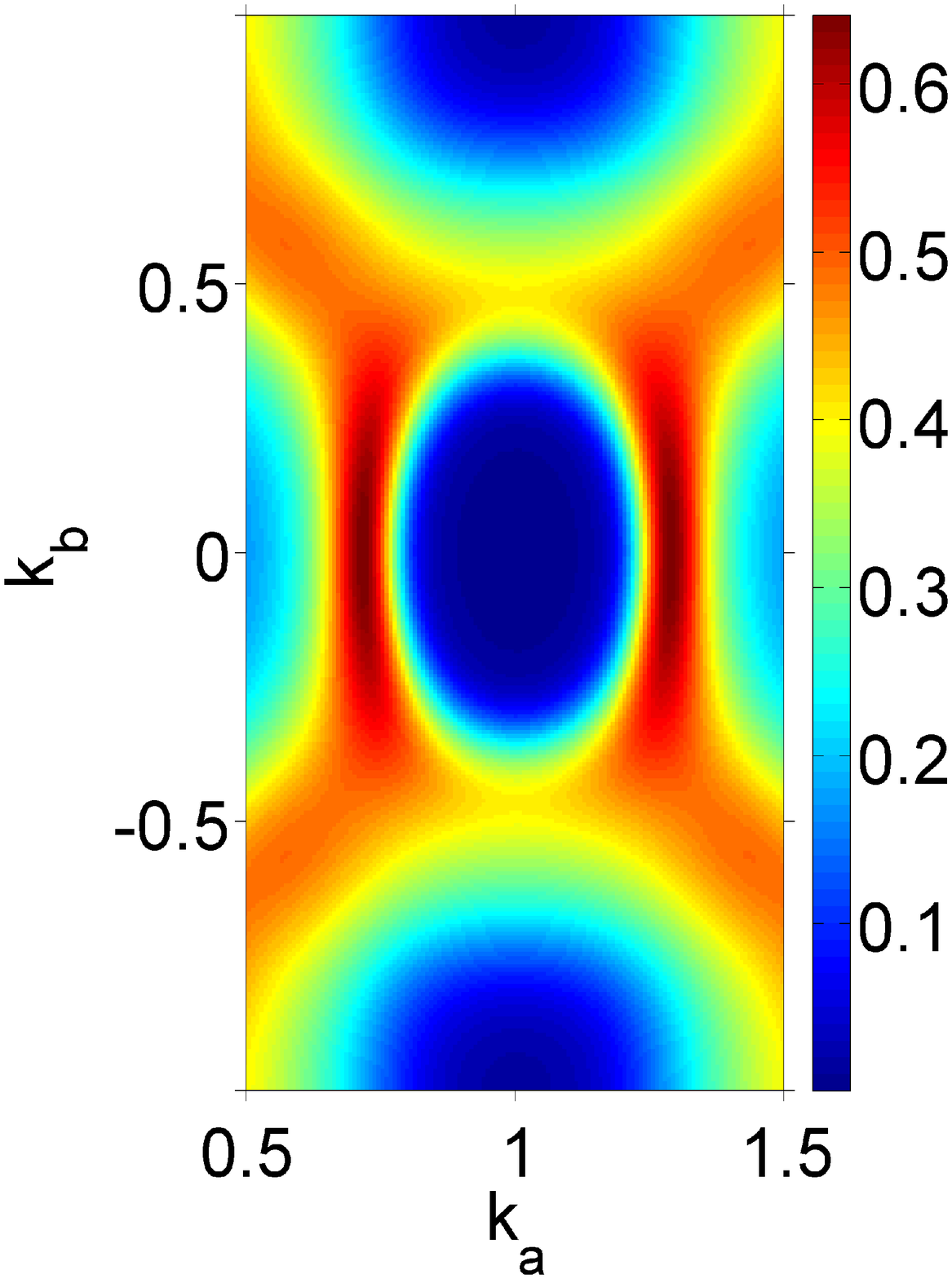}} \subfigure[
  $E=175\pm15$ meV,
  $k_c=5.2$]{\includegraphics[width=.5\columnwidth]{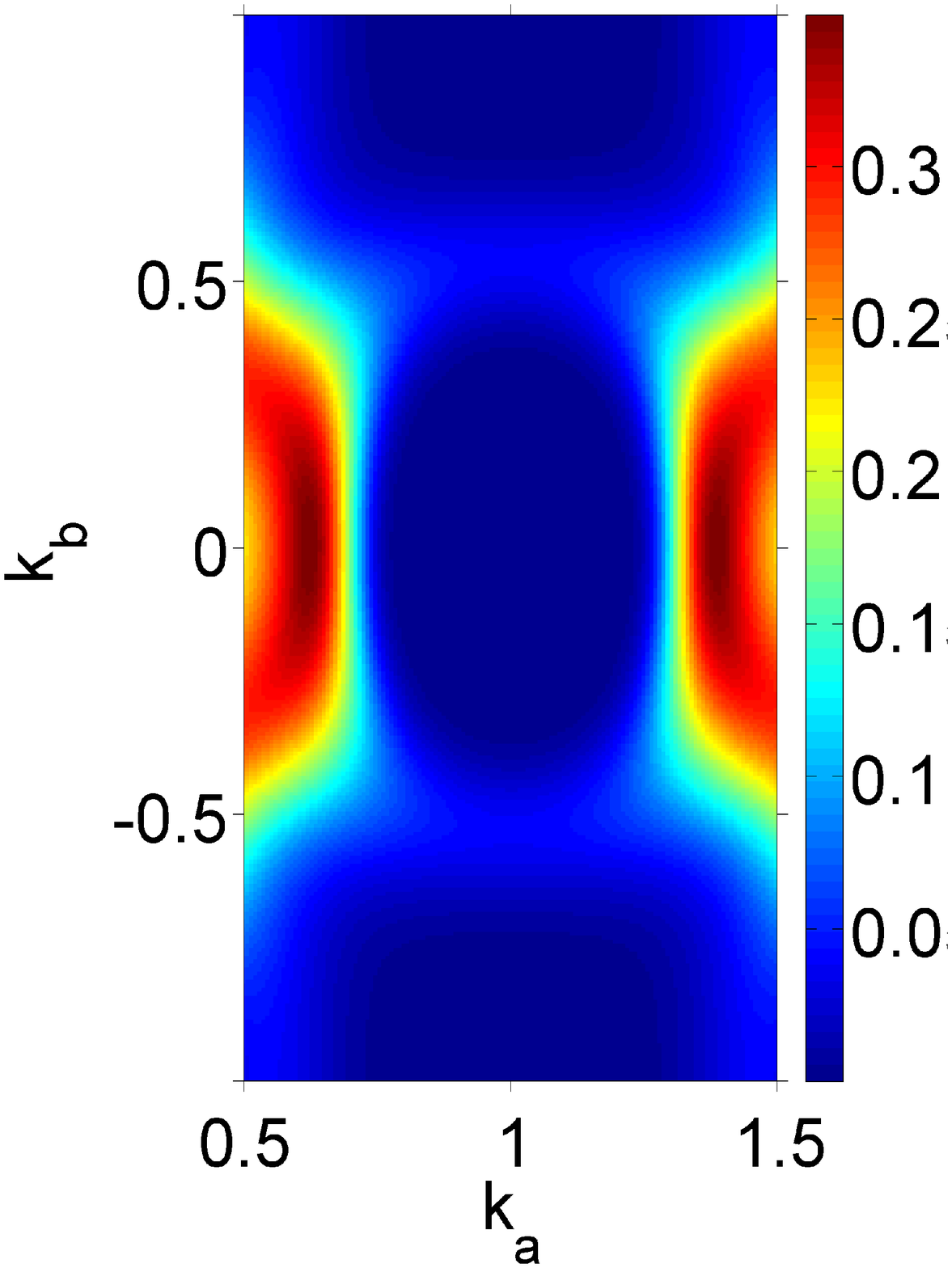}}
  \caption{(Color online) Constant-energy cuts of the dynamical structure
    factor per spin \eqref{eq:BQ_Sxx} integrated over the given energy interval
    for a twinned sample for $S=1$ and the
    	relative biquadratic exchange $\nu=0.5$. Other parameters are 
    	given in Tab.\ \ref{tab:BQ_results}. The range of $k_{a,b}$ and the values of $k_c$
    and $E$ have been chosen to match the rendering of the experimental data \cite{zhao09}.  
    	The reciprocal lattice vectors are given in units $\pi$/lattice
        constant. Note that the weights in the given energy intervals are dimensionless
according to formula \eqref{eq:BQ_Sxx}.}
  \label{fig:intensities_S1}
\end{figure*}

\begin{figure*}
  \centering \subfigure[ $E=48\pm6$ meV,
  $k_c=3$]{\includegraphics[width=.5\columnwidth]{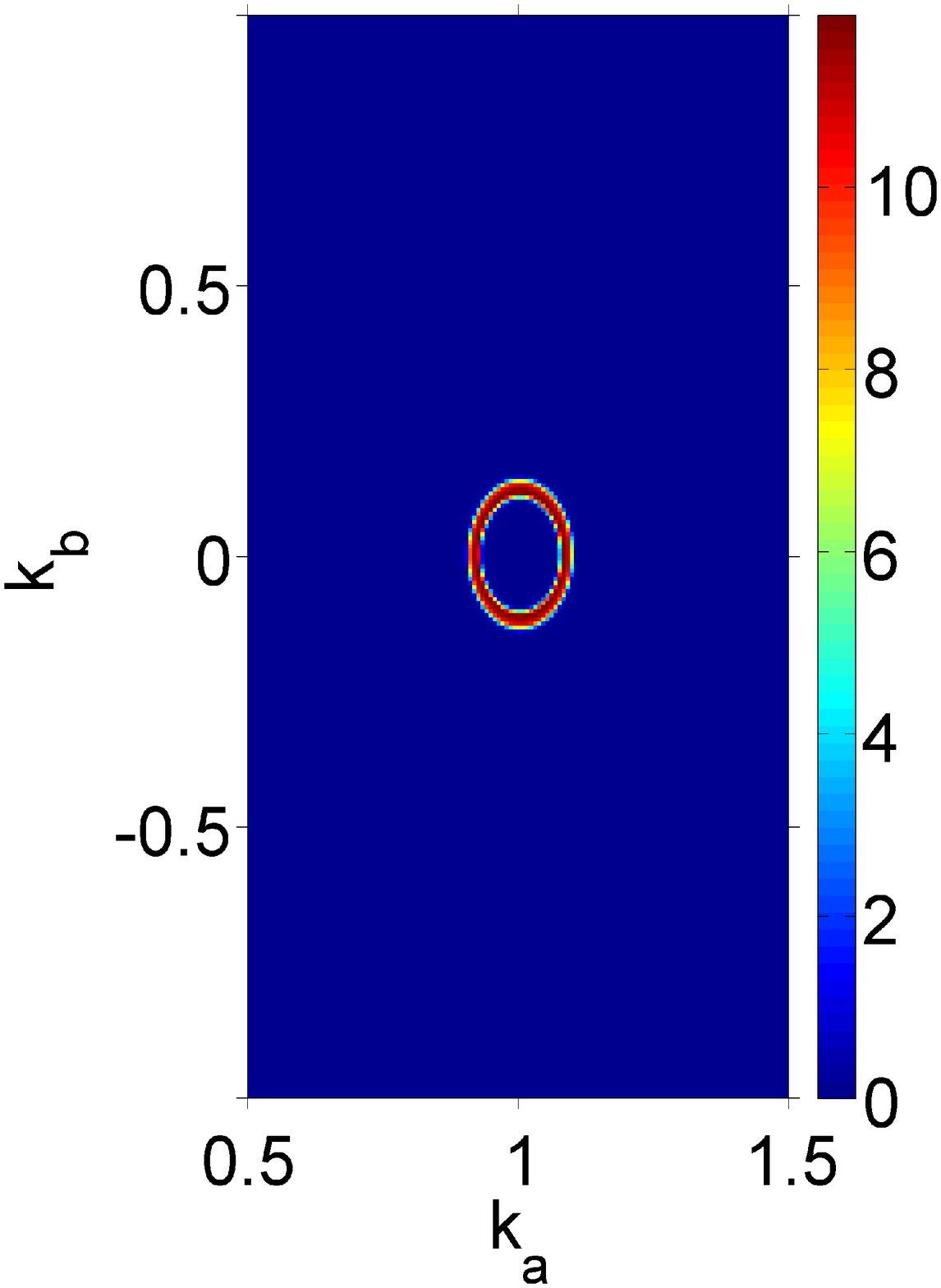}} \subfigure[
  $E=65\pm4$ meV,
  $k_c=3$]{\includegraphics[width=.5\columnwidth]{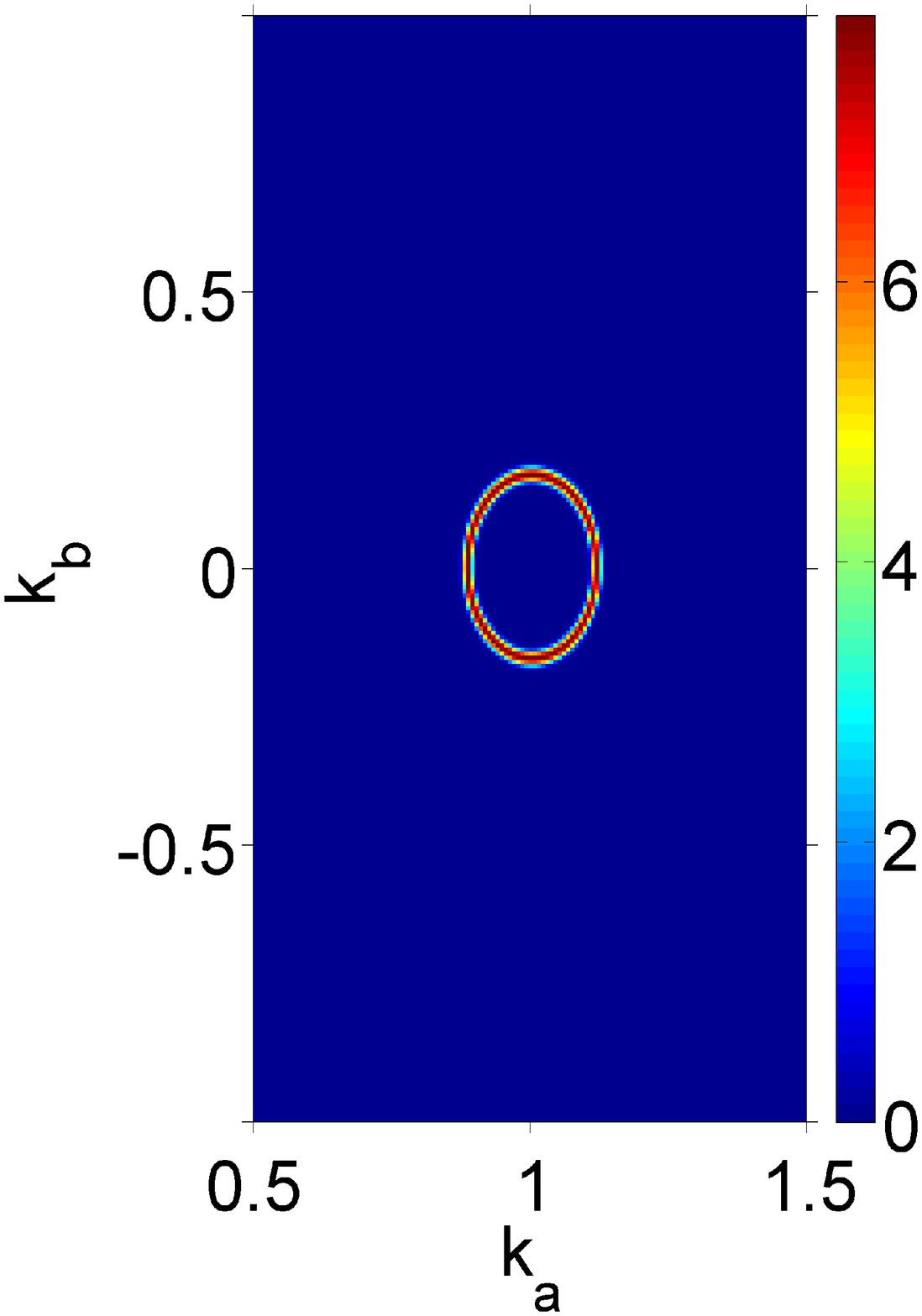}} \subfigure[
  $E=100\pm10$ meV,
  $k_c=3.5$]{\includegraphics[width=.5\columnwidth]{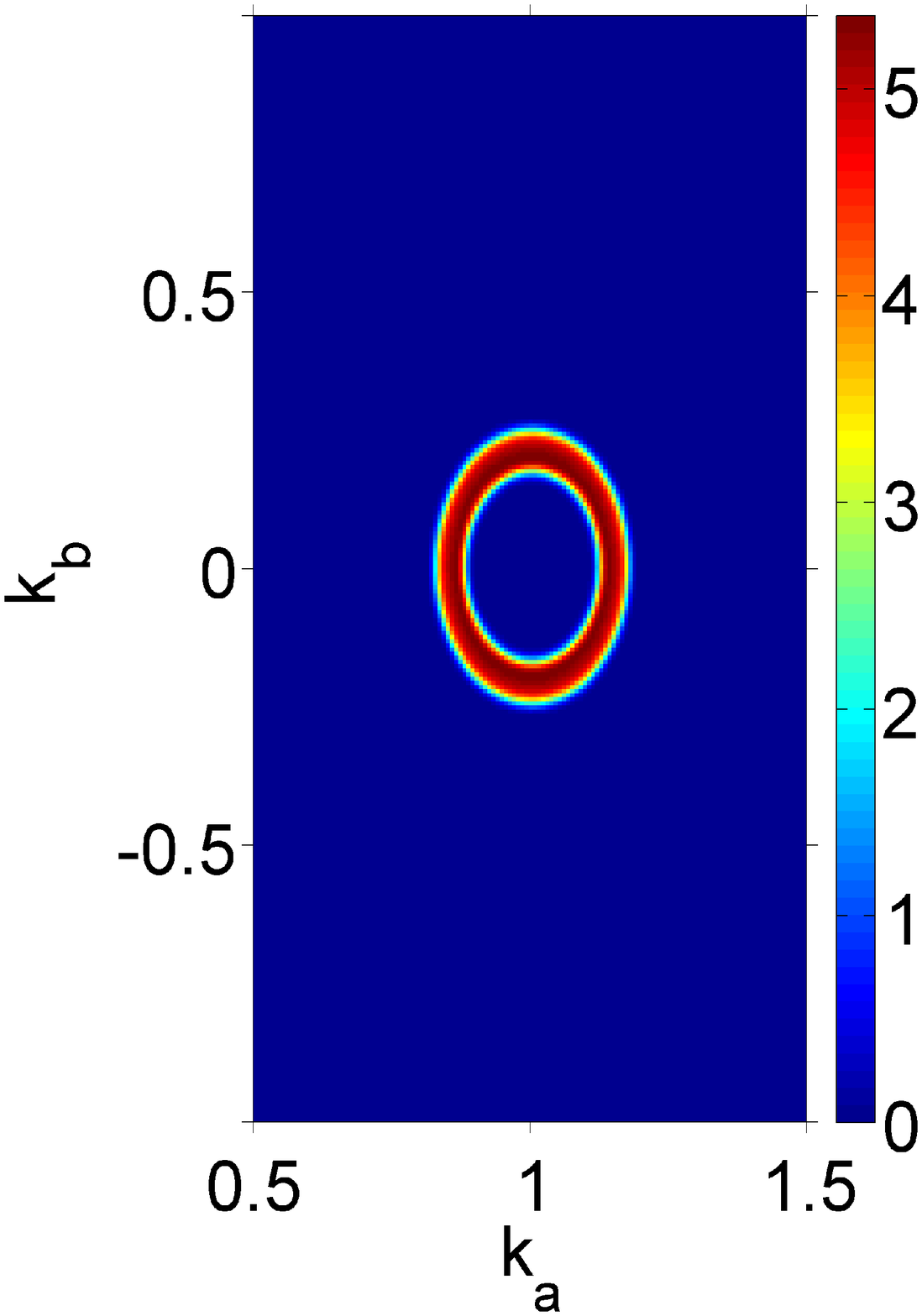}}
  \subfigure[ $E=115\pm10$ meV, $k_c=4$]{\includegraphics[width=.5\columnwidth]{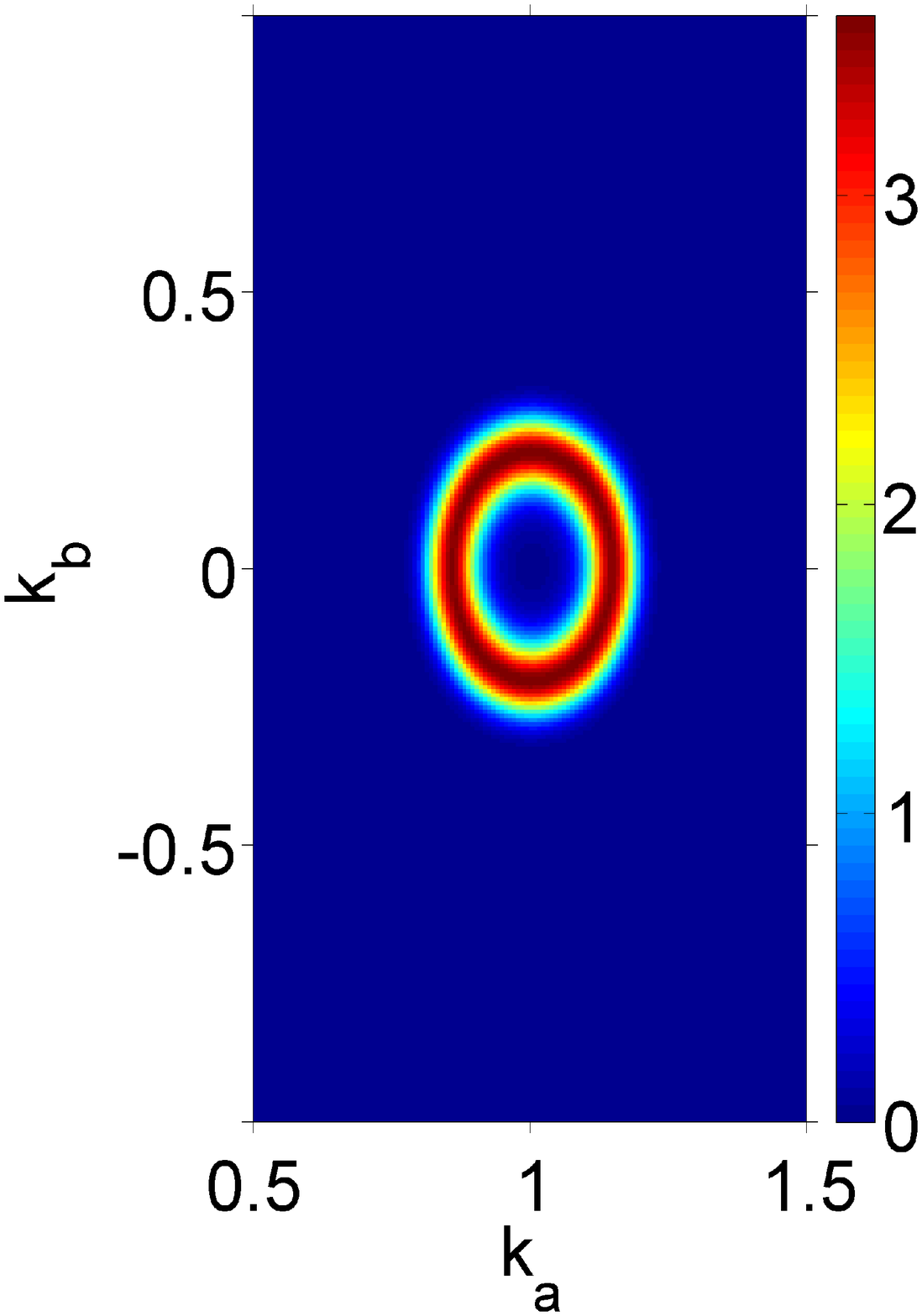}} \\
  \subfigure[ $E=137\pm15$ meV,
  $k_c=4$]{\includegraphics[width=.5\columnwidth]{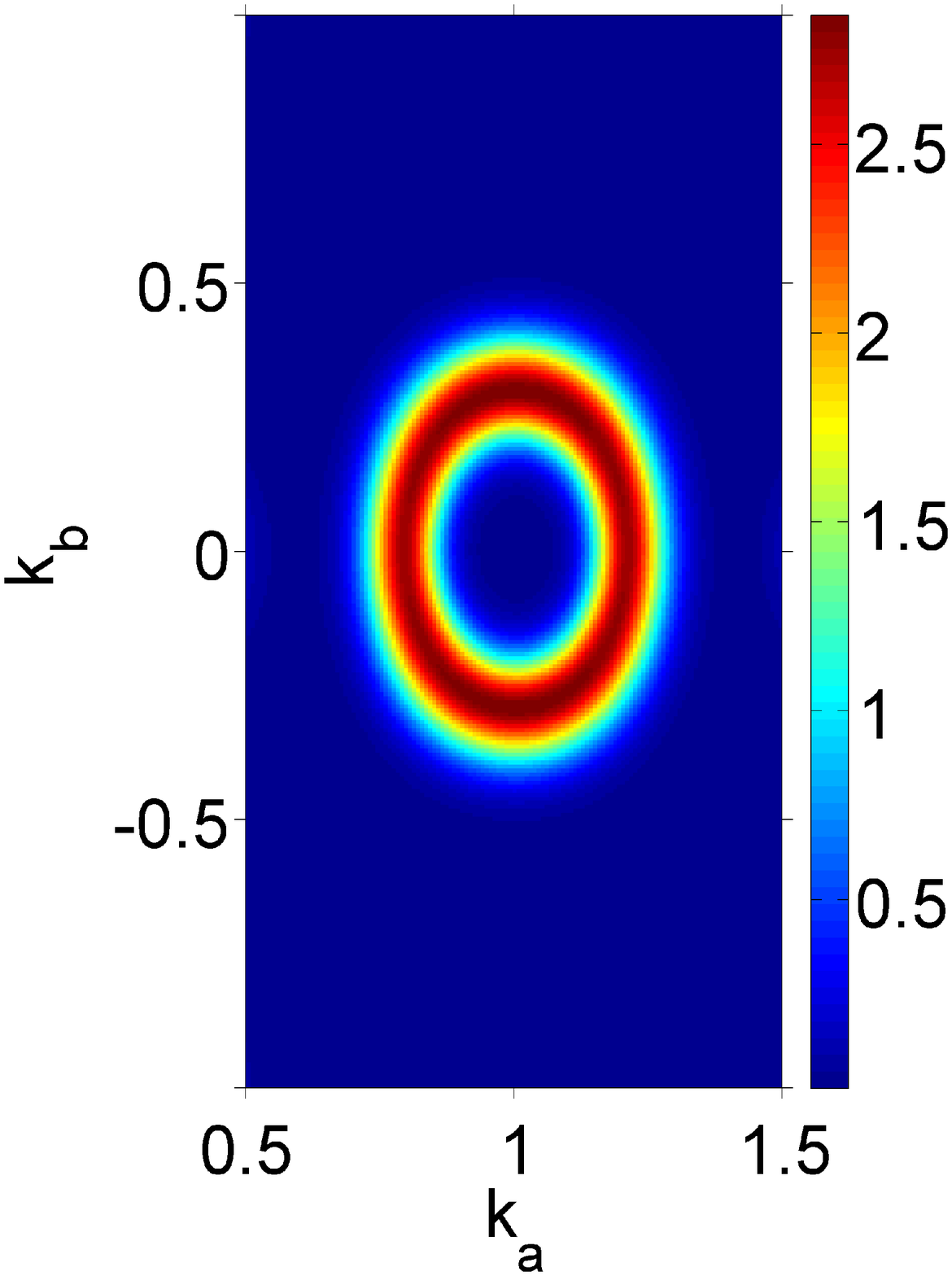}} \subfigure[
  $E=135\pm10$ meV,
  $k_c=4.5$]{\includegraphics[width=.5\columnwidth]{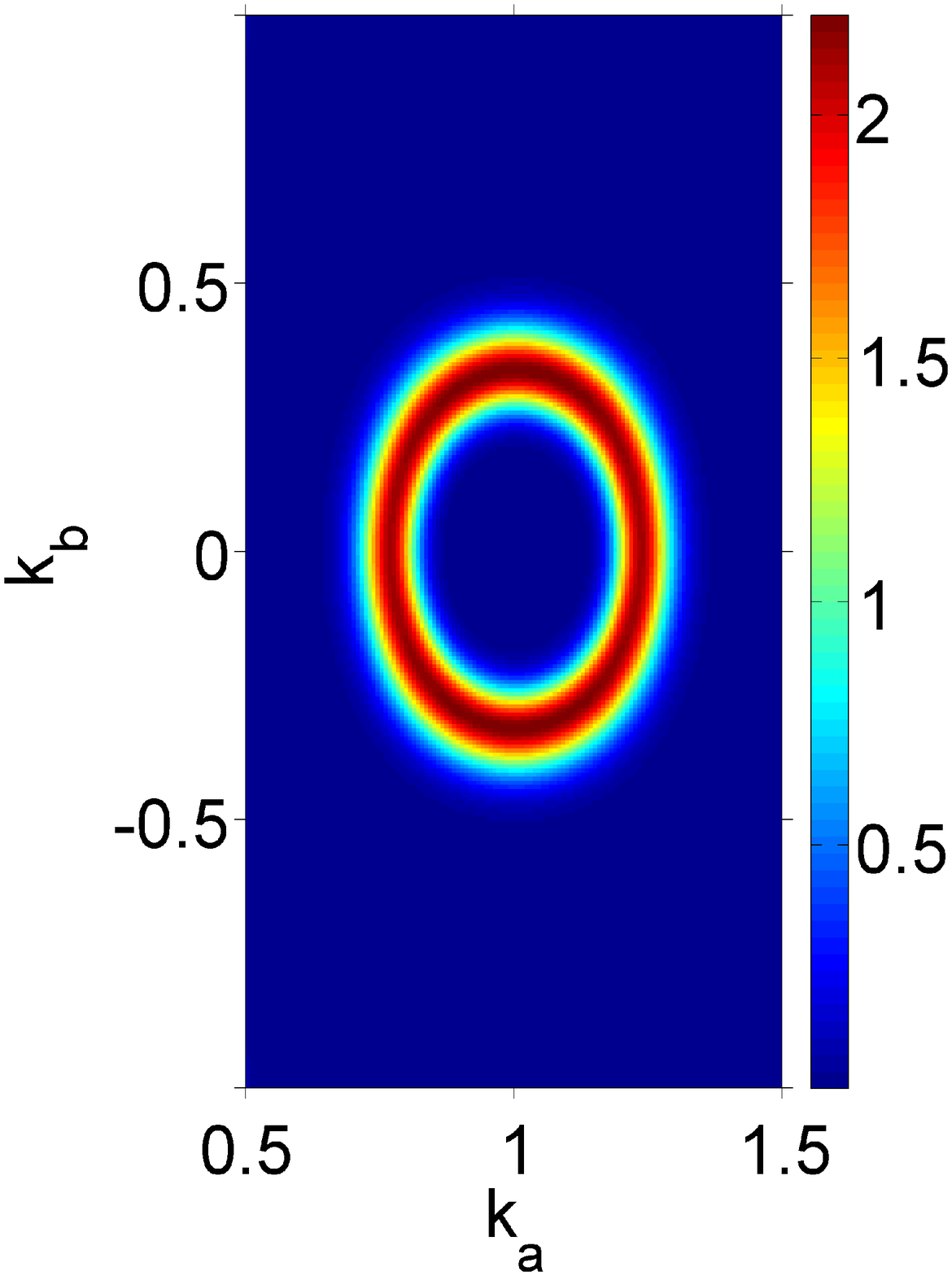}}
  \subfigure[ $E=144\pm15$ meV,
  $k_c=5$]{\includegraphics[width=.5\columnwidth]{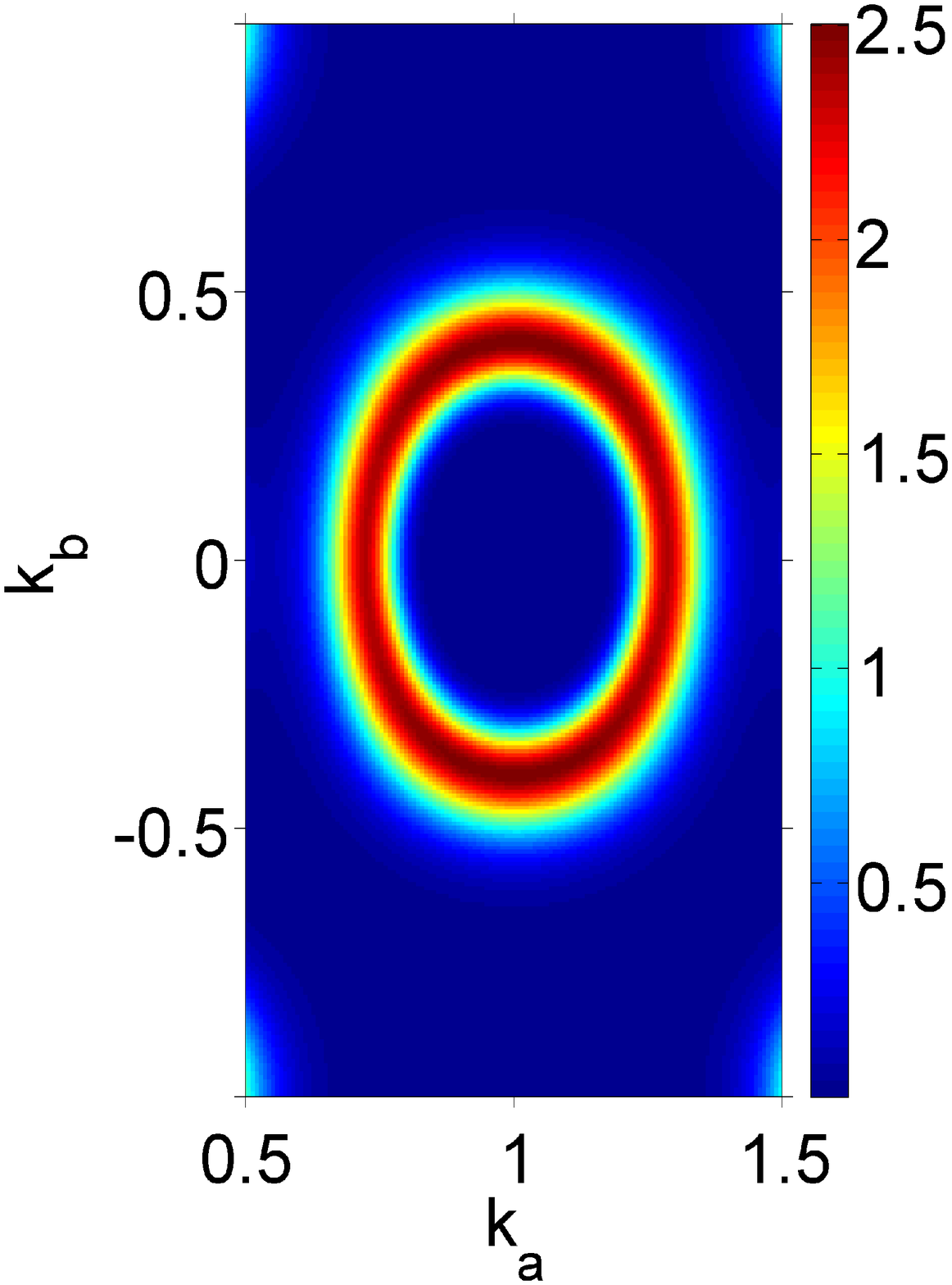}} \subfigure[
  $E=175\pm15$ meV,
  $k_c=5.2$]{\includegraphics[width=.5\columnwidth]{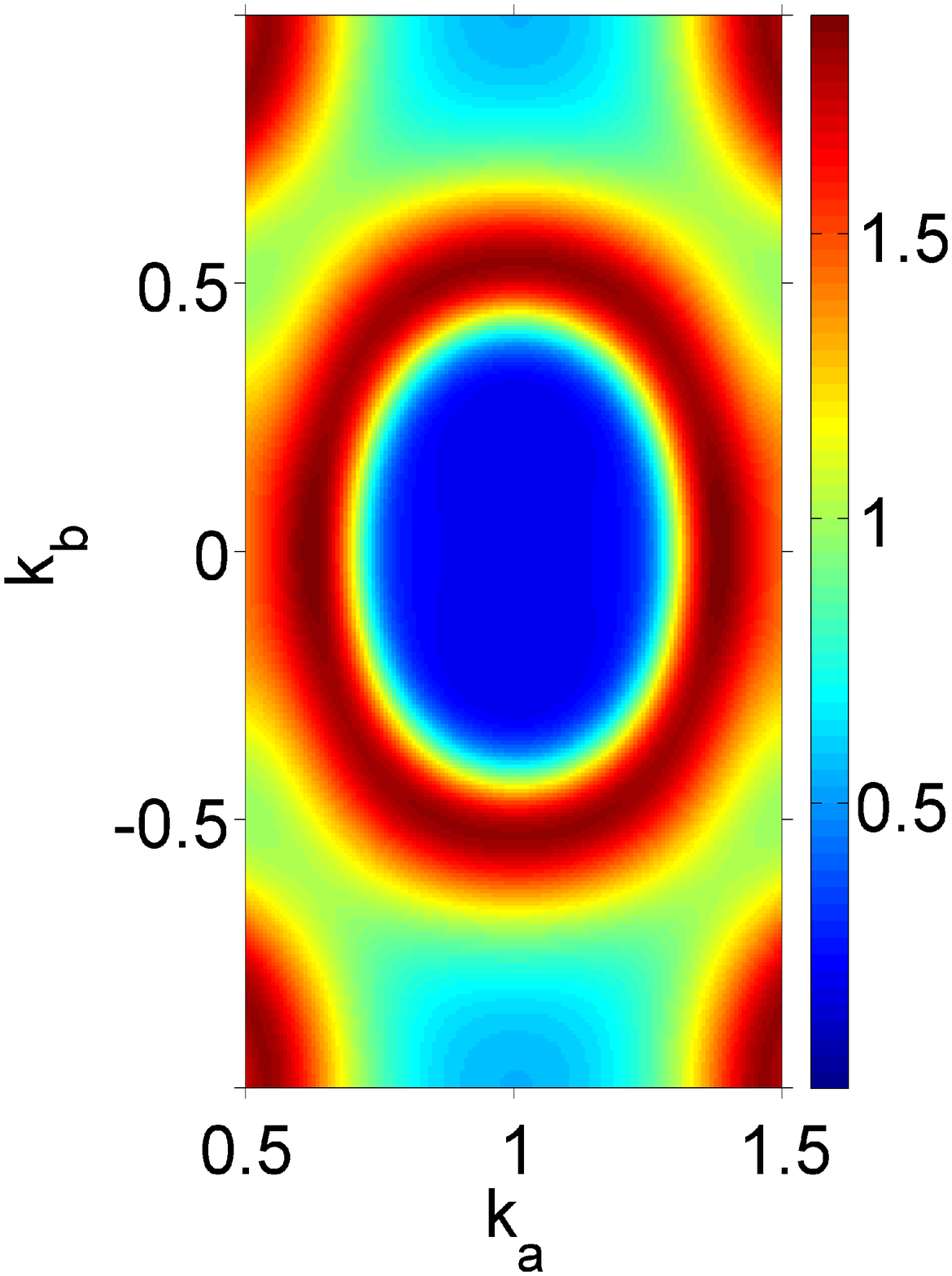}}
  \caption{Same as in the previous Fig.\ \ref{fig:intensities_S1}, but for $S=2$. 
  	The parameters are given in Tab.\ \ref{tab:S_higher}.}
  \label{fig:intensities_S2}
\end{figure*}

Constant-energy cuts are computed from \eqref{eq:BQ_Sxx} for equally weighted twinned domains.
This means that the spatially anisotropic result from \eqref{eq:BQ_Sxx} is superposed
for one choice of the crystallographic $a$ and $b$ directions and for the
swapped choice $a\leftrightarrow b$ so that the resulting superposition is spatially isotropic
in the sense that there is no difference between the $a$ and the $b$ direction.
The results for $S=1$ are shown in Fig.\ \ref{fig:intensities_S1}; those
for $S=2$ in Fig.\ \ref{fig:intensities_S2}. They are to be compared
with the experimental findings in Ref.\ \onlinecite{zhao09}.

For low energies, concentric rings emerge from the magnetic ordering
vector $Q=(1,0,3)$. They display a certain ellipticity which is not surprising
in view of the spatial anisotropy of the spin order. 
The rings increase in size for higher energies. 
For $S=1$ (Fig.\ \ref{fig:intensities_S1}), less intensive spots $(1,1,k_c)$ appear additionally 
for $E=100$ meV and $E=115$ meV which merge with the concentric spin-wave rings for higher energies. 
For $S=2$  (Fig.\ \ref{fig:intensities_S2}), the circular signatures of the dispersion cones
persist up to 144meV; only for even higher energies significant additional features occur. In
general, one observes the trend of decreasing intensity for increasing energy
because $S(\mathbf{k},\omega)\propto 1/\omega_\mathbf{k}$, see Eq.\ \eqref{eq:BQ_Sxx}.  

The comparison to the experimental data \cite{zhao09} shows that the $S=2$ results
match better which is not surprising since they reproduce the dispersion
everywhere, see Fig.\ \ref{fig:BQ_dispersions_higher_S}. The experimental data
displays also rings of high scattering intensities which increase very much 
like the theoretical results do. But the experimental results are broader
indicating a finite life-time of the magnetic excitations. This effect is
lacking in our model for two reasons: (i) The spin-only model does not include
any Landau damping due to the decay of magnons into electronic particle-hole pairs.
The consideration of this effect would require to pass to a doped $t$-$J$ type
of model or to switch completely to an itinerant approach, see for instance
Ref.\ \onlinecite{knoll10b}. (ii) Even in the framework of a spin-only
model the scattering of magnons from other magnons will lead to life-time
effects which are not taken into account by a mean-field approach.

Hence, the agreement of Fig.\ \ref{fig:intensities_S2} to the experimental data
is encouraging, in particular in view of the neglect of damping effects.
We conclude that the description of the magnetic excitations of undoped
pnictides within a model of localized spins is possible if a significant
biquadratic exchange is included and $S=2$ is chosen.

\section{Conclusions}

In this paper, we obtained the following results:\\
First, we pointed out that the instability of the columnar phase,
i.e., the phase with spin stripes, can become unstable in two 
qualitatively different ways in three dimensions. Either the
sublattice magnetization vanishes so that the long-range order
parameter vanishes (magnetization-driven scenario).
Or the smallest spin-wave velocity vanishes (velocity-driven scenario). 
The latter scenario is not possible in two dimensions because
a vanishing velocity implies a logarithmically diverging correction
of the quantum corrections of the magnetization so that the
magnetization will always vanish before the velocity becomes zero.
But the subtlety of a logarithmic divergence makes it numerically difficult
to detect the vanishing magnetization, in particular for large values
of the spin. \tr{We presume that the magnetization-driven scenario
leads to a weak first order transition to a possible intermediate phase
and that the velocity-driven scenario indicates a strong first
order transition to the N\'eel phase.}

Second, we discussed the importance of a biquadratic exchange.
The motivation is the very distinctive spatial anisotropy of
the dispersion in the pnictides which cannot be induced
by the small distortive anisotropy. In order to be able to
treat biquadratic exchange on the same footing as we treated
nearest-neighbor Heisenberg Hamiltonians before we studied
two mean-field approaches to biquadratic exchange. One is
based on the Dyson-Maleev representation, the other on the
Schwinger boson representation. Surprisingly, even at
zero temperature these mean-field theories provide distinctively 
different results. We tested both approaches
for the N\'eel phase of the nearest-neighbor Heisenberg on the square
lattice against series expansion data and exact diagonalization.
We established that the Dyson-Maleev mean-field theory provides
reliable results within about 10\% of the contribution of
the biquadratic exchange.

Third, we applied the developed mean-field approach to the
columnar phase of the $J_1$-$  J_{\text{bq}}$-$J_2$-$J_c$ Heisenberg model
in three dimensions. The aim was to explain the 
magnetic dispersions  observed in the undoped iron pnictides,
parent compounds of a novel class of superconductors.
It turned out that a biquadratic exchange indeed enhances
the spatial anisotropy of the spin-wave velocities.
But the effect is reduced by quantum corrections so that
it is not large enough for $S=1$ to match the experimental situation.
Only for larger values of $S$ an almost vanishing
effective coupling in $b$ direction along the stripes of parallel spins
is possible. For $S=3/2$, however, the required exchange
couplings are unlikely. For instance, the biquadratic exchange
would have to be larger than the nearest neighbor exchange.
But for $S=2$ perfect agreement of the dispersion is obtained without
invoking any spatial anisotropy of the model itself.
Note that $S=2$ corresponds to the chemical valence Fe$^{2+}$
which follows from an ionic balance of charge in the pnictides.
This valence implies four holes in the d-shell so that
the maximum spin according to Hund's rule is $S=2$.
We provided also results for the dynamic structure factor
for the matching model at $S=2$ and for $S=1$ for comparison.

Of course, there are open points which are beyond the scope
of the present article. First, there is the relatively
small staggered moments seen in experiment which indicate
rather smaller than larger spin values. \tr{The
complicated local electronic situation and the residual
itineracy of the charges are likely candidates to cause
this discrepancy.} But we mention
that recent experiments also found large moments in related
substances which are also superconducting.
Second, the spin-only model considered here does not
include the important charge degrees of freedom which
generate Landau damping and the bad metallic
behavior of the externally undoped parent pnictides. 
So life-time effects are not treated.

Further research is called for: 
\tr{On the level of localized spin models the investigation of
the recently found high-spin substances appears to be interesting
but may require the more challenging treatment of canted
magnetic order. On the level of itinerant charges the influence of doped
charges has to be studied. These charges} can be externally 
doped or internally self-doped
between the bands. Furthermore, it would be interesting
to learn more about the values for biquadratic exchanges
from density-functional calculations.

\acknowledgments

The authors are indebted to Jaan Oitmaa for providing
the series expansion data in Fig.\ \ref{fig:2DNeel_comparison}, 
to Carsten Raas for providing
the exact diagonalization data in Fig.\ \ref{fig:2DNeel_comparison}
and to Alexander Yaresko for providing the data of Fig.\ 
\ref{fig:intro_angle_dep}. One of us (DS) gratefully acknowledges 
the financial support of the NRW Forschungsschule 
``Forschung mit Synchrotronstrahlung 
in den Nano- und Biowissenschaften''.

\bigskip

\appendix

\section{Dyson-Maleev representation}

\subsection{Antiferromagnetic exchange}

For antiparallel orientation of coupled spins, 
the Hamiltonian \eqref{eq:H_Neel}  is given in Dyson-Maleev representation by
\begin{align}
  H^\text{afm}&=H_\text{bl}^\text{afm}+H_\text{bq}^\text{afm},\label{eq:H_Neel_DM}
\end{align}
with
\begin{widetext}
  \begin{subequations}
    \begin{align}
      H^\text{afm}_\text{bl}&=J\sum\limits_{\langle i,j\rangle}
      \biggl\{-S^2+S\left(\hat{n}_i+\hat{n}_j+
      b^\dagger_ib^\dagger_j+b^{\phantom{\dagger}}_ib^{\phantom{\dagger}}_j\right)-
      \hat{n}_i\hat{n}_j-\frac{1}{2}\left(b^\dagger_i\hat{n}_ib^\dagger_j+
      b^{\phantom{\dagger}}_i\hat{n}_jb^{\phantom{\dagger}}_j\right)\biggr\}
      \\
      \begin{split}
        H^\text{afm}_\text{bq}&=
        -  J_{\text{bq}}\sum\limits_{\langle i,j\rangle}\biggl\{ S^4 -2S^3\left(\hat{n}_i+\hat{n}_j+\hat{I}_0\right)\biggr. 
        \\
        &\quad+S^2\left[\hat{n}^2_i+\hat{n}_j^2+4\hat{n}_i\hat{n}_j+
        \hat{I}^2_0+\left(\hat{n}_i+\hat{n}_j\right)\hat{I}_0+\hat{I}_0\left(\hat{n}_i+
        \hat{n}_j\right)+\hat{I}'_0\right]
        \\
        &\quad-S\left[2\left(\hat{n}_i\hat{n}_j^2+\hat{n}_i^2\hat{n}_j\right)+
        \hat{n}_i\hat{n}_j\hat{I}_0+\hat{I}_0\hat{n}_i\hat{n}_j+
        \frac{1}{2}\bigl(\hat{I}^{\phantom{\dagger}}_0\hat{I}'_0+\hat{I}'_0\hat{I}_0^{\phantom{\dagger}}+
        \left(\hat{n}_i+\hat{n}_j\right)\hat{I}'_0+\hat{I}_0'\left(\hat{n}_i+\hat{n}_j\right)\bigr)\right]
        \\
        &\biggl.\quad+\hat{n}_i^2\hat{n}_j^2+\frac{1}{4}\hat{I}_0'^2+
        \frac{1}{2}\left(\hat{n}_i\hat{n}_j\hat{I}_0'+\hat{I}_0'\hat{n}_i\hat{n}_j\right)\biggr\}
      \end{split}
    \end{align}
  \end{subequations}
\end{widetext}
where
\begin{subequations}
  \begin{align}
    \hat{I}_0&:=b^\dagger_ib^\dagger_j+b^{\phantom{\dagger}}_ib^{\phantom{\dagger}}_j \\
    \hat{I}_0'&:=b^\dagger_i\hat{n}_i^{\phantom{\dagger}}
    b^\dagger_j+b_i^{\phantom{\dagger}}\hat{n}_j^{\phantom{\dagger}} b_j^{\phantom{\dagger}}
  \end{align}
\end{subequations}
with $i,j$ being the two coupled sites. 
For the mean-field decoupling the reader is referred
to Eq.\ \eqref{eq:H_col3D_MF}.

\subsection{Ferromagnetic exchange}

For parallel orientation of coupled spins, the Hamiltonian \eqref{eq:H_Neel} in Dyson-Maleev
representation reads
\begin{align}
  H^\text{fm}&=H_\text{bl}^\text{fm}+H_\text{bq}^\text{fm}
\end{align}
where
\begin{widetext}
  \begin{subequations}
    \begin{align}
      H_\text{bl}^\text{fm}&=J\sum\limits_{\langle i,j\rangle}
      \left\{S^2-S\left(\hat{n}_i^{\phantom{\dagger}}+\hat{n}_j^{\phantom{\dagger}}-
      b^\dagger_ib^{\phantom{\dagger}}_j-b^\dagger_jb^{\phantom{\dagger}}_i\right)+
      \hat{n}_i^{\phantom{\dagger}}\hat{n}_j^{\phantom{\dagger}}-\frac{1}{2}\left(b^\dagger_i\hat{n}_i^{\phantom{\dagger}}
          b^{\phantom{\dagger}}_j+b^\dagger_j\hat{n}_j^{\phantom{\dagger}}
          b^{\phantom{\dagger}}_i\right)\right\} \\
      \begin{split}
        H_\text{bq}^\text{fm}&=-  J_{\text{bq}}\sum\limits_{\langle i,j\rangle} \biggl\{S^4-2S^3\left(\hat{n}_i+
        \hat{n}_j-\hat{F}_0\right)\biggr. \\
        &\quad+S^2\left[\hat{n}_i^2+\hat{n}_j^2+4\hat{n}_i\hat{n}_j+
        \hat{F}^2_0-\left(\hat{n}_i+\hat{n}_j\right)\hat{F}_0-\hat{F}_0\left(\hat{n}_i+
        \hat{n}_j\right)-\hat{F}'_0\right]
        \\
        &\quad+S\left[-2\left(\hat{n}_i\hat{n}_j^2+\hat{n}_i^2\hat{n}_j\right)+
        \hat{n}_i\hat{n}_j\hat{F}_0+\hat{F}_0\hat{n}_i\hat{n}_j+
        \frac{1}{2}\bigl(-\hat{F}^{\phantom{\dagger}}_0\hat{F}_0'-\hat{F}_0'\hat{F}^{\phantom{\dagger}}_0+
        \left(\hat{n}_i+\hat{n}_j\right)\hat{F}_0'+\hat{F}_0'\left(\hat{n}_i+\hat{n}_j\right)\bigr)\right]
        \\
        &\quad\biggl.+\hat{n}^2_i\hat{n}_j^2+\frac{1}{4}\hat{F}_0'^2-
        \frac{1}{2}\left(\hat{n}^{\phantom{\dagger}}_i\hat{n}^{\phantom{\dagger}}_j\hat{F}'_0+
        \hat{F}'_0\hat{n}^{\phantom{\dagger}}_i\hat{n}^{\phantom{\dagger}}_j\right)\biggr\}
      \end{split}
    \end{align}
  \end{subequations}
\end{widetext}
with
\begin{subequations}
  \begin{align}
    \hat{F}_0&:=b^\dagger_ib^{\phantom{\dagger}}_j+b^\dagger_jb^{\phantom{\dagger}}_i \\
    \hat{F}'_0&:=b^\dagger_i\hat{n}_i^{\phantom{\dagger}}
    b^{\phantom{\dagger}}_j+b^\dagger_j\hat{n}_j^{\phantom{\dagger}} b^{\phantom{\dagger}}_i.
  \end{align}
\end{subequations}
For the mean-field decoupling the reader is referred
to Eq.\ \eqref{eq:H_col3D_MF}.

\bibliographystyle{apsrev}

\begin{thebibliography}{73}
\expandafter\ifx\csname natexlab\endcsname\relax\def\natexlab#1{#1}\fi
\expandafter\ifx\csname bibnamefont\endcsname\relax
  \def\bibnamefont#1{#1}\fi
\expandafter\ifx\csname bibfnamefont\endcsname\relax
  \def\bibfnamefont#1{#1}\fi
\expandafter\ifx\csname citenamefont\endcsname\relax
  \def\citenamefont#1{#1}\fi
\expandafter\ifx\csname url\endcsname\relax
  \def\url#1{\texttt{#1}}\fi
\expandafter\ifx\csname urlprefix\endcsname\relax\def\urlprefix{URL }\fi
\providecommand{\bibinfo}[2]{#2}
\providecommand{\eprint}[2][]{\url{#2}}

\bibitem[{\citenamefont{Lacroix et~al.}(2011)\citenamefont{Lacroix, Mendels,
  and Mila}}]{lacro11}
\bibinfo{editor}{\bibfnamefont{C.}~\bibnamefont{Lacroix}},
  \bibinfo{editor}{\bibfnamefont{P.}~\bibnamefont{Mendels}}, \bibnamefont{and}
  \bibinfo{editor}{\bibfnamefont{F.}~\bibnamefont{Mila}}, eds.,
  \emph{\bibinfo{title}{Introduction to Frustrated Magnetism}}, vol.
  \bibinfo{volume}{164} of \emph{\bibinfo{series}{Springer Series in
  Solid-State Sciences}} (\bibinfo{publisher}{Springer},
  \bibinfo{address}{Berlin}, \bibinfo{year}{2011}).

\bibitem[{\citenamefont{Xu and Ting}(1990)}]{xu90}
\bibinfo{author}{\bibfnamefont{J.~H.} \bibnamefont{Xu}} \bibnamefont{and}
  \bibinfo{author}{\bibfnamefont{C.~S.} \bibnamefont{Ting}},
  \bibinfo{journal}{Phys. Rev. B} \textbf{\bibinfo{volume}{42}},
  \bibinfo{pages}{6861} (\bibinfo{year}{1990}).

\bibitem[{\citenamefont{Irkhin et~al.}(1992)\citenamefont{Irkhin, Katanin, and
  Katsnelson}}]{irkhi92}
\bibinfo{author}{\bibfnamefont{V.~Y.} \bibnamefont{Irkhin}},
  \bibinfo{author}{\bibfnamefont{A.~A.} \bibnamefont{Katanin}},
  \bibnamefont{and} \bibinfo{author}{\bibfnamefont{M.~I.}
  \bibnamefont{Katsnelson}}, \bibinfo{journal}{J. Phys.: Condens. Mattter}
  \textbf{\bibinfo{volume}{4}}, \bibinfo{pages}{5227} (\bibinfo{year}{1992}).

\bibitem[{\citenamefont{Singh et~al.}(1999)\citenamefont{Singh, Weihong, Hamer,
  and Oitmaa}}]{singh99b}
\bibinfo{author}{\bibfnamefont{R.~R.~P.} \bibnamefont{Singh}},
  \bibinfo{author}{\bibfnamefont{Z.}~\bibnamefont{Weihong}},
  \bibinfo{author}{\bibfnamefont{C.~J.} \bibnamefont{Hamer}}, \bibnamefont{and}
  \bibinfo{author}{\bibfnamefont{J.}~\bibnamefont{Oitmaa}},
  \bibinfo{journal}{Phys. Rev. B} \textbf{\bibinfo{volume}{60}},
  \bibinfo{pages}{7278} (\bibinfo{year}{1999}).

\bibitem[{\citenamefont{Kotov et~al.}(1999)\citenamefont{Kotov, Oitmaa,
  Sushkov, and Zheng}}]{kotov99b}
\bibinfo{author}{\bibfnamefont{V.~N.} \bibnamefont{Kotov}},
  \bibinfo{author}{\bibfnamefont{J.}~\bibnamefont{Oitmaa}},
  \bibinfo{author}{\bibfnamefont{O.~P.} \bibnamefont{Sushkov}},
  \bibnamefont{and} \bibinfo{author}{\bibfnamefont{W.}~\bibnamefont{Zheng}},
  \bibinfo{journal}{Phys. Rev. B} \textbf{\bibinfo{volume}{60}},
  \bibinfo{pages}{14613} (\bibinfo{year}{1999}).

\bibitem[{\citenamefont{Sushkov et~al.}(2001)\citenamefont{Sushkov, Oitmaa, and
  Weihong}}]{sushk01}
\bibinfo{author}{\bibfnamefont{O.~P.} \bibnamefont{Sushkov}},
  \bibinfo{author}{\bibfnamefont{J.}~\bibnamefont{Oitmaa}}, \bibnamefont{and}
  \bibinfo{author}{\bibfnamefont{Z.}~\bibnamefont{Weihong}},
  \bibinfo{journal}{Phys. Rev. B} \textbf{\bibinfo{volume}{63}},
  \bibinfo{pages}{104420} (\bibinfo{year}{2001}).

\bibitem[{\citenamefont{Singh et~al.}(2003)\citenamefont{Singh, Zheng, Oitmaa,
  Sushkov, and Hamer}}]{singh03}
\bibinfo{author}{\bibfnamefont{R.~R.~P.} \bibnamefont{Singh}},
  \bibinfo{author}{\bibfnamefont{W.}~\bibnamefont{Zheng}},
  \bibinfo{author}{\bibfnamefont{J.}~\bibnamefont{Oitmaa}},
  \bibinfo{author}{\bibfnamefont{O.~P.} \bibnamefont{Sushkov}},
  \bibnamefont{and} \bibinfo{author}{\bibfnamefont{C.~J.} \bibnamefont{Hamer}},
  \bibinfo{journal}{Phys. Rev. Lett.} \textbf{\bibinfo{volume}{91}},
  \bibinfo{pages}{017201} (\bibinfo{year}{2003}).

\bibitem[{\citenamefont{Bishop et~al.}(2008)\citenamefont{Bishop, Li, Darradi,
  and Richter}}]{bisho08b}
\bibinfo{author}{\bibfnamefont{R.~F.} \bibnamefont{Bishop}},
  \bibinfo{author}{\bibfnamefont{P.~H.~Y.} \bibnamefont{Li}},
  \bibinfo{author}{\bibfnamefont{R.}~\bibnamefont{Darradi}}, \bibnamefont{and}
  \bibinfo{author}{\bibfnamefont{J.}~\bibnamefont{Richter}},
  \bibinfo{journal}{Europhys. Lett.} \textbf{\bibinfo{volume}{83}},
  \bibinfo{pages}{47004} (\bibinfo{year}{2008}).

\bibitem[{\citenamefont{Uhrig et~al.}(2009)\citenamefont{Uhrig, Holt, Oitmaa,
  Sushkov, and Singh}}]{uhrig09a}
\bibinfo{author}{\bibfnamefont{G.~S.} \bibnamefont{Uhrig}},
  \bibinfo{author}{\bibfnamefont{M.}~\bibnamefont{Holt}},
  \bibinfo{author}{\bibfnamefont{J.}~\bibnamefont{Oitmaa}},
  \bibinfo{author}{\bibfnamefont{O.~P.} \bibnamefont{Sushkov}},
  \bibnamefont{and} \bibinfo{author}{\bibfnamefont{R.~R.~P.}
  \bibnamefont{Singh}}, \bibinfo{journal}{Phys. Rev. B}
  \textbf{\bibinfo{volume}{79}}, \bibinfo{pages}{092416}
  (\bibinfo{year}{2009}).

\bibitem[{\citenamefont{Majumdar}(2010)}]{majum10}
\bibinfo{author}{\bibfnamefont{K.}~\bibnamefont{Majumdar}},
  \bibinfo{journal}{Phys. Rev. B} \textbf{\bibinfo{volume}{82}},
  \bibinfo{pages}{144407} (\bibinfo{year}{2010}).

\bibitem[{\citenamefont{Schmalfu\ss{} et~al.}(2006)\citenamefont{Schmalfu\ss{},
  Darradi, Richter, Schulenburg, and Ihle}}]{schma06}
\bibinfo{author}{\bibfnamefont{D.}~\bibnamefont{Schmalfu\ss{}}},
  \bibinfo{author}{\bibfnamefont{R.}~\bibnamefont{Darradi}},
  \bibinfo{author}{\bibfnamefont{J.}~\bibnamefont{Richter}},
  \bibinfo{author}{\bibfnamefont{J.}~\bibnamefont{Schulenburg}},
  \bibnamefont{and} \bibinfo{author}{\bibfnamefont{D.}~\bibnamefont{Ihle}},
  \bibinfo{journal}{Phys. Rev. Lett.} \textbf{\bibinfo{volume}{97}},
  \bibinfo{pages}{157201} (\bibinfo{year}{2006}).

\bibitem[{\citenamefont{Yao and Carlson}(2010)}]{yao10}
\bibinfo{author}{\bibfnamefont{D.-X.} \bibnamefont{Yao}} \bibnamefont{and}
  \bibinfo{author}{\bibfnamefont{E.~W.} \bibnamefont{Carlson}},
  \bibinfo{journal}{Front. Phys. China} \textbf{\bibinfo{volume}{5}},
  \bibinfo{pages}{166} (\bibinfo{year}{2010}), \eprint{0910.2528v1}.

\bibitem[{\citenamefont{Majumdar}(2011{\natexlab{a}})}]{majum11a}
\bibinfo{author}{\bibfnamefont{K.}~\bibnamefont{Majumdar}},
  \bibinfo{journal}{J. Phys.: Condens. Mattter} \textbf{\bibinfo{volume}{23}},
  \bibinfo{pages}{046001} (\bibinfo{year}{2011}{\natexlab{a}}).

\bibitem[{\citenamefont{Majumdar}(2011{\natexlab{b}})}]{majum11b}
\bibinfo{author}{\bibfnamefont{K.}~\bibnamefont{Majumdar}},
  \bibinfo{journal}{J. Phys.: Condens. Mattter} \textbf{\bibinfo{volume}{23}},
  \bibinfo{pages}{116004} (\bibinfo{year}{2011}{\natexlab{b}}).

\bibitem[{\citenamefont{Holt et~al.}(2011)\citenamefont{Holt, Sushkov, Stanek,
  and Uhrig}}]{holt11}
\bibinfo{author}{\bibfnamefont{M.}~\bibnamefont{Holt}},
  \bibinfo{author}{\bibfnamefont{O.~P.} \bibnamefont{Sushkov}},
  \bibinfo{author}{\bibfnamefont{D.}~\bibnamefont{Stanek}}, \bibnamefont{and}
  \bibinfo{author}{\bibfnamefont{G.~S.} \bibnamefont{Uhrig}},
  \bibinfo{journal}{Phys. Rev. B} \textbf{\bibinfo{volume}{83}},
  \bibinfo{pages}{144528} (\bibinfo{year}{2011}).

\bibitem[{\citenamefont{Betts and Oitmaa}(1977)}]{betts77}
\bibinfo{author}{\bibfnamefont{D.~D.} \bibnamefont{Betts}} \bibnamefont{and}
  \bibinfo{author}{\bibfnamefont{J.}~\bibnamefont{Oitmaa}},
  \bibinfo{journal}{Phys. Lett.} \textbf{\bibinfo{volume}{62A}},
  \bibinfo{pages}{277} (\bibinfo{year}{1977}).

\bibitem[{\citenamefont{Manousakis}(1991)}]{manou91}
\bibinfo{author}{\bibfnamefont{E.}~\bibnamefont{Manousakis}},
  \bibinfo{journal}{Rev. Mod. Phys.} \textbf{\bibinfo{volume}{63}},
  \bibinfo{pages}{1} (\bibinfo{year}{1991}).

\bibitem[{\citenamefont{Auerbach}(1994)}]{auerb94}
\bibinfo{author}{\bibfnamefont{A.}~\bibnamefont{Auerbach}},
  \emph{\bibinfo{title}{Interacting Electrons and Quantum Magnetism}}, Graduate
  Texts in Contemporary Physics (\bibinfo{publisher}{Springer},
  \bibinfo{address}{New York}, \bibinfo{year}{1994}).

\bibitem[{\citenamefont{Chandra and Doucot}(1988)}]{chand88}
\bibinfo{author}{\bibfnamefont{P.}~\bibnamefont{Chandra}} \bibnamefont{and}
  \bibinfo{author}{\bibfnamefont{B.}~\bibnamefont{Doucot}},
  \bibinfo{journal}{Phys. Rev. B} \textbf{\bibinfo{volume}{38}},
  \bibinfo{pages}{9335} (\bibinfo{year}{1988}).

\bibitem[{\citenamefont{Yao and Carlson}(2008)}]{yao08}
\bibinfo{author}{\bibfnamefont{D.-X.} \bibnamefont{Yao}} \bibnamefont{and}
  \bibinfo{author}{\bibfnamefont{E.~W.} \bibnamefont{Carlson}},
  \bibinfo{journal}{Phys. Rev. B} \textbf{\bibinfo{volume}{78}},
  \bibinfo{pages}{052507} (\bibinfo{year}{2008}).

\bibitem[{\citenamefont{Applegate et~al.}(2010)\citenamefont{Applegate, Oitmaa,
  and Singh}}]{apple10}
\bibinfo{author}{\bibfnamefont{R.}~\bibnamefont{Applegate}},
  \bibinfo{author}{\bibfnamefont{J.}~\bibnamefont{Oitmaa}}, \bibnamefont{and}
  \bibinfo{author}{\bibfnamefont{R.~R.~P.} \bibnamefont{Singh}},
  \bibinfo{journal}{Phys. Rev. B} \textbf{\bibinfo{volume}{81}},
  \bibinfo{pages}{024505} (\bibinfo{year}{2010}).

\bibitem[{\citenamefont{Schmidt et~al.}(2010)\citenamefont{Schmidt, Siahatgar,
  and Thalmeier}}]{schmi10}
\bibinfo{author}{\bibfnamefont{B.}~\bibnamefont{Schmidt}},
  \bibinfo{author}{\bibfnamefont{M.}~\bibnamefont{Siahatgar}},
  \bibnamefont{and}
  \bibinfo{author}{\bibfnamefont{P.}~\bibnamefont{Thalmeier}},
  \bibinfo{journal}{Phys. Rev. B} \textbf{\bibinfo{volume}{81}},
  \bibinfo{pages}{165101} (\bibinfo{year}{2010}).

\bibitem[{\citenamefont{Smerald and Shannon}(2010)}]{smera10}
\bibinfo{author}{\bibfnamefont{A.}~\bibnamefont{Smerald}} \bibnamefont{and}
  \bibinfo{author}{\bibfnamefont{N.}~\bibnamefont{Shannon}},
  \bibinfo{journal}{Europhys. Lett.} \textbf{\bibinfo{volume}{92}},
  \bibinfo{pages}{47005} (\bibinfo{year}{2010}).

\bibitem[{\citenamefont{Chandra et~al.}(1990)\citenamefont{Chandra, Coleman,
  and Larkin}}]{chand90}
\bibinfo{author}{\bibfnamefont{P.}~\bibnamefont{Chandra}},
  \bibinfo{author}{\bibfnamefont{P.}~\bibnamefont{Coleman}}, \bibnamefont{and}
  \bibinfo{author}{\bibfnamefont{A.~I.} \bibnamefont{Larkin}},
  \bibinfo{journal}{Phys. Rev. Lett.} \textbf{\bibinfo{volume}{64}},
  \bibinfo{pages}{88} (\bibinfo{year}{1990}).

\bibitem[{\citenamefont{Mermin and Wagner}(1966)}]{mermi66}
\bibinfo{author}{\bibfnamefont{N.~D.} \bibnamefont{Mermin}} \bibnamefont{and}
  \bibinfo{author}{\bibfnamefont{H.}~\bibnamefont{Wagner}},
  \bibinfo{journal}{Phys. Rev. Lett.} \textbf{\bibinfo{volume}{17}},
  \bibinfo{pages}{1133} (\bibinfo{year}{1966}).

\bibitem[{\citenamefont{Weber et~al.}(2003)\citenamefont{Weber, Capriotti,
  Misguich, Becca, Elhajal, and Mila}}]{weber03}
\bibinfo{author}{\bibfnamefont{C.}~\bibnamefont{Weber}},
  \bibinfo{author}{\bibfnamefont{L.}~\bibnamefont{Capriotti}},
  \bibinfo{author}{\bibfnamefont{G.}~\bibnamefont{Misguich}},
  \bibinfo{author}{\bibfnamefont{F.}~\bibnamefont{Becca}},
  \bibinfo{author}{\bibfnamefont{M.}~\bibnamefont{Elhajal}}, \bibnamefont{and}
  \bibinfo{author}{\bibfnamefont{F.}~\bibnamefont{Mila}},
  \bibinfo{journal}{Phys. Rev. Lett.} \textbf{\bibinfo{volume}{91}},
  \bibinfo{pages}{177202} (\bibinfo{year}{2003}).

\bibitem[{\citenamefont{Capriotti et~al.}(2004)\citenamefont{Capriotti, Fubini,
  Roscilde, and Tognetti}}]{capri04}
\bibinfo{author}{\bibfnamefont{L.}~\bibnamefont{Capriotti}},
  \bibinfo{author}{\bibfnamefont{A.}~\bibnamefont{Fubini}},
  \bibinfo{author}{\bibfnamefont{T.}~\bibnamefont{Roscilde}}, \bibnamefont{and}
  \bibinfo{author}{\bibfnamefont{V.}~\bibnamefont{Tognetti}},
  \bibinfo{journal}{Phys. Rev. Lett.} \textbf{\bibinfo{volume}{92}},
  \bibinfo{pages}{157202} (\bibinfo{year}{2004}).

\bibitem[{\citenamefont{Xu et~al.}(2008)\citenamefont{Xu, M\"uller, and
  Sachdev}}]{xu08}
\bibinfo{author}{\bibfnamefont{C.}~\bibnamefont{Xu}},
  \bibinfo{author}{\bibfnamefont{M.}~\bibnamefont{M\"uller}}, \bibnamefont{and}
  \bibinfo{author}{\bibfnamefont{S.}~\bibnamefont{Sachdev}},
  \bibinfo{journal}{Phys. Rev. B} \textbf{\bibinfo{volume}{78}},
  \bibinfo{pages}{020501(R)} (\bibinfo{year}{2008}).

\bibitem[{\citenamefont{Shannon et~al.}(2006)\citenamefont{Shannon, Momoi, and
  Sindzingre}}]{shann06}
\bibinfo{author}{\bibfnamefont{N.}~\bibnamefont{Shannon}},
  \bibinfo{author}{\bibfnamefont{T.}~\bibnamefont{Momoi}}, \bibnamefont{and}
  \bibinfo{author}{\bibfnamefont{P.}~\bibnamefont{Sindzingre}},
  \bibinfo{journal}{Phys. Rev. Lett.} \textbf{\bibinfo{volume}{96}},
  \bibinfo{pages}{027213} (\bibinfo{year}{2006}).

\bibitem[{\citenamefont{Kamihara et~al.}(2008)\citenamefont{Kamihara, Watanabe,
  Hirano, and Hosono}}]{kamih08}
\bibinfo{author}{\bibfnamefont{Y.}~\bibnamefont{Kamihara}},
  \bibinfo{author}{\bibfnamefont{T.}~\bibnamefont{Watanabe}},
  \bibinfo{author}{\bibfnamefont{M.}~\bibnamefont{Hirano}}, \bibnamefont{and}
  \bibinfo{author}{\bibfnamefont{H.}~\bibnamefont{Hosono}},
  \bibinfo{journal}{J. Am. Chem. Soc.} \textbf{\bibinfo{volume}{130}},
  \bibinfo{pages}{3296} (\bibinfo{year}{2008}).

\bibitem[{\citenamefont{{de la Cruz} et~al.}(2008)\citenamefont{{de la Cruz},
  Huang, Lynn, Li, {Ratcliff II}, Zarestky, Mook, Chen, Luo, Wang
  et~al.}}]{cruz08}
\bibinfo{author}{\bibfnamefont{C.}~\bibnamefont{{de la Cruz}}},
  \bibinfo{author}{\bibfnamefont{Q.}~\bibnamefont{Huang}},
  \bibinfo{author}{\bibfnamefont{J.~W.} \bibnamefont{Lynn}},
  \bibinfo{author}{\bibfnamefont{J.}~\bibnamefont{Li}},
  \bibinfo{author}{\bibfnamefont{W.}~\bibnamefont{{Ratcliff II}}},
  \bibinfo{author}{\bibfnamefont{J.~L.} \bibnamefont{Zarestky}},
  \bibinfo{author}{\bibfnamefont{H.~A.} \bibnamefont{Mook}},
  \bibinfo{author}{\bibfnamefont{G.~F.} \bibnamefont{Chen}},
  \bibinfo{author}{\bibfnamefont{J.~L.} \bibnamefont{Luo}},
  \bibinfo{author}{\bibfnamefont{N.~L.} \bibnamefont{Wang}},
  \bibnamefont{et~al.}, \bibinfo{journal}{Nature}
  \textbf{\bibinfo{volume}{453}}, \bibinfo{pages}{899} (\bibinfo{year}{2008}).

\bibitem[{\citenamefont{Diallo et~al.}(2009)\citenamefont{Diallo, Antropov,
  Perring, Broholm, Pulikkotil, Ni, Bud'ko, Canfield, Kreyssig, Goldman
  et~al.}}]{diall09}
\bibinfo{author}{\bibfnamefont{S.~O.} \bibnamefont{Diallo}},
  \bibinfo{author}{\bibfnamefont{V.~P.} \bibnamefont{Antropov}},
  \bibinfo{author}{\bibfnamefont{T.~G.} \bibnamefont{Perring}},
  \bibinfo{author}{\bibfnamefont{C.}~\bibnamefont{Broholm}},
  \bibinfo{author}{\bibfnamefont{J.~J.} \bibnamefont{Pulikkotil}},
  \bibinfo{author}{\bibfnamefont{N.}~\bibnamefont{Ni}},
  \bibinfo{author}{\bibfnamefont{S.~L.} \bibnamefont{Bud'ko}},
  \bibinfo{author}{\bibfnamefont{P.~C.} \bibnamefont{Canfield}},
  \bibinfo{author}{\bibfnamefont{A.}~\bibnamefont{Kreyssig}},
  \bibinfo{author}{\bibfnamefont{A.~I.} \bibnamefont{Goldman}},
  \bibnamefont{et~al.}, \bibinfo{journal}{Phys. Rev. Lett.}
  \textbf{\bibinfo{volume}{102}}, \bibinfo{pages}{187206}
  (\bibinfo{year}{2009}).

\bibitem[{\citenamefont{Zhao et~al.}(2009)\citenamefont{Zhao, Adroja, Yao,
  Bewley, Li, Wang, Wu, Chen, Hu, and Dai}}]{zhao09}
\bibinfo{author}{\bibfnamefont{J.}~\bibnamefont{Zhao}},
  \bibinfo{author}{\bibfnamefont{D.~T.} \bibnamefont{Adroja}},
  \bibinfo{author}{\bibfnamefont{D.-X.} \bibnamefont{Yao}},
  \bibinfo{author}{\bibfnamefont{R.}~\bibnamefont{Bewley}},
  \bibinfo{author}{\bibfnamefont{S.}~\bibnamefont{Li}},
  \bibinfo{author}{\bibfnamefont{X.~F.} \bibnamefont{Wang}},
  \bibinfo{author}{\bibfnamefont{G.}~\bibnamefont{Wu}},
  \bibinfo{author}{\bibfnamefont{X.~H.} \bibnamefont{Chen}},
  \bibinfo{author}{\bibfnamefont{J.}~\bibnamefont{Hu}}, \bibnamefont{and}
  \bibinfo{author}{\bibfnamefont{P.}~\bibnamefont{Dai}}, \bibinfo{journal}{Nat.
  Phys.} \textbf{\bibinfo{volume}{5}}, \bibinfo{pages}{555}
  (\bibinfo{year}{2009}).

\bibitem[{\citenamefont{Ren et~al.}(2008)\citenamefont{Ren, Che, Dong, Yang,
  Lu, Yi, Shen, Li, Sun, Zhou et~al.}}]{ren08}
\bibinfo{author}{\bibfnamefont{Z.~A.} \bibnamefont{Ren}},
  \bibinfo{author}{\bibfnamefont{G.-C.} \bibnamefont{Che}},
  \bibinfo{author}{\bibfnamefont{X.-L.} \bibnamefont{Dong}},
  \bibinfo{author}{\bibfnamefont{J.}~\bibnamefont{Yang}},
  \bibinfo{author}{\bibfnamefont{W.}~\bibnamefont{Lu}},
  \bibinfo{author}{\bibfnamefont{W.}~\bibnamefont{Yi}},
  \bibinfo{author}{\bibfnamefont{X.-L.} \bibnamefont{Shen}},
  \bibinfo{author}{\bibfnamefont{Z.-C.} \bibnamefont{Li}},
  \bibinfo{author}{\bibfnamefont{L.-L.} \bibnamefont{Sun}},
  \bibinfo{author}{\bibfnamefont{F.}~\bibnamefont{Zhou}}, \bibnamefont{et~al.},
  \bibinfo{journal}{Europhys. Lett.} \textbf{\bibinfo{volume}{83}},
  \bibinfo{pages}{17002} (\bibinfo{year}{2008}).

\bibitem[{\citenamefont{Yin et~al.}(2008)\citenamefont{Yin, Leb{\`e}gue, Han,
  Neal, Savrasov, and Pickett}}]{yin08}
\bibinfo{author}{\bibfnamefont{Z.~P.} \bibnamefont{Yin}},
  \bibinfo{author}{\bibfnamefont{S.}~\bibnamefont{Leb{\`e}gue}},
  \bibinfo{author}{\bibfnamefont{M.~J.} \bibnamefont{Han}},
  \bibinfo{author}{\bibfnamefont{B.~P.} \bibnamefont{Neal}},
  \bibinfo{author}{\bibfnamefont{S.~Y.} \bibnamefont{Savrasov}},
  \bibnamefont{and} \bibinfo{author}{\bibfnamefont{W.~E.}
  \bibnamefont{Pickett}}, \bibinfo{journal}{Phys. Rev. Lett.}
  \textbf{\bibinfo{volume}{101}}, \bibinfo{pages}{047001}
  (\bibinfo{year}{2008}).

\bibitem[{\citenamefont{Ewings et~al.}(2008)\citenamefont{Ewings, Perring,
  Bewley, Guidi, Pitcher, Parker, Clarke, and Boothroyd}}]{ewing08}
\bibinfo{author}{\bibfnamefont{R.~A.} \bibnamefont{Ewings}},
  \bibinfo{author}{\bibfnamefont{T.~G.} \bibnamefont{Perring}},
  \bibinfo{author}{\bibfnamefont{R.~I.} \bibnamefont{Bewley}},
  \bibinfo{author}{\bibfnamefont{T.}~\bibnamefont{Guidi}},
  \bibinfo{author}{\bibfnamefont{M.~J.} \bibnamefont{Pitcher}},
  \bibinfo{author}{\bibfnamefont{D.~R.} \bibnamefont{Parker}},
  \bibinfo{author}{\bibfnamefont{S.~J.} \bibnamefont{Clarke}},
  \bibnamefont{and} \bibinfo{author}{\bibfnamefont{A.~T.}
  \bibnamefont{Boothroyd}}, \bibinfo{journal}{Phys. Rev. B}
  \textbf{\bibinfo{volume}{78}}, \bibinfo{pages}{220501}
  (\bibinfo{year}{2008}).

\bibitem[{\citenamefont{Zhao et~al.}(2008{\natexlab{a}})\citenamefont{Zhao,
  {Ratcliff II}, Lynn, Chen, Luo, Wang, Hu, and Dai}}]{zhao08c}
\bibinfo{author}{\bibfnamefont{J.}~\bibnamefont{Zhao}},
  \bibinfo{author}{\bibfnamefont{W.}~\bibnamefont{{Ratcliff II}}},
  \bibinfo{author}{\bibfnamefont{J.~W.} \bibnamefont{Lynn}},
  \bibinfo{author}{\bibfnamefont{G.~F.} \bibnamefont{Chen}},
  \bibinfo{author}{\bibfnamefont{J.~L.} \bibnamefont{Luo}},
  \bibinfo{author}{\bibfnamefont{N.~L.} \bibnamefont{Wang}},
  \bibinfo{author}{\bibfnamefont{J.}~\bibnamefont{Hu}}, \bibnamefont{and}
  \bibinfo{author}{\bibfnamefont{P.}~\bibnamefont{Dai}},
  \bibinfo{journal}{Phys. Rev. B} \textbf{\bibinfo{volume}{78}},
  \bibinfo{pages}{140504} (\bibinfo{year}{2008}{\natexlab{a}}).

\bibitem[{\citenamefont{McQueeney et~al.}(2008)\citenamefont{McQueeney, Diallo,
  Antropov, Samolyuk, Broholm, Ni, Nandi, Yethiraj, Zarestky, Pulikkotil
  et~al.}}]{mcque08}
\bibinfo{author}{\bibfnamefont{R.~J.} \bibnamefont{McQueeney}},
  \bibinfo{author}{\bibfnamefont{S.~O.} \bibnamefont{Diallo}},
  \bibinfo{author}{\bibfnamefont{V.~P.} \bibnamefont{Antropov}},
  \bibinfo{author}{\bibfnamefont{G.~D.} \bibnamefont{Samolyuk}},
  \bibinfo{author}{\bibfnamefont{C.}~\bibnamefont{Broholm}},
  \bibinfo{author}{\bibfnamefont{N.}~\bibnamefont{Ni}},
  \bibinfo{author}{\bibfnamefont{S.}~\bibnamefont{Nandi}},
  \bibinfo{author}{\bibfnamefont{M.}~\bibnamefont{Yethiraj}},
  \bibinfo{author}{\bibfnamefont{J.~L.} \bibnamefont{Zarestky}},
  \bibinfo{author}{\bibfnamefont{J.~J.} \bibnamefont{Pulikkotil}},
  \bibnamefont{et~al.}, \bibinfo{journal}{Phys. Rev. Lett.}
  \textbf{\bibinfo{volume}{101}}, \bibinfo{pages}{227205}
  (\bibinfo{year}{2008}).

\bibitem[{\citenamefont{Si and Abrahams}(2008)}]{si08}
\bibinfo{author}{\bibfnamefont{Q.}~\bibnamefont{Si}} \bibnamefont{and}
  \bibinfo{author}{\bibfnamefont{E.}~\bibnamefont{Abrahams}},
  \bibinfo{journal}{Phys. Rev. Lett.} \textbf{\bibinfo{volume}{101}},
  \bibinfo{pages}{076401} (\bibinfo{year}{2008}).

\bibitem[{\citenamefont{Ong et~al.}(2009)\citenamefont{Ong, Uhrig, and
  Sushkov}}]{ong09}
\bibinfo{author}{\bibfnamefont{A.}~\bibnamefont{Ong}},
  \bibinfo{author}{\bibfnamefont{G.~S.} \bibnamefont{Uhrig}}, \bibnamefont{and}
  \bibinfo{author}{\bibfnamefont{O.~P.} \bibnamefont{Sushkov}},
  \bibinfo{journal}{Phys. Rev. B} \textbf{\bibinfo{volume}{80}},
  \bibinfo{pages}{014514} (\bibinfo{year}{2009}).

\bibitem[{\citenamefont{Yildirim}(2008)}]{yildr08}
\bibinfo{author}{\bibfnamefont{T.}~\bibnamefont{Yildirim}},
  \bibinfo{journal}{Phys. Rev. Lett.} \textbf{\bibinfo{volume}{101}},
  \bibinfo{pages}{057010} (\bibinfo{year}{2008}).

\bibitem[{\citenamefont{Wu et~al.}(2008)\citenamefont{Wu, Phillips, and
  Neto}}]{wu08a}
\bibinfo{author}{\bibfnamefont{J.}~\bibnamefont{Wu}},
  \bibinfo{author}{\bibfnamefont{P.}~\bibnamefont{Phillips}}, \bibnamefont{and}
  \bibinfo{author}{\bibfnamefont{A.~H.~C.} \bibnamefont{Neto}},
  \bibinfo{journal}{Phys. Rev. Lett.} \textbf{\bibinfo{volume}{101}},
  \bibinfo{pages}{126401} (\bibinfo{year}{2008}).

\bibitem[{\citenamefont{Lv et~al.}(2010)\citenamefont{Lv, Kr\"uger, and
  Phillips}}]{lv10}
\bibinfo{author}{\bibfnamefont{W.}~\bibnamefont{Lv}},
  \bibinfo{author}{\bibfnamefont{F.}~\bibnamefont{Kr\"uger}}, \bibnamefont{and}
  \bibinfo{author}{\bibfnamefont{P.}~\bibnamefont{Phillips}},
  \bibinfo{journal}{Phys. Rev. B} \textbf{\bibinfo{volume}{82}},
  \bibinfo{pages}{045125} (\bibinfo{year}{2010}).

\bibitem[{\citenamefont{Pulikkotil et~al.}(2010)\citenamefont{Pulikkotil, Ke,
  {van Schilfgaarde}, Kotani, and Antropov}}]{pulik10}
\bibinfo{author}{\bibfnamefont{J.~J.} \bibnamefont{Pulikkotil}},
  \bibinfo{author}{\bibfnamefont{L.}~\bibnamefont{Ke}},
  \bibinfo{author}{\bibfnamefont{M.}~\bibnamefont{{van Schilfgaarde}}},
  \bibinfo{author}{\bibfnamefont{T.}~\bibnamefont{Kotani}}, \bibnamefont{and}
  \bibinfo{author}{\bibfnamefont{V.~P.} \bibnamefont{Antropov}},
  \bibinfo{journal}{Supercond. Sci. Technol.} \textbf{\bibinfo{volume}{23}},
  \bibinfo{pages}{054012} (\bibinfo{year}{2010}).

\bibitem[{\citenamefont{Zhao et~al.}(2008{\natexlab{b}})\citenamefont{Zhao,
  Yao, Li, Hong, Chen, Chang, {Ratcliff II}, Lynn, Mook, Chen
  et~al.}}]{zhao08a}
\bibinfo{author}{\bibfnamefont{J.}~\bibnamefont{Zhao}},
  \bibinfo{author}{\bibfnamefont{D.-X.} \bibnamefont{Yao}},
  \bibinfo{author}{\bibfnamefont{S.}~\bibnamefont{Li}},
  \bibinfo{author}{\bibfnamefont{T.}~\bibnamefont{Hong}},
  \bibinfo{author}{\bibfnamefont{Y.}~\bibnamefont{Chen}},
  \bibinfo{author}{\bibfnamefont{S.}~\bibnamefont{Chang}},
  \bibinfo{author}{\bibfnamefont{W.}~\bibnamefont{{Ratcliff II}}},
  \bibinfo{author}{\bibfnamefont{J.~W.} \bibnamefont{Lynn}},
  \bibinfo{author}{\bibfnamefont{H.~A.} \bibnamefont{Mook}},
  \bibinfo{author}{\bibfnamefont{G.~F.} \bibnamefont{Chen}},
  \bibnamefont{et~al.}, \bibinfo{journal}{Phys. Rev. Lett.}
  \textbf{\bibinfo{volume}{101}}, \bibinfo{pages}{167203}
  (\bibinfo{year}{2008}{\natexlab{b}}).

\bibitem[{\citenamefont{Han et~al.}(2009)\citenamefont{Han, Yin, Pickett, and
  Savrasov}}]{han09}
\bibinfo{author}{\bibfnamefont{M.~J.} \bibnamefont{Han}},
  \bibinfo{author}{\bibfnamefont{Q.}~\bibnamefont{Yin}},
  \bibinfo{author}{\bibfnamefont{W.~E.} \bibnamefont{Pickett}},
  \bibnamefont{and} \bibinfo{author}{\bibfnamefont{S.~Y.}
  \bibnamefont{Savrasov}}, \bibinfo{journal}{Phys. Rev. Lett.}
  \textbf{\bibinfo{volume}{102}}, \bibinfo{pages}{107003}
  (\bibinfo{year}{2009}).

\bibitem[{\citenamefont{Singh}(2009)}]{singh09a}
\bibinfo{author}{\bibfnamefont{R.~R.~P.} \bibnamefont{Singh}},
  \bibinfo{journal}{arXiv:0903.4408}  (\bibinfo{year}{2009}).

\bibitem[{\citenamefont{Girardeau and Popovic-Boziv}(1977)}]{girar77a}
\bibinfo{author}{\bibfnamefont{M.~D.} \bibnamefont{Girardeau}}
  \bibnamefont{and}
  \bibinfo{author}{\bibfnamefont{M.}~\bibnamefont{Popovic-Boziv}},
  \bibinfo{journal}{J. Phys. C} \textbf{\bibinfo{volume}{10}},
  \bibinfo{pages}{2471} (\bibinfo{year}{1977}).

\bibitem[{\citenamefont{Mila and Zhang}(2000)}]{mila00}
\bibinfo{author}{\bibfnamefont{F.}~\bibnamefont{Mila}} \bibnamefont{and}
  \bibinfo{author}{\bibfnamefont{F.-C.} \bibnamefont{Zhang}},
  \bibinfo{journal}{Eur. Phys. J. B} \textbf{\bibinfo{volume}{16}},
  \bibinfo{pages}{7} (\bibinfo{year}{2000}).

\bibitem[{\citenamefont{Hamerla et~al.}(2010)\citenamefont{Hamerla, Duffe, and
  Uhrig}}]{hamer10}
\bibinfo{author}{\bibfnamefont{S.~A.} \bibnamefont{Hamerla}},
  \bibinfo{author}{\bibfnamefont{S.}~\bibnamefont{Duffe}}, \bibnamefont{and}
  \bibinfo{author}{\bibfnamefont{G.~S.} \bibnamefont{Uhrig}},
  \bibinfo{journal}{Phys. Rev. B} \textbf{\bibinfo{volume}{82}},
  \bibinfo{pages}{235117} (\bibinfo{year}{2010}).

\bibitem[{\citenamefont{Yaresko et~al.}(2009)\citenamefont{Yaresko, Liu,
  Antonov, and Andersen}}]{yares09}
\bibinfo{author}{\bibfnamefont{A.~N.} \bibnamefont{Yaresko}},
  \bibinfo{author}{\bibfnamefont{G.-Q.} \bibnamefont{Liu}},
  \bibinfo{author}{\bibfnamefont{V.~N.} \bibnamefont{Antonov}},
  \bibnamefont{and} \bibinfo{author}{\bibfnamefont{O.~K.}
  \bibnamefont{Andersen}}, \bibinfo{journal}{Phys. Rev. B}
  \textbf{\bibinfo{volume}{79}}, \bibinfo{pages}{144421}
  (\bibinfo{year}{2009}).

\bibitem[{\citenamefont{Graser et~al.}(2010)\citenamefont{Graser, Kemper,
  Maier, Cheng, Hirschfeld, and Scalapino}}]{grase10}
\bibinfo{author}{\bibfnamefont{S.}~\bibnamefont{Graser}},
  \bibinfo{author}{\bibfnamefont{A.~F.} \bibnamefont{Kemper}},
  \bibinfo{author}{\bibfnamefont{T.~A.} \bibnamefont{Maier}},
  \bibinfo{author}{\bibfnamefont{H.-P.} \bibnamefont{Cheng}},
  \bibinfo{author}{\bibfnamefont{P.~J.} \bibnamefont{Hirschfeld}},
  \bibnamefont{and} \bibinfo{author}{\bibfnamefont{D.~J.}
  \bibnamefont{Scalapino}}, \bibinfo{journal}{Phys. Rev. B}
  \textbf{\bibinfo{volume}{81}}, \bibinfo{pages}{214503}
  (\bibinfo{year}{2010}).

\bibitem[{\citenamefont{Schickling et~al.}(2011)\citenamefont{Schickling,
  Gebhard, and B\"unemann}}]{schic11}
\bibinfo{author}{\bibfnamefont{T.}~\bibnamefont{Schickling}},
  \bibinfo{author}{\bibfnamefont{F.}~\bibnamefont{Gebhard}}, \bibnamefont{and}
  \bibinfo{author}{\bibfnamefont{J.}~\bibnamefont{B\"unemann}},
  \bibinfo{journal}{Phys. Rev. Lett.} \textbf{\bibinfo{volume}{106}},
  \bibinfo{pages}{146402} (\bibinfo{year}{2011}).

\bibitem[{\citenamefont{Dyson}(1956{\natexlab{a}})}]{dyson56a}
\bibinfo{author}{\bibfnamefont{F.~J.} \bibnamefont{Dyson}},
  \bibinfo{journal}{Phys. Rev.} \textbf{\bibinfo{volume}{102}},
  \bibinfo{pages}{1217} (\bibinfo{year}{1956}{\natexlab{a}}).

\bibitem[{\citenamefont{Dyson}(1956{\natexlab{b}})}]{dyson56b}
\bibinfo{author}{\bibfnamefont{F.~J.} \bibnamefont{Dyson}},
  \bibinfo{journal}{Phys. Rev.} \textbf{\bibinfo{volume}{102}},
  \bibinfo{pages}{1230} (\bibinfo{year}{1956}{\natexlab{b}}).

\bibitem[{\citenamefont{Maleev}(1958)}]{malee58}
\bibinfo{author}{\bibfnamefont{S.}~\bibnamefont{Maleev}},
  \bibinfo{journal}{Sov. Phys. JETP} \textbf{\bibinfo{volume}{6}},
  \bibinfo{pages}{776} (\bibinfo{year}{1958}).

\bibitem[{\citenamefont{Fetter and Walecka}(1971)}]{fette71}
\bibinfo{author}{\bibfnamefont{A.~L.} \bibnamefont{Fetter}} \bibnamefont{and}
  \bibinfo{author}{\bibfnamefont{J.~D.} \bibnamefont{Walecka}},
  \emph{\bibinfo{title}{Quantum Theory of Many-Particle Systems}},
  International Series in Pure and Applied Physics
  (\bibinfo{publisher}{McGraw-Hill, New York}, \bibinfo{year}{1971}).

\bibitem[{\citenamefont{Lange}(1966)}]{lange66}
\bibinfo{author}{\bibfnamefont{R.~V.} \bibnamefont{Lange}},
  \bibinfo{journal}{Phys. Rev.} \textbf{\bibinfo{volume}{146}},
  \bibinfo{pages}{301} (\bibinfo{year}{1966}).

\bibitem[{\citenamefont{Auerbach and Arovas}(1988)}]{auerb88}
\bibinfo{author}{\bibfnamefont{A.}~\bibnamefont{Auerbach}} \bibnamefont{and}
  \bibinfo{author}{\bibfnamefont{D.~P.} \bibnamefont{Arovas}},
  \bibinfo{journal}{Phys. Rev. Lett.} \textbf{\bibinfo{volume}{61}},
  \bibinfo{pages}{617} (\bibinfo{year}{1988}).

\bibitem[{\citenamefont{Arovas and Auerbach}(1988)}]{arova88}
\bibinfo{author}{\bibfnamefont{D.~P.} \bibnamefont{Arovas}} \bibnamefont{and}
  \bibinfo{author}{\bibfnamefont{A.}~\bibnamefont{Auerbach}},
  \bibinfo{journal}{Phys. Rev. B} \textbf{\bibinfo{volume}{38}},
  \bibinfo{pages}{316} (\bibinfo{year}{1988}).

\bibitem[{\citenamefont{Raas}(2009)}]{Raas2009}
\bibinfo{author}{\bibfnamefont{C.}~\bibnamefont{Raas}},
  \bibinfo{howpublished}{Private communication} (\bibinfo{year}{2009}).

\bibitem[{\citenamefont{Oitmaa et~al.}(2006)\citenamefont{Oitmaa, Hamer, and
  Zheng}}]{oitma06}
\bibinfo{author}{\bibfnamefont{J.}~\bibnamefont{Oitmaa}},
  \bibinfo{author}{\bibfnamefont{C.~J.} \bibnamefont{Hamer}}, \bibnamefont{and}
  \bibinfo{author}{\bibfnamefont{W.}~\bibnamefont{Zheng}},
  \emph{\bibinfo{title}{Series Expansion Methods for Strongly Interacting
  Lattice Models}} (\bibinfo{publisher}{Cambridge Univ. Press},
  \bibinfo{address}{Cambridge}, \bibinfo{year}{2006}).

\bibitem[{\citenamefont{Oitmaa}(2009)}]{Oitmaa2009}
\bibinfo{author}{\bibfnamefont{J.}~\bibnamefont{Oitmaa}},
  \bibinfo{howpublished}{Private communication} (\bibinfo{year}{2009}).

\bibitem[{\citenamefont{Stanek}(2010)}]{stane10}
\bibinfo{author}{\bibfnamefont{D.}~\bibnamefont{Stanek}},
  \bibinfo{type}{Diploma thesis}, \bibinfo{school}{TU Dortmund}
  (\bibinfo{year}{2010}),
  \urlprefix\url{http://t1.physik.tu-dortmund.de/uhrig/diploma.html}.

\bibitem[{\citenamefont{Raghu et~al.}(2008)\citenamefont{Raghu, Qi, Liu,
  Scalapino, and Zhang}}]{raghu08}
\bibinfo{author}{\bibfnamefont{S.}~\bibnamefont{Raghu}},
  \bibinfo{author}{\bibfnamefont{X.-L.} \bibnamefont{Qi}},
  \bibinfo{author}{\bibfnamefont{C.-X.} \bibnamefont{Liu}},
  \bibinfo{author}{\bibfnamefont{D.~J.} \bibnamefont{Scalapino}},
  \bibnamefont{and} \bibinfo{author}{\bibfnamefont{S.-C.} \bibnamefont{Zhang}},
  \bibinfo{journal}{Phys. Rev. B} \textbf{\bibinfo{volume}{77}},
  \bibinfo{pages}{220503} (\bibinfo{year}{2008}).

\bibitem[{\citenamefont{Lee et~al.}(2010{\natexlab{a}})\citenamefont{Lee,
  Zhang, Jeschke, Valent{\'i}, and Monien}}]{lee10a}
\bibinfo{author}{\bibfnamefont{H.}~\bibnamefont{Lee}},
  \bibinfo{author}{\bibfnamefont{Y.-Z.} \bibnamefont{Zhang}},
  \bibinfo{author}{\bibfnamefont{H.~O.} \bibnamefont{Jeschke}},
  \bibinfo{author}{\bibfnamefont{R.}~\bibnamefont{Valent{\'i}}},
  \bibnamefont{and} \bibinfo{author}{\bibfnamefont{H.}~\bibnamefont{Monien}},
  \bibinfo{journal}{Phys. Rev. Lett.} \textbf{\bibinfo{volume}{104}},
  \bibinfo{pages}{026402} (\bibinfo{year}{2010}{\natexlab{a}}).

\bibitem[{\citenamefont{Lee et~al.}(2010{\natexlab{b}})\citenamefont{Lee,
  Zhang, Jeschke, and Valent{\'i}}}]{lee10b}
\bibinfo{author}{\bibfnamefont{H.}~\bibnamefont{Lee}},
  \bibinfo{author}{\bibfnamefont{Y.-Z.} \bibnamefont{Zhang}},
  \bibinfo{author}{\bibfnamefont{H.~O.} \bibnamefont{Jeschke}},
  \bibnamefont{and}
  \bibinfo{author}{\bibfnamefont{R.}~\bibnamefont{Valent{\'i}}},
  \bibinfo{journal}{Phys. Rev. B} \textbf{\bibinfo{volume}{81}},
  \bibinfo{pages}{220506} (\bibinfo{year}{2010}{\natexlab{b}}).

\bibitem[{\citenamefont{Bao et~al.}(2011)\citenamefont{Bao, Huang, Chen, Green,
  Wang, He, Wang, and Qiu}}]{bao11a}
\bibinfo{author}{\bibfnamefont{W.}~\bibnamefont{Bao}},
  \bibinfo{author}{\bibfnamefont{Q.}~\bibnamefont{Huang}},
  \bibinfo{author}{\bibfnamefont{G.~F.} \bibnamefont{Chen}},
  \bibinfo{author}{\bibfnamefont{M.~A.} \bibnamefont{Green}},
  \bibinfo{author}{\bibfnamefont{D.~M.} \bibnamefont{Wang}},
  \bibinfo{author}{\bibfnamefont{J.~B.} \bibnamefont{He}},
  \bibinfo{author}{\bibfnamefont{X.~Q.} \bibnamefont{Wang}}, \bibnamefont{and}
  \bibinfo{author}{\bibfnamefont{Y.}~\bibnamefont{Qiu}},
  \bibinfo{journal}{1102.0830}  (\bibinfo{year}{2011}).

\bibitem[{\citenamefont{Li et~al.}(2011)\citenamefont{Li, Ma, Pang, and
  Li}}]{li11a}
\bibinfo{author}{\bibfnamefont{Z.}~\bibnamefont{Li}},
  \bibinfo{author}{\bibfnamefont{X.}~\bibnamefont{Ma}},
  \bibinfo{author}{\bibfnamefont{H.}~\bibnamefont{Pang}}, \bibnamefont{and}
  \bibinfo{author}{\bibfnamefont{F.}~\bibnamefont{Li}}, p.
  \bibinfo{pages}{1103.0098} (\bibinfo{year}{2011}).

\bibitem[{\citenamefont{Pomjakushin et~al.}(2011)\citenamefont{Pomjakushin,
  Pomjakushina, Krzton-Maziopa, Conder, and Shermadini}}]{pomja11a}
\bibinfo{author}{\bibfnamefont{V.~Y.} \bibnamefont{Pomjakushin}},
  \bibinfo{author}{\bibfnamefont{E.~V.} \bibnamefont{Pomjakushina}},
  \bibinfo{author}{\bibfnamefont{A.}~\bibnamefont{Krzton-Maziopa}},
  \bibinfo{author}{\bibfnamefont{K.}~\bibnamefont{Conder}}, \bibnamefont{and}
  \bibinfo{author}{\bibfnamefont{Z.}~\bibnamefont{Shermadini}}, p.
  \bibinfo{pages}{1102.3380} (\bibinfo{year}{2011}).

\bibitem[{\citenamefont{Ryan et~al.}(2011)\citenamefont{Ryan,
  Rowan-Weetaluktuk, Cadogan, Hu, Straszheim, Bud�ko, and
  Canfield}}]{ryan11a}
\bibinfo{author}{\bibfnamefont{D.}~\bibnamefont{Ryan}},
  \bibinfo{author}{\bibfnamefont{W.}~\bibnamefont{Rowan-Weetaluktuk}},
  \bibinfo{author}{\bibfnamefont{J.}~\bibnamefont{Cadogan}},
  \bibinfo{author}{\bibfnamefont{R.}~\bibnamefont{Hu}},
  \bibinfo{author}{\bibfnamefont{W.}~\bibnamefont{Straszheim}},
  \bibinfo{author}{\bibfnamefont{S.}~\bibnamefont{Bud�ko}}, \bibnamefont{and}
  \bibinfo{author}{\bibfnamefont{P.}~\bibnamefont{Canfield}}, p.
  \bibinfo{pages}{1103.0059} (\bibinfo{year}{2011}).

\bibitem[{\citenamefont{Knolle et~al.}(2010)\citenamefont{Knolle, Eremin,
  Chubukov, and Moessner}}]{knoll10b}
\bibinfo{author}{\bibfnamefont{J.}~\bibnamefont{Knolle}},
  \bibinfo{author}{\bibfnamefont{I.}~\bibnamefont{Eremin}},
  \bibinfo{author}{\bibfnamefont{A.~V.} \bibnamefont{Chubukov}},
  \bibnamefont{and} \bibinfo{author}{\bibfnamefont{R.}~\bibnamefont{Moessner}},
  \bibinfo{journal}{Phys. Rev. B} \textbf{\bibinfo{volume}{81}},
  \bibinfo{pages}{140506(R)} (\bibinfo{year}{2010}).

\bibitem[{\citenamefont{Goswami et~al.}(2011)\citenamefont{Goswami, Yu, Si, and
  Abrahams}}]{goswa11}
\bibinfo{author}{\bibfnamefont{P.}~\bibnamefont{Goswami}},
  \bibinfo{author}{\bibfnamefont{R.}~\bibnamefont{Yu}},
  \bibinfo{author}{\bibfnamefont{Q.}~\bibnamefont{Si}}, \bibnamefont{and}
  \bibinfo{author}{\bibfnamefont{E.}~\bibnamefont{Abrahams}}, p.
  \bibinfo{pages}{1009.1111} (\bibinfo{year}{2011}).

\end{thebibliography}

\end{document}